\documentstyle[aps,epsfig]{revtex}

\begin{document}

{\large

\begin{center}
{\bf {\Large
Quantization and renormalization of the manifest left-right symmetric
model \\
of electroweak interactions.}}
\end{center}

\vspace*{3cm}
\begin{center}
{\Large
P. Duka$^1$, J. Gluza$^{1,2}$ and M. Zra\l ek$^1$.} \\
\vspace*{0.5cm}
$^1$ Department of Field Theory and Particle Physics \\
Institute of Physics, University of Silesia\\
Uniwersytecka 4, PL-40-007 Katowice, Poland \\
\vspace*{0.3cm}
$^2$ DESY Zeuthen, Platanenalle 6, 15738 Zeuthen, Germany
\end{center}
}

\vspace*{3cm}

\hspace*{8cm}  {\large {\bf Abstract }}
\begin{abstract}
{\large Quantization and renormalization of the left-right symmetric model is the
main purpose of the paper. First the model at tree level with a Higgs sector
containing one bidoublet and two triplets is precisely discussed. Then the
canonical quantization and Faddeev-Popov Lagrangian are carried out
('t Hooft gauge). The BRST symmetry is discussed.
Subsequently the on mass shell
renormalization is performed and, as a test of consistency, the
renormalization of the $Z N_i N_j$ vertex is analyzed.}
\end{abstract}

%PACS number(s): 13.15.-f, 12.15.Cc,14.60.Gh

\baselineskip 6mm
\newpage

\textwidth 18cm
\textheight 21cm

%%%%%%%%%%%%%%%%%%%%%%%%%%%%%%%%%%%%%%%%%%%%%%%%%%%%%%%%%%%%%%%%%%%%%%%%%%%
%\section{Introduction}
\ \ \
%%%%%%%%%%%%%%%%%%%%%%%%%%%%%%%%%%%%%%%%%%%%%%%%%%%%%%%%%%%%%%%%%%%%%%%%%%%%
%\section{Full Lagrangian for the l-r symmetric model}
%%%%%%%%%%%%%%%%%%%%%%%%%%%%%%%%%%%%%%%%%%%%%%%%%%%%%%%%%%%%%%%%%%%%%%%%%%%%
\part*{I Introduction.}

The Standard Model (SM) agrees   with all ever done experiments on both classical
and quantum levels \cite{holik}. Any announcement of detection of a new,
non-standard signal has, sooner or later, died out.
For instance, world-wide discussed discrepancies at LEP
 have disappeared when more data have been collected and
analyzed \cite{altarelli}.
However, recent Superkamiokande (SK) atmospheric observations \cite{supk1}
show (indirectly) strong
evidence for neutrino mass. This has been recognized as a first
 positive sign of
 physics beyond the orthodox SM.
Nonzero neutrino masses
are quite natural in the electroweak theory.
With the
known left-handed neutrinos and the standard Higgs doublet massive neutrinos
appear from higher dimension operators acting at the Planck scale
\cite{shr}.
However, expected effects of quantum gravity in SM induce Majorana masses,
few order of magnitudes smaller than we need to explain SK data \cite{pati1}.
In these circumstances right-handed neutrinos seem to be
inevitable in the theory and open the route to higher unification e.g. via
left-right symmetric gauge groups \cite{pati1,pati2}.
Whether such models are responsible for the SK signal
is still an open question.
Putting aside the problem of neutrino masses, left right models have  many
other
interesting and nonstandard features. Actually,
the theoretical concept of the left-right symmetry in physics at very high
energies has already been invented almost 25 years
ago simultaneously with the idea of GUT models \cite{pati2,georgi}.

The smallest gauge group which implements the hypothesis of left-right
symmetry of weak interactions is

\begin{equation}
\label{symmetry}
SU(2)_L \otimes SU(2)_R \otimes U(1)_{B-L}.
\end{equation}

This gauge group can be understood as a second step (after the SM)
in unifying fundamental interactions and
 unravels several puzzles of the SM.

First of all it restores quark-lepton symmetry. In the SM both left-handed
neutrinos and charged leptons of the same flavour are elements of the same
object with respect to the SU(2) gauge group. The same happens with
left-handed up and down quarks. However, right-handed fields i.e. quarks and
leptons are not treated in the same way: they are singlet fields
with respect to the
SU(2) gauge group.
%\footnote{Traditionally there are no right-handed neutrinos in the
%SM. We can safely resign from this
%assumption.}.
In the left-right symmetric models both left-handed and right-handed fields
are treated in the same way - they form doublets under SU(2)  left- and
SU(2) right- gauge groups.
Moreover, violation of the space
inversion symmetry is not an ad hoc assumption but follows from
vacuum structure of the theory.
At high energy parity is conserved, but at some
energy scale $v_R$ (connected with the masses of heavy gauge bosons)
the space inversion symmetry is broken spontaneously.
Observed near maximal parity
violation at low energies is explained by the large difference between
masses of heavy and light gauge bosons.
The suppression of the right-handed charged currents (not
visible up to now) is linked with the high-low  (left-right)
energy gap, too.

Phenomena mentioned above are generic to the left-right symmetric  unified
models. There are also other important features of the model which appear
naturally and are able to solve some unanswered SM's problems.
Among them the following issues can be specified:
\begin{itemize}
\item[i)] Small masses of light neutrinos.

In the conventional version of the L-R model \cite{pati3}
we have three light and three heavy (presumably Majorana) neutrinos.
Small masses of known neutrinos can be explained
via so-called seesaw mechanism \cite{yanagida}.

\item[ii)] Physical interpretation of the $U(1)$ generator as the B-L quantum
number \cite{bl}

 The SM has an anomaly free global B-L symmetry. The
right-handed neutrinos create the B-L symmetry to be local and anomaly free.
Such a symmetry is gaugeable which makes it possible to replace the arbitrary
weak hypercharge Y of the SM by better known and invented earlier
B(aryon) and L(epton) numbers. Then the electric charges
of particles are connected
with known quantities, eigenvalues of the left ($T_{3L}$) and right
($T_{3R}$)
generators of the $SU_L(2)$ and $SU_R(2)$ groups, respectively in the way:

\begin{equation}
\label{q}
Q=T_{3L}+T_{3R}+\frac{B-L}{2}.
\end{equation}

\item[iii)] Charge quantization \cite{babu},

If neutrinos are
Majorana particles (Dirac neutrinos are not so natural) then
the B-L number is violated and the requirement that the theory must be anomaly
free (which is necessary for renormalizability) leads to charge
quantization.

\item[iv)] Understanding of the smallness of CP violation in the quark sector
\cite{sup}.

In some versions of the L-R models phases responsible for CP violation are
connected with the right-handed quark sector. Then the smallness of the CP
effect is related directly to the suppression of the V+A current:
the bigger $M_{W_2}$, the smaller CP violation.
%$\eta_{+-}=A(K_L^0 \rightarrow \pi^+ \pi^-)/A(K_S^0 \rightarrow \pi^+ \pi^-)
%$
%and
%$\eta_{+-}=A(K_L^0 \rightarrow \pi^+ \pi^- \pi^0)/A(K_S^0 \rightarrow \pi^+ \p%i^-\pi^0)$

\item[v)] Solution of the strong CP problem \cite{strongcp}.

As it was mentioned above mirror symmetry is an exact  symmetry in the L-R models
before spontaneous breakdown. Then the strong CP parameter $\Theta$ changes
under parity to $-\Theta$, thus in the L-R models $\Theta=0$ naturally.
Small, finite contributions to $\Theta$ arise at two loop level.
\end{itemize}

Not all puzzles of the SM are solved. Some of them still remain. The large
number of parameters (masses and mixing angles) and the fine-tuning problems
(the naturality problem and the cosmological constant problem) wait for
disentanglement. Supersymmetric version of the model finds the answer to
some questions (e.q. naturalness problem) but still more of them wait
for an explanation. We may only hope that all these doubts will be
explained in the future.

Plenty of different L-R symmetric models based on the gauge group Eq. (1)
have appeared in the literature \cite{theor}.
They have usually been used to find out new
phenomenological phenomena connected with
\begin{itemize}
\item     extra gauge bosons  at lepton \cite{gau1,rep,snow} and hadron
        colliders \cite{snow,gau2,buch};
\item heavy neutrinos at $e^+e^-$ \cite{rep,buch,heavy1,heavy1a,mzr},
$e^- \gamma$ \cite{mzr} and hadron colliders \cite{heavy2};
\item  doubly \cite{hunt,doubly}, singly charged \cite{hunt,singly,oln}
and neutral     \cite{hunt,oln,neutr}
       Higgs particles;
\item lepton number violating processes at both high
      \cite{high} and low energies \cite{viol};
\item CP effects in Higgs \cite{coc}, lepton \cite{cplep} and quark
\cite{sup},\cite{qlep} sectors;
\item electromagnetic properties \cite{e-m} and nature \cite{nat} of neutrinos;
\item  flavour changing neutral processes (FCNC) \cite{fcnc};
\item Kaon \cite{kaon,maal} and B \cite{b} physics.
\end{itemize}

We are aware of only few  papers where radiative corrections to
the LR model have been used \cite{sok,pil}. Typical non standard
one loop diagrams have been  used very rarely (finite box
diagrams) \cite{maal}. Wherever needed, only the SM virtual
corrections have been considered \cite{kaon,rai}. Such a
(misleading) procedure has been  justified by the argument that
corrections to small non standard tree level diagrams are
negligible (see for contrary arguments in \cite{mich,new}).
However, the experimental results (for example from LEP I and II)
are getting better and better. We also do not know the role of the
Higgs sector in the case of strongly broken `custodial' symmetry.
Moreover, it was found \cite{sok} that in left-right models heavy
boson scalars do not decouple in low-energy phenomena. In such
circumstances a natural question arises whether non standard
corrections are really negligible. To answer that question it is
necessary to calculate them. To make it properly the model has to
be renormalized (see \cite{mich} for another renormalization
approach to the LR model).

Since 1971 we know \cite{hooft} how to build a renormalizable
gauge field theory. The L-R gauge model satisfies all its
requirements so it is a renormalizable model. Although
theoretically understood, the problem is how to make it in
practice, especially when much more new quantities in comparison
to the SM must be worked out. The problem is not only with the
larger number of Feynman diagrams which have to be added for each
quantity. New important renormalization constants such as new
heavy gauge and Higgs fields appear. There are also two new mixing
angles between charged and neutral gauge bosons. Mixing matrices
in quark \cite{sack} and lepton \cite{kniehl} sectors have to be
renormalized. Neutrinos are Majorana particles so little more
complicated techniques have to be applied \cite{gl5}.

At the beginning of the development of the L-R models
the breaking of the gauge symmetry
was implemented by choosing two Higgs doublets $\chi_L$(2,0,2) and
$\chi_R$(0,2,2) and the bidoublet $\phi$(2,2,0) \cite{mohap}.
Later on it was realized  that choosing, instead of two doublets,
two Higgs triplets
$\Delta_L$(3,1,2) and $\Delta_R$(1,3,2) the smallness of neutrino masses
may be connected with the large scale of right-handed symmetry
breaking \cite{senj,deshpande}.
For many different purposes also other Higgs sectors were considered.
It made it possible to understand a small mass for neutrinos by
radiative corrections, to have
see-saw mechanism for Dirac neutrinos, or to eliminate the FCNC in quark
interaction. There were introduced Higgs sectors with
(i) bidoublet, doublets and triplets\cite{bdt},
(ii) doublets and triplets \cite{dt},
(iii) doublets and singlets \cite{ds}.
We will focus here on the model where two triplets and bidoublet form the Higgs
sector and where P symmetry prior to symmetry breakdown is assumed (for models
without explicit P symmetry, see e.g.   \cite{langacker,olness}).
This is a realization of the so called Manifest (or quasi-manifest)
L-R symmetric model where
the gauge couplings $g_L$ and $g_R$ of the $SU(2)_L$ and $SU(2)_R$
subgroups, respectively, are equal \cite{pati,mohapatra1,beg}
and Yukawa couplings form hermitian matrices \cite{man}.

At the tree level this minimal (manifest or quasi-manifest)
L-R symmetric model has been analyzed extensively from both theoretical
\cite{grimus}
and phenomenological point of views \cite{rai},\cite{pol1},\cite{gluza}.

In order to be consistent with observed phenomena, the symmetry breaking
pattern in our model should satisfy inequalities: $v_R >>
\sqrt{\kappa_1^2+\kappa_2^2} >>v_L$ where $v_R,\; \kappa_1,\;
\kappa_2,\;v_L$ are the VEVs of appropriate fields (see Chapter 1).
To avoid fine-tuning problems with the model parameters it was shown
that $v_L=0$ should be imposed
\cite{deshpande,gunion}.  In what follows we also require vanishing of
VEV for the left-handed neutral triplet field, $<\Delta_L>_0=0$.  The
most general form of the L-R symmetric Yukawa potential is worked out,
however, with one assumption that there is no hard CP violation in the Higgs
sector \cite{deshpande}.

In the next Chapter we describe in full details the model
at the tree level. First, the Lagrangian in the weak
basis before diagonalization is discussed. Then the Higgs potential
is described. The fermion and boson mass matrices are diagonalized,
the L-R model Lagrangian in the physical basis and CP symmetry are considered.
Presented
couplings among the physical fields are very convenient in practical
model applications. In Chapter 3 the procedure of the L-R
model quantization is presented. The linear gauge fixing terms are
introduced (the 'tHooft gauge).
The Goldstone bosons and the Faddeev-Popov ghosts are defined.
Next BRST transformations for all physical and unphysical fields
are introduced and afterwards the
 Faddeev-Popov ghost Lagrangian is given.
In Chapter 4 the renormalization procedure is presented.
All renormalization constants for the fields,
masses, mixing parameters are defined. The renormalization of the charged
and neutral current mixing matrices is described. Finally in
Chapter 5, to check the consistency of the procedures the renomalization
of the $Z_1N_iN_j$ vertex is performed.
At the beginning the general scheme and
calculations of necessary renormalization constants are presented.
Then, the cancellation of the infinite part of the $Z_1N_iN_j$ vertex
is analyzed.

\part*{II Model at the tree level.}

We write down below the full Lagrangian at the tree level which
we use in the paper. First we present the Lagrangian in a weak
unphysical base. Then, after mass matrices diagonalization,
the Lagrangian in the physical base is given.

\section*{ 2.1 Structure of the Lagrangian of the theory.}

\subsection*{2.1.1 Fermion and gauge fields.}

In the $SU(2)_L \otimes SU(2)_R \otimes U(1)_{B-L}$ gauge model quarks (Q)
and leptons (L) are placed in doublets \cite{grimus}

\begin{equation}
\label{L}
L_{iL}=\left( \matrix{ \nu_i' \cr l_i' } \right)_L \;\;:\;(2,1,-1),\;\;\;\;
L_{iR}=\left( \matrix{ \nu_i' \cr l_i' } \right)_R :\;\;\;(1,2,-1),
\end{equation}
\begin{equation}
\label{Q}
Q_{iL}=\left( \matrix{ u_i' \cr d_i' } \right)_L \;\;:\;(2,1,1/3),\;\;\;\;
Q_{iR}=\left( \matrix{ u_i' \cr d_i' } \right)_R :\;\;\;(1,2,1/3).
\end{equation}

i=1,2,3 runs over number of generations. The numbers ($d_L, d_R,
Y$) in parenthesis characterize the $SU(2)_L$, $SU(2)_R$ and
$U(1)_{B-L}$ representation. $d_{L,R}$ denote dimensions of the
$SU(2)_L$ and $SU(2)_R$ representation, $Y=B-L$. The quantum
numbers for $U(1)_{B-L}$ gauge group are connected with charges of
the particles by Eq.(\ref{q}). The fermion fields (Eq. (\ref{L})
and (\ref{Q})) have the following gauge transformations
$(\Psi_{L,R}=(Q,L)_{L,R})$

\begin{eqnarray}
\label{psigauge}
\Psi_L^{\prime}& =& \left[ e^{-ig^{\prime}\frac{Y}{2}\Theta (x)}
e^{-ig_L\frac{\vec{\tau}}{2}\vec{\Theta}(x)}
\right] \Psi_L,
\nonumber \\
\Psi_R^{\prime}& =& \left[e^{-ig^{\prime}\frac{Y}{2}\Theta (x)}
e^{-ig_R\frac{\vec{\tau}}{2}\vec{\Theta}(x)} \right] \Psi_R.
\end{eqnarray}

Gauge invariance is achieved when 7 gauge fields ${(\vec{W}_{L,R})}_{\mu},
B_{\mu}$ are introduced. Then, appropriate covariant derivatives for $%
\Psi_{L,R}$ fields are

\begin{eqnarray}
D_{\mu}\Psi_L& =& \left( \partial_{\mu}-ig_L\frac{\vec{\tau}}{2}\vec{W}%
_{L\mu} -ig^{\prime}\frac{Y}{2}B_{\mu} \right) \Psi_L, \\ \nonumber
D_{\mu}\Psi_R& =& \left( \partial_{\mu}-ig_R\frac{\vec{\tau}}{2}\vec{W}%
_{R\mu} -ig^{\prime}\frac{Y}{2}B_{\mu} \right) \Psi_R,
\end{eqnarray}

with the following gauge fields transformations

\begin{eqnarray}
\left( \frac{\vec{\tau}}{2}\vec{W}_{L,R}\right)_{\mu}^{\prime}&=&
e^{-ig_{L,R}\frac{\vec{\tau}}{2}\vec{\Theta_{L,R}}(x)}
\left( \frac{\vec{\tau}}{2}
\vec{W}_{L,R}\right)_{\mu}e^{ig_{L,R}\frac{\vec{\tau}}{2}
\vec{\Theta_{L,R}}(x)}
\nonumber \\
&-& \frac{1}{ig_{L,R}}e^{-ig_{L,R}\frac{\vec{\tau}}{2}\vec{\Theta_{L,R}}(x)}
\partial_{\mu} \left( e^{ig_{L,R}\frac{\vec{\tau}}{2}
\vec{\Theta_{L,R}}(x)} \right),
\\
\label{B}
B_{\mu}^{\prime}&=&B_{\mu}-\frac{1}{ig}e^{-ig^{\prime}\frac{Y}{2}\vec{\Theta}%
(x)}\partial_{\mu} \left( e^{ig^{\prime}\frac{Y}{2}\vec{\Theta}(x)} \right).
\end{eqnarray}

We start with the three different couplings $g_{L}$, $g_{R}$ and g'.
Transformations defined in Eqs. (\ref{psigauge}-\ref{B}) allow consistently
to build up two kinds of invariant interactions. The first one gives
the fermion-gauge interaction
\begin{equation}
\label{l1}
L_f=\sum\limits_{\Psi=(Q),(L)}\bar{\Psi}_L\gamma^{\mu} \left( i\partial_{\mu}
+g_L\frac{\vec{\tau}}{2}\vec{W}_{L\mu}+g^{\prime}\frac{Y}{2}B_{\mu} \right)
\Psi_L + \left( L \rightarrow R \right).
\end{equation}
The second gives the gauge-gauge one
\begin{equation}
\label{l2}
L_g=-\frac{1}{4}W_{Li}^{\mu\nu}W_{Li\mu\nu}- \frac{1}{4}W_{Ri}^{\mu\nu}
W_{Ri\mu\nu}-\frac{1}{4}B^{\mu\nu}B_{\mu\nu},
\end{equation}
where
\begin{eqnarray}
W_{1L,R}^{\mu\nu}&=&\partial^{\mu}W_{1L,R}^{\nu}-\partial^{\nu}W_{1L,R}^{\mu}
+g_{L,R}(W_{2L,R}^{\mu}W_{3L,R}^{\nu}-W_{3L,R}^{\mu}W_{2L,R}^{\nu}) ,
\nonumber \\
W_{2L,R}^{\mu\nu}&=& \partial^{\mu}W_{2L,R}^{\nu}-\partial^{\nu}W_{2L,R}^{\mu}
+g_{L,R}(W_{3L,R}^{\mu}W_{1L,R}^{\nu}-W_{1L,R}^{\mu}W_{3L,R}^{\nu})  ,
\nonumber \\
W_{3L,R}^{\mu\nu}&=&\partial^{\mu}W_{3L,R}^{\nu}-\partial^{\nu}W_{3L,R}^{\mu}
+g_{L,R}(W_{1L,R}^{\mu}W_{2L,R}^{\nu}-W_{2L,R}^{\mu}W_{1L,R}^{\nu})   ,
\nonumber \\
B^{\mu\nu}&=&\partial^{\mu}B^{\nu}-\partial^{\nu}B^{\mu}.
\end{eqnarray}

Until now fermion and gauge fields are massless - mass terms which could
appear are not gauge invariant. To make them massive we have to
introduce scalar fields - Higgs particles and apply the spontaneous
symmetry breaking mechanism.
%%%%%%%%%%%%%%%%%%%%%%%%%%%%%%%%%%%%%%%%%%%%%%%%%%%%%%%%%%%%%%%%%%%%%%%%%%%

\subsection*{2.1.2 Higgs sector of the theory.}

%\setcounter{equation}{0}
%\renewcommand{\theequation}{1.2.\arabic{equation}}

%%%%%%%%%%%%%%%%%%%%%%%%%%%%%%%%%%%%%%%%%%%%%%%%%%%%%%%%%%%%%%%%%%%%%%%%%%%%
In order to produce fermion mass matrices we need to introduce only one
Higgs multiplet - the so called bidoublet \cite{mohapatra1}
(again, as in the fermion case, in parenthesis quantum numbers
$(d_{L},d_{R},B-L)$ are given)

\begin{equation}
\label{bidublet}
\phi = \left( \matrix{ \phi_1^0 & \phi_1^+ \cr \phi_2^- & \phi_2^0 \cr}
\right)\;\;\;\;: \left( 2,2,0 \right),
\end{equation}

with gauge transformation

\begin{equation}
\label{phigauge}
\phi^{\prime}= e^{-ig_L\frac{\vec{\tau}}{2}\vec{\Theta}(x)} \;\phi\;
e^{ig_R\frac{\vec{\tau}}{2}\vec{\Theta}(x)}.
\end{equation}

Consequently the most general Yukawa Lagrangian is given by

\begin{equation}
\label{yukava} L_Y=-\sum\limits_{i,j}\bar{L}_{iL} \left(
(h_l)_{ij} \phi+ (\tilde{h_l})_{ij}\tilde{\phi} \right) L_{jR}-
\sum\limits_{i,j}\bar{Q}_{iL} \left( (h_q)_{ij}\phi+
(\tilde{h}_q)_{ij}\tilde{\phi} \right) Q_{jR}+h.c. ,
\end{equation}

where

\begin{equation}
\tilde{\phi}  =  \tau_2\phi^{\ast}\tau_2=
\left(
\begin{array}{cc}
\phi_2^{0\ast} & -\phi_2^+ \\
-\phi_1^- & \phi_1^{0\ast}
\end{array}
\right)
\end{equation}

is also the object with quantum numbers $(2,2,0)$.
However, bidoublet $\phi$ is not sufficient for breaking $SU(2)_L \otimes
SU(2)_R \otimes U(1)_{B-L}$ to the $U(1)_{em}$, the only symmetry which
remains. There are many other possible ways to break the symmetry
\cite{mohapatra2}. Among them the most popular ones are models
with Higgs triplets $(\Delta_L$ $\sim (3,1,2)$,
$\Delta_R$ $\sim (1,3,2) )$ in addition to the
bidoublet \cite{mohapatra1}

\begin{equation}
\label{triplet}
\Delta_{L,R}= \left( \matrix{ \delta_{L,R}^+/\sqrt{2} & \delta_{L,R}^{++}
\cr \delta_{L,R}^0 & -\delta_{L,R}^+/\sqrt{2} \cr } \right).
\end{equation}

%We describe in the paper a model with one bidoublet $(\phi)$ and two
%triplets $(\Delta_L,\; \Delta_R)$ of Higgs fields.
The gauge transformations for the triplets fields are the following

\begin{equation}
\label{deltagauge}
\Delta '_{L,R}=e^{-ig'\frac{Y}{2}\Theta (x)}
e^{-ig_{L,R}\frac{\vec{\tau}}{2}\vec{\Theta}(x)}\;\Delta_{L,R}\;
e^{ig_{L,R}\frac{\vec{\tau}}{2}\vec{\Theta}(x)}.
\end{equation}

Now we can build the kinetic part of the Higgs bosons Lagrangian

\begin{equation}
\label{tr} L^{kin}_{Higgs}= Tr \left[ \left( D_{\mu} \Delta_L
\right)^{\dagger} \left( D^{\mu} \Delta_L \right) \right] + Tr
\left[ \left( D_{\mu} \Delta_R \right)^{\dagger} \left( D^{\mu}
\Delta_R \right) \right] + Tr \left[ \left( D_{\mu} \phi
\right)^{\dagger} \left( D^{\mu} \phi \right) \right],
\end{equation}

where

\begin{eqnarray}
D_{\mu}\phi &=&\partial_{\mu} \phi -ig_L \vec{W}_{L\mu} \frac{\vec{\tau}}{2}
\phi +ig_R \phi \frac{\vec{\tau}}{2} \vec{W}_{R\mu} , \nonumber \\
D_{\mu}\Delta_{L,R}&=& \partial_{\mu}\Delta_{L,R}-ig_{L,R}
\left[ \frac{\vec{\tau}} {2}
\vec{W}_{L,R\mu}, \Delta_{L,R} \right]-ig^{\prime}B_{\mu}\Delta_{L,R}.
\end{eqnarray}

In L-R symmetric models every L field has an R counterpart. This allows
the definition of the space inversion symmetry
$(Px=(t,-\vec{x}), \;\varepsilon (\mu)=1 \;for\; \mu =0\;and\;-1\; for\;
\mu =1,2,3)$;

\begin{eqnarray}
\label{Ppsi}
\psi_{L,R}(x) &\rightarrow &  \psi_{R,L}(Px), \\
\vec{W}^\mu_{L,R}(x) &\rightarrow & \varepsilon (\mu) \vec{W}^\mu_{R,L}(Px),
\\
B^\mu (x) &\rightarrow & \varepsilon (\mu) B^\mu (Px),
\end{eqnarray}

and

\begin{eqnarray}
\label{delta1}
\Delta_{L,R}(x) \rightarrow \Delta_{R,L}(Px), \\
\label{phi1}
\phi (x) \rightarrow \phi^\dagger (Px).
\end{eqnarray}

From now on, we will consider a model which is symmetric under the above
parity transformation (Eq.(\ref{Ppsi} - \ref{phi1})).
The parts of the Lagrangian density which we have already discussed
( Eqs. (\ref{l1}), (\ref{l2}), (\ref{yukava}),
(\ref{tr})) have the P symmetry if $g_L =g_R \equiv g$,
$h_{l,q}=h^\dagger_{l,q}$, $\tilde{h}_{l,q}=\tilde{h}^\dagger _{l,q}$.
%%%%%%%%%%%%%%%%%%%%%%%%%%%%%%%%%%%%%%%%%%%%%%%%%%%%%%%%%%%%%%%%%%%%%%%%%%

\subsection*{2.1.3 The Higgs potential.}

%\setcounter{equation}{0}
%\renewcommand{\theequation}{1.3.\arabic{equation}}

%%%%%%%%%%%%%%%%%%%%%%%%%%%%%%%%%%%%%%%%%%%%%%%%%%%%%%%%%%%%%%%%%%%%%%%%%%%

In order to break the left-right symmetry to the final $U(1)_{em}$ the Higgs
potential is introduced. The most general potential for one bidoublet
$\phi$ (\ref{bidublet}) and two triplets $\Delta_{L,R}$ (\ref{triplet})
which have the left-right symmetry (Eq.(\ref{delta1} - \ref{phi1}))
was introduced in \cite{deshpande}

\begin{eqnarray}
\label{potencjal}
V(\phi ,\Delta_L,\Delta_R)=
&-&\mu_1^2\left(Tr\left[\phi^{\dagger}\phi \right] \right) -
\mu_2^2\left(Tr\left[\tilde{\phi}\phi^{\dagger} \right] + Tr\left[\tilde{\phi%
}^{\dagger} \phi \right] \right)
-\mu_3^2 \left( Tr \left[ \Delta_L \Delta_L^{\dagger} \right] + Tr \left[
\Delta_R \Delta_R^{\dagger} \right] \right)  \nonumber \\
&+&\lambda_1 \left( \left( Tr \left[ \phi \phi^{\dagger} \right] \right)^2
\right)+ \lambda_2 \left( \left( Tr \left[ \tilde{\phi} \phi^{\dagger}
\right] \right)^2 + \left( Tr \left[ \tilde{\phi}^{\dagger} \phi \right]
\right)^2 \right) + \lambda_3 \left( Tr \left[ \tilde{\phi} \phi^{\dagger} \right] Tr \left[
\tilde{\phi}^{\dagger} \phi \right] \right)  \nonumber \\
&+& \lambda_4 \left( Tr \left[ \phi \phi^{\dagger} \right] \left( Tr \left[
\tilde{\phi} \phi^{\dagger} \right] + Tr \left[ \tilde{\phi}^{\dagger} \phi
\right] \right) \right)  + \rho_1 \left( \left( Tr \left[ \Delta_L \Delta_L^{\dagger} \right]
\right)^2+\left( Tr \left[ \Delta_R \Delta_R^{\dagger} \right] \right)^2
\right)  \nonumber \\
&+&\rho_2 \left( Tr \left[ \Delta_L \Delta_L \right] Tr \left[
\Delta_L^{\dagger} \Delta_L^{\dagger} \right] + Tr \left[ \Delta_R \Delta_R
\right] Tr \left[ \Delta_R^{\dagger} \Delta_R^{\dagger} \right] \right)
+ \rho_3 \left( Tr \left[ \Delta_L \Delta_L^{\dagger} \right] Tr \left[
\Delta_R \Delta_R^{\dagger} \right] \right)  \nonumber \\
&+&\rho_4 \left( Tr \left[ \Delta_L \Delta_L \right] Tr \left[
\Delta_R^{\dagger} \Delta_R^{\dagger} \right] + Tr \left[ \Delta_L^{\dagger}
\Delta_L^{\dagger} \right] Tr \left[ \Delta_R \Delta_R \right] \right)
\nonumber \\
&+& \alpha_1 \left( Tr \left[ \phi \phi^{\dagger} \right] \left( Tr \left[
\Delta_L \Delta_L^{\dagger} \right] + Tr \left[ \Delta_R \Delta_R^{\dagger}
\right] \right) \right)  +\alpha_2 \left( Tr \left[ \phi \tilde{\phi}^{\dagger} \right] Tr \left[
\Delta_R \Delta_R^{\dagger} \right]+ Tr \left[ \phi^{\dagger} \tilde{\phi}
\right] Tr \left[ \Delta_L \Delta_L^{\dagger} \right] \right)  \nonumber \\
&+&\alpha^\ast_2
\left( Tr \left[ \phi^{\dagger} \tilde{\phi} \right] Tr \left[
\Delta_R \Delta_R^{\dagger} \right]+ Tr \left[ \tilde{\phi}^{\dagger} \phi
\right] Tr \left[ \Delta_L \Delta_L^{\dagger} \right] \right)
+ \alpha_3 \left( Tr \left[ \phi \phi^{\dagger} \Delta_L
\Delta_L^{\dagger} \right] +Tr \left[ \phi^{\dagger} \phi \Delta_R
\Delta_R^{\dagger} \right] \right)  \nonumber \\
&+& \beta_1 \left( Tr \left[ \phi \Delta_R \phi^{\dagger} \Delta_L^{\dagger}
\right] +Tr \left[ \phi^{\dagger} \Delta_L \phi \Delta_R^{\dagger} \right]
\right)  + \beta_2 \left( Tr \left[ \tilde{\phi} \Delta_R \phi^{\dagger}
\Delta_L^{\dagger} \right] +Tr \left[ {\tilde{\phi}}^{\dagger} \Delta_L \phi
\Delta_R^{\dagger} \right] \right)  \nonumber \\
&+& \beta_3 \left( Tr \left[ \phi \Delta_R {\tilde{\phi}}^{\dagger}
\Delta_L^{\dagger} \right]
+ Tr \left[ \phi^{\dagger} \Delta_L \tilde{\phi}
\Delta_R^{\dagger} \right] \right).
\end{eqnarray}

From the parity transformation (Eq.(\ref{Ppsi} - \ref{phi1}) it
follows also that all terms in the potential are self-conjugate
except the $\alpha_2$ terms. This means that all parameters
$\mu_i$, $\lambda_i$, $\rho_i$ and $\beta_i$ are real. Only
$\alpha_2$ can be complex. To avoid the explicit CP symmetry
breaking in the Higgs sector also $\alpha_2$ is taken to be real (
ref. \cite{deshpande}). The neutral Higgs fields $\phi^0_1$,
$\phi^0_2$, $\delta^0_L$ and $\delta^0_R$ acquire VEV for which
the potential Eq.(\ref{potencjal}) has minimum

\begin{equation}
\label{vev}
<\phi>=\left( \matrix{ \kappa_1/\sqrt{2} & 0 \cr 0 & \kappa_2/\sqrt{2} }
\right) \;\;\;,\;\;\; <\Delta_{L,R} > = \left( \matrix{ 0 & 0 \cr
v_{L,R}/\sqrt{2} & 0 } \right).
\end{equation}

The VEV's for $\phi^0_{1,2}$, $\delta^0_{L,R}$ can be complex,
but the freedom of gauge symmetry transformation gives
a chance to make two of them real. Usually $v_{R}$ and $\kappa_1$
are taken to be real and $v_{L}$ and $\kappa_2$ remain complex

\begin{equation}
v_{L}=|v_{L}|e^{i\Theta_L},\;\;\;\;\;\; \kappa_2=|\kappa_2|e^{i\Theta_2}.
\end{equation}

There are six minimization conditions

\begin{equation}
\label{pochodne}
\frac{\partial V}{\partial v_R}=\frac{\partial V}{\partial \kappa_1}=
\frac{\partial V}{\partial |v_L|}=\frac{\partial V}{\partial \Theta_L}=
\frac{\partial V}{\partial |\kappa_2|}=\frac{\partial V}{\partial \Theta_2}=0
\end{equation}

which connect the VEV's with the Higgs potential parameters.
From equation (\ref{pochodne}) it follows that
in order  to avoid fine-tuning and  to satisfy
the phenomenological requirements,
the $\Theta_2$ phase, $|v_L|$ and $\beta_i$ parameters in the potential
must vanish \cite{deshpande}

\begin{equation}
\Theta_2=v_L=\beta_1=\beta_2=\beta_3=0.
\end{equation}

In what follows we consider the Higgs potential  in the form given
by Eq.(\ref{potencjal}) with $\beta_1=\beta_2=\beta_3=0$ without
explicit CP symmetry breaking ($\alpha_2=\alpha^\ast_2)$ where
spontaneous CP symmetry breaking also does not appear
($\Theta_2=|v_L|=0$). After SSB when $\phi$ acquires VEV
(Eq.(\ref{vev})), the full gauge symmetry Eq.(\ref{symmetry}) is
broken to $U(1)_Q\otimes U(1)_{B-L}$. The one $U(1)_{B-L}$ group
follows from the $B-L=0$ attribution for $\phi$,
($(B-L)<\phi>=0$). The other $U(1)_Q$ follows from the remaining
gauge symmetry given by the charge operator $\hat{Q}$

\begin{equation}
\hat{Q}<\phi>=\left[\frac{1}{2} \tau_3,<\phi>\right]=0.
\end{equation}

Next from SSB for right-handed triplet it follows that

\begin{equation}
U(1)_Q\otimes U(1)_{B-L}
\stackrel{\Delta_R \rightarrow \;<\Delta_R>}{\longrightarrow}
U(1)_Q,
\end{equation}

which is possible to see from

\begin{equation}
(B-L)<\Delta_R>=2<\Delta_R> \neq 0 ,
\end{equation}

and

\begin{equation}
\hat{Q}(<\Delta_R>)=\left[\frac{1}{2} \tau_3,<\Delta_R>\right]
+\frac{B-L}{2}<\Delta_R>=0.
\end{equation}

It is worth to stress that the SSB for bidoublet $\phi $ and
right - handed triplet $\Delta_R$ is sufficient to get the correct
final gauge group $U(1)_Q \equiv U(1)_{em}$.
The left - handed triplet $\Delta_L$ does not have to be spontaneously broken
$(v_L=0)$.

%%%%%%%%%%%%%%%%%%%%%%%%%%%%%%%%%%%%%%%%%%%%%%%%%%%%%%%%%%%%%%%%%%%%%%%

\section*{2.2 Lagrangian in the physical basis.}

In the previous section we have introduced the full Lagrangian with scalar,
fermion and gauge fields. Now we will find transformations of these fields to
physical basis.
%%%%%%%%%%%%%%%%%%%%%%%%%%%%%%%%%%%%%%%%%%%%%%%%%%%%%%%%%%%%%%%%%%%%%%%

\subsection*{2.2.1 The physical fields.}

We consider the diagonalization of the mass matrices separately for gauge
bosons, for fermions and for Higgs particles. The CP symmetry
can be broken only in the fermion sector. Some comments about it
are also given.

\subsubsection*{\bf 2.2.1.1 Gauge bosons.}

After SSB (Eq.(\ref{vev})) charged and neutral gauge bosons mass matrices
$\tilde{M}_W^2$ and $\tilde{M}_0^2$ are obtained from $L^{kin}_{Higgs}$
(Eq.(\ref{tr})). In the
$W_{L,R}^{\mu \pm}=\frac{1}{\sqrt{2}}(W^{1\mu}_{L,R} \mp iW^{2\mu}_{L,R})$
and $(W^{\mu}_{3L}, W^{\mu}_{3R},B^{\mu})$ basis we have

\begin{eqnarray}
\label{lm}
L_M &=&
\left(
W_L^{+\mu},\;W_R^{+\mu}
\right)
\tilde{M}_W^2
\left(
\begin{array}{c}
W^-_{L\mu} \\
W^-_{R\mu}
\end{array}
\right)+
h.c. +
\frac{1}{2}
\left(
W^\mu_{3L},\;W^\mu_{3R},\;B^\mu
\right)
\tilde{M}_0^2
\left(
\begin{array}{c}
W_{3L\mu} \\
W_{3R\mu} \\
B_\mu
\end{array}
\right),
\end{eqnarray}

\begin{eqnarray}
\tilde{M}_W^2&=&\frac{g^2}{4} \left( \matrix{ \kappa_+^2 &
-2\kappa_1\kappa_2 \cr -2\kappa_1\kappa_2 & \kappa_+^2+2v_R^2 } \right) ,
\end{eqnarray}

and

\begin{eqnarray}
\tilde{M}_0^2&=&\frac{1}{2}\left( \matrix{ \frac{g^2}{2} \kappa_+^2 &
-\frac{g^2}{2}\kappa_+^2 & 0 \cr -\frac{g^2}{2}\kappa_+^2 &
\frac{g^2}{2}(\kappa_+^2+4v_R^2) & -2gg'v_R^2 \cr 0 & -2gg'v_R^2 &
2g'^2v_R^2 } \right),
\end{eqnarray}

where $\kappa_+ =\sqrt{\kappa_1^2 +\kappa_2^2}$.

The symmetric mass matrices are diagonalized by the orthogonal
transformations

\begin{equation}
\label{fiz1}
\left(
\begin{array}{c}
W^{\pm}_{L} \\
W^{\pm}_{R}
\end{array}
\right)=\left(
\begin{array}{cc}
cos\xi & sin\xi \\
-sin\xi & cos\xi
\end{array}
\right)=\left(
\begin{array}{c}
W^{\pm}_{1} \\
W^{\pm}_{2}
\end{array}
\right),
\end{equation}

and

\begin{displaymath}
\left(
\begin{array}{c}
W_{3L} \\
W_{3R} \\
B
\end{array}
\right)=\left(
\begin{array}{ccc}
c_{W}c & c_{W}s & s_{W} \\
-s_{W}s_{M}c - c_{M}s & -s_{W}s_{M}s +c_{M}c & c_{W}s_{M} \\
-s_{W}c_{M}c + s_{M}s & - s_{W}c_{M}s - s_{M}c &  c_{W}c_{M}
\end{array}
\right)\left(
\begin{array}{c}
Z_{1} \\
Z_{2} \\
A
\end{array}
\right)\equiv
\end{displaymath}

\begin{equation}
\label{fiz2}
\left(
\begin{array}{ccc}
x_{1} & x_{2} & x_{3} \\
y_{1} & y_{2} & y_{3} \\
v_{1} & v_{2} & v_{3}
\end{array}
\right)\left(
\begin{array}{c}
Z_{1} \\
Z_{2} \\
A
\end{array}
\right) ,
\end{equation}

where

\begin{eqnarray*}
g &=& \frac{e}{sin{\Theta_W}},\;\;g^{\prime}=
\frac{e}{\sqrt{\cos{2\Theta_W}}},\;\;
c_W=\cos \Theta_W,\;\;s_W=\sin {\Theta_W}, \\
c_M &=& \frac{\sqrt{\cos{2\Theta_W}}}{\cos{\Theta_W}},\;\;
s_M=tg\Theta_W,\;\; s=\sin{\phi},\;\;c=\cos{\phi}.
\end{eqnarray*}

From the relations above it is easy to see that $x_3$ and $y_3$
in the matrix Eq.(\ref{fiz2}) are equal, i.e. $x_3=y_3$. \\
Masses of the physical gauge bosons are the following

\begin{eqnarray}
M^2_{W_{1,2}} &=& \frac{g^2}{4}
\left[
\kappa^2_+ +v_R^2 \mp \sqrt{v_R^4+4 \kappa^2_1\kappa^2_2}
\right], \\
M^2_{Z_{1,2}} &=& \frac{1}{4}
\left\{
\left[
g^2 \kappa^2_+ +2v_R^2
\left(
g^2+g'^2
\right)
\right]
\mp \sqrt{
\left[
g^2 \kappa^2_+ +2v_R^2
\left(
g^2+g'^2
\right)
\right]^2-
4g^2
\left(
g^2+2g'^2
\right)
\kappa^2_+ v_R^2 }
\right\}.
\end{eqnarray}

The mixing angles are given by

\begin{eqnarray}
\tan{2\xi}=-\frac{2\kappa_1 \kappa_2}{v_R^2},\;\;\;
\sin{2\phi}=-\frac{g^2 \kappa^2_+ \sqrt{\cos{2\Theta_W}}}{2\cos^2{\Theta_W}
\left(
M^2_{Z_2}-M^2_{Z_1}
\right)}.
\end{eqnarray}

Experimental data gives bounds on the additional gauge boson
masses $M_{W_2},\;M_{Z_2} $ and the
$ \rho =\frac{M^2_{W_1}}{M^2_{Z_1}\cos^2{\Theta_W}}$
parameter which are satisfied only if
$v_R>>\kappa_+$.
In this approximation there is (to the first order in
$\frac{\kappa_+}{v_R}$ and $\frac{\sqrt{\kappa_1 \kappa_2}}{v_R}$)

\begin{eqnarray}
M^2_{W_1} & \simeq & \frac{g^2}{4} \kappa^2_+ \left(
1-\frac{(\kappa_1 \kappa_2)^2}
{\kappa_+^2v_R^2} \right) ,\;\;
M^2_{W_2}  \simeq  \frac{g^2 v_R^2}{2},\nonumber \\
M^2_{Z_1} & \simeq & \frac{g^2 \kappa^2_+}{4\cos^2{\Theta_W}} \left( 1-
\frac{\cos^2{2 \Theta_W} \kappa_+^2}{2 \cos^4{\Theta_W}v_R^2} \right) ,\;\;
M^2_{Z_2}  \simeq   \frac{v_R^2 g^2 \cos^2{\Theta_W}}{\cos{2\Theta_W}},
\;\;\sin{2 \phi } \simeq
-\frac{\kappa^2_+\left(\cos{2\Theta_W}\right)^{\frac{3}{2}}}
{2 v_R^2 \cos^4{\Theta_W}}.
\end{eqnarray}

\subsubsection*{\bf 2.2.1.2 Fermions.}

\paragraph*{\bf a. Quarks.}

The most general Yukawa interaction for quarks is given by
Eq.(\ref{yukava}). After SSB the mass Lagrangian is

\begin{equation}
\label{mass}
L^q_{mass}=
-\bar{U}^\prime_L
M_u
U^\prime_R
-\bar{D}^\prime_L
M_d
D^\prime_R
+h.c.\;,
\end{equation}

where

\begin{equation}
\label{m}
M_u=\frac{1}{\sqrt{2}}\left( h_q \kappa_1+{\tilde{h}}_{q}\kappa_2
\right),\;\; M_d=\frac{1}{\sqrt{2}} \left( \tilde{h}_q \kappa_1+h_{q}
\kappa_2 \right),
\end{equation}

and $U^\prime_{L,R}$, $D^\prime_{L,R}$ are three dimensional vectors
built of the weak quark fields (e.g. $\bar{U}^\prime_{L}=
\left(
\bar{u}^{\prime}_L \bar{c}^{\prime}_L \bar{t}^{\prime}_L
\right)$).

In the considered model $\kappa_1$ and $\kappa_2$ are real, thus from relation
Eq.(\ref{m}) follows that quark mass matrices are hermitian

\begin{equation}
M_u=M^\dagger_u,\;\;\;M_d=M^\dagger_d .
\end{equation}

These matrices are diagonalized by a biunitary transformation

\begin{equation}
\label{ud} U^{\prime}_{L,R}=V^{u}_{L,R} U_{L,R},\;\;\;
D^{\prime}_{L,R}=V^{d}_{L,R} D_{L,R} ,
\end{equation}

and

\begin{eqnarray}
\label{trans}
V_L^{u\dagger } M_u V_R^{u}=diag(m_u,\;m_c,\;m_t), \\ \nonumber
V_L^{d\dagger } M_d V_R^{d}=diag(m_d,\;m_s,\;m_c).
\end{eqnarray}

If all diagonal mass matrix elements are positive after the unitary
transformation

\begin{equation}
\label{vlr}
V^i_L=V^i_R\;\;for\;\; i=u,d\;,
\end{equation}

the Manifest Left-Right Symmetric model (MLRS) is realized
\cite{deshpande}. If after the unitary transformation some masses are
negative, then to make them positive we must modify Eq.(\ref{vlr}), namely

\begin{equation}
\label{positive} V_R^{(i)}=V_L^{(i)} W^{(i)}\;\;for \;\;i=u,d ,
\end{equation}

where
$W^{(i)}=diag\left(\varepsilon_1^{(i)},\varepsilon_2^{(i)},
\varepsilon_3^{(i)}\right)$ and $\varepsilon_k^{(i)}=+1$ if $m_k>0$
and
$\varepsilon_k^{(i)}=-1$ for $m_k<0$. We refer to the relation
Eq.(\ref{positive}) as Quasi Manifest L-R Symmetry (QMLRS) \cite{deshpande}.

\paragraph*{\bf b. Leptons.}

The Yukawa interaction for leptons is more complicated. Leptons have
both Dirac and Majorana Yukawa couplings. The most general Lagrangian
invariant under the gauge transformation
(Eqs.(\ref{psigauge}),(\ref{phigauge}),(\ref{deltagauge}))
and the discrete L-R
symmetry operation (Eq.(\ref{Ppsi} - \ref{phi1})) is \cite{deshpande}

\begin{equation}
\label{yuklep}
L^{lepton}_{Yukawa}=
-\left\{
\bar{L}_L \left[ h_l\phi + \tilde{h}_l\tilde{\phi } \right] L_R   +h.c.
\right\}-
\bar{L}^c_R \Sigma_L h_M L_L-\bar{L}^c_L \Sigma_R h_M L_R \;+h.c.\;,
\end{equation}

where

\begin{equation}
\Sigma_{L,R}=i\tau_2 \Delta_{L,R}= \left( \matrix{ \delta_{L,R}^0
& -\delta_{L,R}^{+}\sqrt{2} \cr -\delta_{L,R}^+\sqrt{2} &
-\delta_{L,R}^{++} \cr } \right),
\end{equation}

and the $ 3 \times 3 $ complex matrix $h_M$ is symmetric $( h_M=h^T_M)$.
After SSB, the neutrino and charged lepton mass Lagrangians are obtained

\begin{equation}
\label{massnu}
L^\nu_{mass}=-\frac{1}{2}
\left(
\bar{n}^{\prime c}_L M_\nu n_R^\prime +\bar{n}^{\prime c}_R
M_{\nu}^{\ast} n^\prime_L
\right),\;\;
n_R^\prime=
\left(
\begin{array}{c}
\nu^{\prime c}_R \\
\nu_R^\prime
\end{array}
\right),\;\;
n_L^\prime=
\left(
\begin{array}{c}
\nu_L^\prime \\
\nu^{\prime c}_L
\end{array}
\right),\;\;
\nu^{\prime c}_{L,R}=C\bar{\nu}^{\prime T}_{R,L},
\end{equation}

\begin{equation}
\label{massl} L^l_{mass}= -\bar{l}_L' M_l l_R' -\bar{l}_R'
M^\dagger_l l_L' ,
\end{equation}

where $\nu^\prime_{L,R}$ $(l_{L,R}')$ are three dimensional vectors
constructed from neutrino (charged lepton) fields and

\begin{eqnarray}
M_\nu&=&
\left(
\begin{array}{cc}
0 & M_D \\
M^T_D & M_R
\end{array}
\right)
=M_\nu^{T},\;\;
M_D=\frac{1}{\sqrt{2}}
\left(
h_l\kappa_1+\tilde{h}_l\kappa_2
\right)
=M^\dagger_D, \\
M_R &=& \sqrt{2} h_M v_R= M^T_R,\;\;
M_l=\frac{1}{\sqrt{2}}
\left(
h_l\kappa_2+\tilde{h}_l\kappa_1
\right)
=M^\dagger_l.
\end{eqnarray}

To find the neutrino and charged lepton mass eigenstates a unitary
transformation must be performed

\begin{equation}
\label{unitr} n_R'=VN_R,\;\;\;n_L'=V^\ast N_L,
\;\;\;l_{L,R}'=V^l_{L,R}l_{L,R},
\end{equation}

where $V\left( V^l_{L,R}\right)$ is unitary $6\times 6\;(3\times3)$ matrix.
After the transformation one gets

\begin{equation}
\label{diag1}
V^T M_\nu V= \left( M_\nu \right)_{diag},\;\;\;V^{l\dagger }_L M_l V^l_R=
(M_l)_{diag},
\end{equation}

where $\left( M_\nu \right)_{diag}$ and $(M_l)_{diag}$ are diagonal, positive
mass matrices.
Similarly like in the quark sector, if $V^l_R=V^l_L$ and
all diagonal masses $(M_l)_{diag}$ are positive,
then Manifest L-R Symmetric model is realized.
If the normal unitary transformation with $V^l_R=V^l_L$ in Eq.(\ref{diag1})
does not give positive $(M_l)_{diag}$, then the modification is necessary
$V^l_R=V^l_L W^l$ and QMLRS model is in the game.

In the physical bases, Lagrangian (Eq.\ref{massnu}, \ref{massl})
is given by

\begin{equation}
L^{lepton}_{mass}=-\frac{1}{2} \bar{N} \left( M_\nu \right)_{diag} N-
\bar{l}(M_l)_{diag} l,
\end{equation}

where

\begin{equation}
N=N_R+N_L=N^c ,\;\;\;l=l_L+l_R.
\end{equation}

\paragraph*{{\bf c. CP symmetry.}}

In the model which we consider (MLRS or QMLRS) any CP violating
phase can appear only in the
quark and lepton mass matrices. The CP symmetry violation or conservation
depends on the specific form of the $M_u$ and $M_d$ for quarks
or $M_l$ and $M_{\nu}$ for leptons.
If they are real the CP symmetry is conserved,
but even if they are complex the CP symmetry can still be satisfied.
This is so
because we can perform a unitary transformation on the fermion fields
without changing the physical observables but redefine the mass matrices.
If there exist quark and lepton bases in which the mass matrices are real
the CP is conserved. And opposite, if in any weak base the mass matrices
are complex the CP symmetry is violated. In the quark sector this
condition is equivalent to existence of a unitary and symmetric matrix
$V_q$ such that

\begin{equation}
\label{Vq}
V^\dagger_q M_u V_q=M^\ast_u, \;\;\;\;V^\dagger_q M_d V_q=M^\ast_d.
\end{equation}

This condition means that in the weak base where one of the mass matrices
(e.q. $M_d$ ) is diagonal the other has phase structure
(for details see e.g. \cite{cplep})

\begin{equation}
\label{Mu} (M_u)_{ij}=|(M_u)_{ij}|
\exp{\left[i\left(\frac{\delta_i-\delta_j}{2}\right)\right]}.
\end{equation}

For n generation in the hermitian matrix $M_u$ only $\frac{n(n-1)}{2}-n+1=
\frac{n^2-3n+2}{2}$ phases do not satisfy the condition \\ ( \ref{Mu} ).
This means that in the MLRS or QMLRS model in the quark sector for three
generation only one phase breaks CP symmetry.
(for two generation CP is conserved). \\
In the lepton sector the CP symmetry is satisfied if there
is a unitary symmetric matrix $V_l$ such that

\begin{equation}
\label{v1} V^\dagger_l M_l V_l=M^\ast_l,\;\;\;V^\dagger_l M_D
V_l=M^\ast_D ,
\end{equation}

and

\begin{equation}
\label{v2}
V^T_l M_R V_l=M^\ast_R.
\end{equation}

Then from Eqs. (\ref{v1}) and (\ref{v2}) it follows that in the
base where $M_l$ is diagonal the $M_D$ and $M_R$ matrices have the
phase structure ( see Ref.\cite{qlep} )

\begin{equation}
\label{MD}
(M_D)_{ij}=|(M_D)_{ij}|
\exp{\left[i\left(\frac{\chi_i-\chi_j}{2}\right)\right]},\;\;\;\;
(M_R)_{ij}=|(M_R)_{ij}|
\exp{\left[-i\left(\frac{\chi_i+\chi_j}{2}\right)\right]}.
\end{equation}

This means that all phases in the hermitian matrix $M_D$ and symmetric
matrix $M_R$ which do not satisfy the condition ( \ref{MD} )
break the CP symmetry. There are $ \frac{n(n-1)}{2}+\frac{n(n+1)}{2}-n=
n(n-1)$ such phases. For $n=3$ generation there are 6 phases which
cause the CP symmetry violation.

\subsubsection*{\bf 2.2.1.3 Higgs particles.}

The Higgs sector is described by 20 degrees of freedom (8 real
fields for bidoublet and $ 6\times 2$ degrees of freedom for two
triplets). After spontaneous symmetry breaking in the Higgs
potential bilinear terms appear which form the mass matrices for
Higgs particles. The precise form of these matrices are given in
Ref. \cite{deshpande}, \cite{gluza}. To get the physical Higgs
particles, the mass matrices have to be diagonalized. The initial
20 degrees of freedom give two charged $G^{\pm}_{L,R}$ and two
neutral $\tilde{G}^{0}_{1,2}$ Goldstone bosons and 14 physical
particles. These physical degrees of freedom produce:

\begin{itemize}
\item[(i)] four neutral scalars with $J^{PC}=0^{++}\;\;
\left( H^0_i\;\;i=0,1,2,3\right),$
\item[(ii)] two neutral pseudoscalars with $J^{PC}=0^{+-}\;\;
\left( A^0_i\;\;i=1,2\right),$
\item[(iii)] two singly charged bosons
$\left( H^{\pm}_i\;\;i=1,2\right),$ and
\item[(iv)] two doubly charged Higgs particles
$\left( \delta_L^{\pm \pm},\;
\delta_R^{\pm \pm}\right)$.
\end{itemize}

\paragraph*{\bf (i) Four neutral scalars.}

The masses of $H^0_0,\;H^0_1$ and $H^0_2$ fields are obtained from
the mass matrix which in the basis

\begin{equation}
\left(
\phi_-^{or},\phi_+^{or},\delta_R^{or}
\right)
\equiv
\left[
\frac{\sqrt{2}}{\kappa_+}Re
\left(
\kappa_1 \phi_1^{0}+\kappa_2 \phi_2^{0\ast}
\right),
\frac{\sqrt{2}}{\kappa_+}Re
\left(
\kappa_1 \phi_2^{0\ast}-\kappa_1 \phi_1^{0}
\right),
\sqrt{2} Re \delta_R^0
\right]
\end{equation}

has the following elements $\left(
\varepsilon =\frac{2\kappa_1 \kappa_2}{\kappa^2_+}\right)$

\begin{eqnarray*}
M_{11} &=& 2\kappa^2_+
\left[
\lambda_1 +\varepsilon^2
\left(
2 \lambda_2 + \lambda_3
\right)+
2 \lambda_4 \varepsilon
\right], \\ \nonumber
M_{12} &=& M_{21} = 2\kappa^2_+\sqrt{1-\varepsilon^2}
\left(
2\lambda_2 + \lambda_3+\lambda_4
\right), \\  \nonumber
M_{13} &=& M_{31} =\frac{1}{2} \kappa_+ v_R
\left[
2\alpha_1+2\alpha_2 \varepsilon +\alpha_3
\left(
1-\sqrt{1-\varepsilon^2}
\right)
\right], \\ \nonumber
M_{22} &=&  2
\left(
2\lambda_2 + \lambda_3
\right)
\kappa^2_+
\left(
1-\varepsilon^2
\right)
+\frac{1}{2} \alpha_3 v^2_R \frac{1}{\sqrt{1-\varepsilon^2}}, \\ \nonumber
M_{23} &=& M_{32} =\frac{1}{2} \kappa_+ v_R
\left[
4\alpha_2 \sqrt{1-\varepsilon^2} + \alpha_3 \varepsilon
\right], \\ \nonumber
M_{23} &=& 2\rho_1 v_R^2.
\end{eqnarray*}

\begin{equation}
\label{mm}
\end{equation}

This mass matrix is diagonalized by the orthogonal transformation

\begin{equation}
\left(
\begin{array}{c}
\phi_-^{or} \\
\phi_+^{or} \\
\delta_R^{or}
\end{array}
\right)=
\left(
\begin{array}{ccc}
a_0 & a_1 & a_2 \\
b_0 & b_1 & b_2 \\
c_0 & c_1 & c_2
\end{array}
\right)
\left(
\begin{array}{c}
H^0_0 \\
H^0_1 \\
H^0_2
\end{array}
\right).
\end{equation}

We will  not present the precise results. However,if  $v_R>> \kappa_+$,
we can find approximately that $H^0_0\simeq \phi_-^{or}$,
$H^0_1\simeq \phi_+^{or}$ and $H^0_2\simeq \delta_R^{or}$ with the masses

\begin{eqnarray}
M^2_{H^0_0} &\simeq & 2\kappa^2_+
\left[
\lambda_1 +\varepsilon^2
\left(
2\lambda_1 +\lambda_3
\right)+
2\lambda_4 \varepsilon
\right], \\ \nonumber
M^2_{H^0_1}&\simeq & \frac{1}{2} \alpha_3 v^2_R
\frac{1}{\sqrt{1-\varepsilon^2}}, \\ \nonumber
M^2_{H^0_2}&\simeq & 2\rho_1 v_R^2.
\end{eqnarray}

The fourth Higgs particle $H^0_3$ is the real part of the $\delta^0_L$ field

\begin{equation}
H^0_3=\sqrt{2} Re(\delta^0_L),
\end{equation}

and has the mass

\begin{equation}
M^2_{H^0_3}= \frac{1}{2} v^2_R
\left(
\rho_3 - 2\rho_1
\right).
\end{equation}

\paragraph*{\bf (ii) Two neutral pseudoscalars.}

Without any approximation the masses of two pseudoscalars
$A^0_1$ and $A^0_2$ are given by

\begin{eqnarray}
M^2_{A^0_1} &=& \frac{1}{2} \alpha_3 v^2_R
\frac{1}{\sqrt{1-\varepsilon^2}}-2\kappa^2_+
\left(
2\lambda_2-\lambda_3
\right),
\\ \nonumber
M^2_{A^0_2} &=& \frac{1}{2}  v^2_R
\left(
\rho_3-2\rho_1
\right).
\end{eqnarray}

\paragraph*{\bf (iii) Two singly charged Higgs particles.}

The masses of the $H_i^\pm \;\;(i=1,2) $ are obtained by diagonalization of
a $2 \times 2$ mass matrix. The result is

\begin{eqnarray}
M^2_{H^\pm_1} &=& \frac{1}{2}  v^2_R
\left(
\rho_3-2\rho_1
\right)+
\frac{1}{4}
\alpha_3 \kappa^2_+ \sqrt{1-\varepsilon^2}, \\ \nonumber
M^2_{H^\pm_2} &=& \frac{1}{2} \alpha_3
\left[
v^2_R \frac{1}{\sqrt{1-\varepsilon^2}}+\frac{1}{2} \kappa^2_+
\sqrt{1-\varepsilon^2}
\right].
\end{eqnarray}

\paragraph*{\bf (iv) Two doubly charged Higgs particles.}

They are simply components of Higgs triplets and

\begin{eqnarray}
M^2_{\delta_L^{\pm \pm}} &=& \frac{1}{2}
\left[
v^2_R
\left(
\rho_3-2\rho_1
\right)+
\alpha_3 \kappa^2_+ \sqrt{1-\varepsilon^2}
\right],  \\ \nonumber
M^2_{\delta_R^{\pm \pm}} &=& 2\rho_2 v^2_R+
\frac{1}{2} \alpha_3 \kappa^2_+ \sqrt{1-\varepsilon^2}.
\end{eqnarray}

\vspace*{0.5cm}

The relations among non physical Higgs particles and the physical ones
and Goldstone bosons are given by \\
$(\kappa_-=\sqrt{\kappa_1^2 -\kappa_2^2})$:

\begin{eqnarray}
\label{rfns}
\phi^0_1&=&\frac{1}{\sqrt{2}\kappa_+}\left[ H_0^0 \left( \kappa_1
a_0-\kappa_2b_0 \right) +H_1^0 \left( \kappa_1a_1-\kappa_2b_1 \right) +
H_2^0 \left( \kappa_1a_2-\kappa_2b_2 \right) +
i\kappa_1 \tilde{G_1^0}-i\kappa_2A_1^0 \right], \nonumber \\
\phi^0_2 & = & \frac{1}{\sqrt{2}\kappa_+}\left[ H_0^0 \left( \kappa_2
a_0+\kappa_1b_0 \right) +H_1^0 \left( \kappa_2a_1+\kappa_1b_1 \right) +
H_2^0 \left( \kappa_2a_2+\kappa_1b_2 \right) -
i\kappa_2 \tilde{G_1^0}-i\kappa_1A_1^0 \right], \nonumber \\
\delta_L^0 &=& \frac{1}{\sqrt{2}} \left( H_3^0+iA_2^0 \right) , \nonumber \\
\delta_R^0 &=& \frac{1}{\sqrt{2}} \left( c_0H_0^0+c_1H_1^0+c_2H_2^0+
i\tilde{G}_2^0\right), \nonumber \\
\phi_1^+&=&\frac{\kappa_1}{\kappa_+\sqrt{1+{\left( \frac{\kappa_-^2}{\sqrt{2}%
\kappa_+v_R} \right)}^2}}H_2^+ -\frac{\kappa_1}{\kappa_+\sqrt{1+{\left(
\frac{\sqrt{2}\kappa_+v_R}{\kappa_-^2} \right)}^2}}G_R^+ - \frac{\kappa_2}{%
\kappa_+}G_L^+ \equiv a_{12}H_2^++a_{1R}G_R^++a_{1L}G_L^+, \nonumber \\
\phi_2^+&=&\frac{\kappa_2}{\kappa_+\sqrt{1+{\left( \frac{\kappa_-^2}{\sqrt{2}%
\kappa_+v_R} \right)}^2}}H_2^+ -\frac{\kappa_2}{\kappa_+\sqrt{1+{\left(
\frac{\sqrt{2}\kappa_+v_R}{\kappa_-^2} \right)}^2}}G_R^+ + \frac{\kappa_1}{%
\kappa_+}G_L^+ \equiv a_{22}H_2^++a_{2R}G_R^++a_{2L}G_L^+, \nonumber \\
\delta_L^+ &=& H_1^+ , \nonumber \\
\delta_R^+&=& \frac{1}{\sqrt{1+{\left( \frac{\kappa_-^2}{\sqrt{2}\kappa_+v_R}
\right)}^2}}G_R^+ + \frac{1}{\sqrt{1+{\left( \frac{\sqrt{2} \kappa_+v_R}{%
\kappa_-^2} \right)}^2}}H_2^+ \equiv a_{RR}G_R^++a_{R2}H_2^+.
\end{eqnarray}

The doubly charged Higgs bosons $\delta_{L,R}^{\pm \pm}$ are
already physical. Combinations of the Goldstone bosons
$G_{L,R}^\pm $ and $\tilde{G}^0_{1,2}$ are absorbed by
$W_{L,R}^\pm $ and $Z_{1,2}$ respectively. Through this mechanism
gauge bosons become massive. In the approximation $v_R >>
\kappa_+$ the relations between physical and unphysical Higgs
particles become much simpler. They are:

\begin{eqnarray}
\phi_1^0&\simeq&\frac{1}{\kappa_+\sqrt{2}}\left[
\kappa_1H_0^0-\kappa_2H_1^0+i\kappa_1 \tilde{ G_1^0}-i\kappa_2A_1^0 \right] ,
\\
\phi_2^0 &\simeq& \frac{1}{\kappa_+\sqrt{2}}\left[
\kappa_2H_0^0+\kappa_1H_1^0-i\kappa_2 \tilde{G_1^0}-i\kappa_1A_1^0 \right],
\\
\delta_R^0 &=& \frac{1}{\sqrt{2}}\left( H_2^0+iG_2^0 \right) , \\
\phi_{1}^+ &\simeq& \frac{\kappa_1}{\kappa_+} H_2^+
-\frac{\kappa_2}{\kappa_+}G_L^+, \\
\phi_{2}^+ &\simeq& \frac{\kappa_2}{\kappa_+} H_2^+
+\frac{\kappa_1}{\kappa_+}G_L^+, \\
\delta_R^+ &\simeq& G_R^+ .
\end{eqnarray}

%%%%%%%%%%%%%%%%%%%%%%%%%%%%%%%%%%%%%%%%%%%%%%%%%%%%%%%%%%%%%%%%%%%%%%%%

\subsection*{2.2.2 Interactions among physical fields.}

\subsubsection*{\bf 2.2.2.1 Gauge boson - fermion interactions.}

The charged and neutral current interactions for fermions are
obtained from Eq.(\ref{l1}).
For quarks the charged current interaction is equal

\begin{equation}
L^{(q)}_{CC}=\frac{g}{\sqrt{2}} \left( \bar{U} \gamma^\mu \left(
P_L \cos{\xi} U_L^{CKM}-P_R \sin{\xi} U_R^{CKM} \right) D
W^+_{1\mu}+ \bar{U} \gamma^\mu \left( P_L \sin{\xi} U_L^{CKM}+P_R
\cos{\xi} U_R^{CKM} \right) D W^+_{2\mu} \right) +h.c.\;,
\end{equation}

where $U=U_L+U_R$, $D=D_L+D_R$ are the quark eigenmass states
(Eq.(\ref{ud})) and the Cabibbo - Kobayashi - Maskawa (CKM)
matrices are given by

\begin{equation}
U_L^{CKM}=V_L^{u\dagger }V_L^d,\;\;\;U_R^{CKM}=V_R^{u\dagger }V_R^d.
\end{equation}

From Eq.(\ref{vlr}) it follows that in the MLRS the CKM  mixing
matrices for left - and right - handed quarks are equal

\begin{equation}
U_R^{CKM}= U_L^{CKM},
\end{equation}

and in agreement with Eq.(\ref{positive}), for QMLRS model

\begin{equation}
\left( U_R^{CKM}\right)_{ij}= \pm \left( U_L^{CKM}\right)_{ij}.
\end{equation}

The neutral current interaction for quarks is given by

\begin{equation}
L^{(q)}_{NC}=\frac{e}{2\sin{\Theta_W}\cos{\Theta_W}}
\sum\limits_{i=up,\;down, }\sum\limits_{j=1,2} \bar{\psi}_i\gamma^\mu
\left[
A^{ji}_L P_L + A^{ji}_R P_R
\right]
\psi_i Z_{j\mu}
+e\sum\limits_{i=up,\;down}Q_i \bar{\psi}_i\gamma^\mu \psi_i A_{\mu}\;.
\end{equation}

The left and right handed couplings $A^{1,2;\;i}_{L,R}$
($i=$ up or down) are the following

\begin{eqnarray}
\label{al1} A_L^{1i} & = &
\cos{\phi}\;g^i_L+\sin{\phi}\;g_L^{\prime i},\\ A_R^{1i} & = &
\cos{\phi}\;g^i_R+\sin{\phi}\;g_R^{\prime i},\\ A_L^{2i} & = &
\sin{\phi}\;g^i_L-\cos{\phi}\;g_L^{\prime i},\\ A_R^{2i} & = &
\sin{\phi}\;g^i_R-\cos{\phi}\;g_R^{\prime i},
\end{eqnarray}

where

\begin{eqnarray}
g^i_L & = & 2 T^L_{3i}-2Q_i\sin^2\Theta , \\
g_L^{\prime i} & = & \frac{2\sin^2\Theta }{\sqrt{\cos 2\Theta }}
\left( Q_i-T^L_{3i}\right), \\
g^i_R & = & -2 Q_i \sin^2 \Theta , \\
\label{grpr}
g_R^{\prime i} & = & \frac{2}{\sqrt{\cos 2\Theta }}
\left( Q_i \sin^2 \Theta - T^R_{3i}\cos^2 \Theta \right).
\end{eqnarray}

For leptons the charged current interaction is given by

\begin{equation}
\label{lcc} L^{(l)}_{CC}=\frac{g}{\sqrt{2}} \left( \bar{N}
\gamma^\mu \left( P_L \cos{\xi} K_L-P_R \sin{\xi} K_R \right) l
W^+_{1\mu}+ \bar{N} \gamma^\mu \left( P_L \sin{\xi} K_L+P_R
\cos{\xi} K_R \right) l W^+_{2\mu} \right) +h.c.\;,
\end{equation}

where $K_L\;\left( K_R \right)$ are $6\times 3$ mixing matrices

\begin{equation}
K_L=V^{\nu \dagger}_L V^l_L,\;\;K_R=V^{\nu \dagger}_R V^l_R.
\end{equation}

The matrices $V^l_{L,R}$ are given by Eq.(\ref{unitr}) and Eq.(\ref{diag1}),
the $6\times 3$
matrices $V^{\nu\dagger}_{L,R}$ are defined by the
V matrix \\
(Eq.(\ref{unitr}), Eq.(\ref{diag1})) in the way

\begin{equation}
\label{vau}
V=
\left(
\begin{array}{c}
V_L^{\nu\ast}\\
V_R^\nu
\end{array}
\right).
\end{equation}

The charged leptons and neutrino neutral current interactions are

\begin{equation}
\label{lnc}
L^{(l)}_{NC}=\frac{e}{2\sin{\Theta_W}\cos{\Theta_W}}
\left[
J_1^{\mu}Z_{1\mu}+
J_2^{\mu}Z_{2\mu}
\right]
+eJ_{EM}^{\mu}A_{\mu}\;,
\end{equation}

where

\begin{equation}
\label{imi}
J^{\mu}_i=\sum\limits_{charged\; leptons}
\bar{l}{\gamma}^{\mu}
\left[
A_L^{il} P_L+A_R^{il}P_R
\right]
l+
\sum\limits_{neutrinos} \bar{N}{\gamma}^{\mu}
\left[
A_L^{i\nu}\Omega_L P_L+A_R^{i\nu}
\Omega_R P_R
\right]
N,
\end{equation}

and

\begin{equation}
J_{EM}^{\mu}=
-\sum\limits_{charged\; leptons} \bar{l} \gamma^{\mu} l.
\end{equation}

The $6\times 6$ mixing matrices $\Omega_{L,R}$ are defined in the way

\begin{equation}
\Omega_L=V_L^{\nu \dagger}V^\nu_L=\Omega^\dagger_L,
\;\;\;\Omega_R=V_R^{\nu \dagger}V^\nu_R=\Omega^\dagger_R.
\end{equation}

The couplings $A^{1,2:i}_{L,R}$ ($i=\nu$ or l) are given by the same
formula as for quarks (Eqs.(\ref{al1}-\ref{grpr})) where we have only put:
$Q_\nu=0$, $T^{L,R}_{3\nu}=+\frac{1}{2}$ for neutrinos and
$Q_l=-1$, $T^{L,R}_{3l}=-\frac{1}{2}$ for charged leptons.

The mixing matrices satisfy a few constrains which are useful in future
applications.
From the relation $V V^{\dagger}=I_{6\times 6}$ it follows that

\begin{equation}
\label{ul}
V^\nu_L V^{\nu \dagger}_L=V^\nu_R V^{\nu \dagger}_R=I_{3\times 3},\;\;
V^\nu_R V^{\nu T}_L=0.
\end{equation}

Then, from the definition of $K_{L,R}$ matrices we have

\begin{equation}
\label{kl} K_L^{\dagger} K_L=K_R^{\dagger} K_R=I_{3\times 3},\;\;
K^T_L K_R=K^T_R K_L=0 ,
\end{equation}

and

\begin{equation}
K_L K_L^{\dagger}= \Omega_L,\;\;K_R K_R^{\dagger}= \Omega_R.
\end{equation}

From the relations Eq.(\ref{ul}) and (\ref{kl}) there is also

\begin{equation}
K^T_R \Omega_L= K_L^T \Omega_R=\Omega_L \Omega^T_R=0 ,
\end{equation}

and

\begin{equation}
\Omega_L K_L=K_L,\;\;\;\Omega_R K_R=K_R.
\end{equation}

\vspace*{0.2cm}
From the relation $V^{\dagger}V=I_{6\times 6} $ we have

\begin{equation}
\Omega^\ast _L+\Omega_R=I_{6\times 6}.
\end{equation}

The relation $V^T M_\nu V= \left( M_\nu \right)_{diag} $ gives

\begin{equation}
V_L^{\nu \dagger} M_D V^\nu_R +\left( V_L^{\nu\dagger} M_D V^\nu_R
\right)^T+
V_R^{\nu T} M_R V^\nu_R=\left( M_\nu \right)_{diag}.
\end{equation}

But from $M_\nu =V^\ast \left( M_\nu \right)_{diag} V^\dagger $ it follows

\begin{eqnarray}
\label{ul1}
V^\nu_L \left( M_\nu \right)_{diag} V^{\nu T}_L &=& 0, \\
V^\nu_L \left( M_\nu \right)_{diag} V^{\nu \dagger}_R &=& M_D, \\
V^\nu_R \left( M_\nu \right)_{diag} V^{\nu T}_R &=& M^\ast_R.
\end{eqnarray}

From Eq.(\ref{ul1}) three other relations follow

\begin{equation}
\Omega_L \left( M_\nu \right)_{diag} \Omega_L^\ast=
\Omega_L \left( M_\nu \right)_{diag} K_L^\ast=
K^\dagger_L \left( M_\nu \right)_{diag} K_L^\ast=0
\end{equation}

All the above relations are useful when studying leptons - Higgs
particles couplings, and give the possibility to express unphysical
couplings by masses $\left( M_\nu \right)_{diag} $ and elements of
mixing matrices $K_{L,R},\; \Omega_{L,R}$.

\subsubsection*{\bf 2.2.2.2 Three and four gauge particles interactions.}

The structure of three and four gauge couplings is very similar to the
analogous couplings in the SM. There are only  different factors. To get
the coupling for physical gauge particles we use the Lagrangian
Eq.(\ref{l2}) where we introduce the physical fields $W^\pm_i$, $Z_i\;
(i=1,2)$ and A (Eqs.(\ref{fiz1}),(\ref{fiz2})).
For three gauge boson interactions we obtained:

\begin{eqnarray}
L_g^{3}&=&
\left\{
-ig
\left(
\partial^{x}_{\gamma} g_{\nu \delta }
-\partial^{x}_{\delta} g_{\nu \gamma }
-\partial^{y}_{\nu} g_{\gamma \delta }
+\partial^{y}_{\delta} g_{\nu \gamma }
+\partial^{z}_{\nu} g_{\gamma \delta }
-\partial^{z}_{\gamma} g_{\nu \delta }
\right)
\left[
\left(
y_1 \sin ^2\xi +x_1 \cos ^2\xi
\right)
\left(
W_1^{-\nu }(x)W_1^{+\gamma }(y)Z_{1}^{\delta }(z)
\right)
\right.
\right.
\nonumber  \\
&+&
\left(
y_2 \sin ^2\xi +x_2 \cos ^2\xi
\right)
\left(
W_1^{-\nu }(x)W_1^{+\gamma }(y)Z_{2}^{\delta }(z)
\right)
+
x_3
\left(
W_1^{-\nu }(x)W_1^{+\gamma }(y)A^{\delta }(z)
\right)
\nonumber  \\
&+&
\left(
y_1 \cos ^2\xi +x_1 \sin ^2\xi
\right)
\left(
W_2^{-\nu }(x)W_2^{+\gamma }(y)Z_{1}^{\delta }(z)
\right)
+
\left(
y_2 \cos ^2\xi +x_2 \sin ^2\xi
\right)
\left(
W_2^{-\nu }(x)W_2^{+\gamma }(y)Z_{2}^{\delta }(z)
\right)
\nonumber  \\
&+&
x_3
\left(
W_2^{-\nu }(x)W_2^{+\gamma }(y)A^{\delta }(z)
\right)
+
\left(
(x_1-y_1) \cos\xi \sin\xi
\right)
\left(
W_1^{-\nu }(x)W_2^{+\gamma }(y)Z_{1}^{\delta }(z)+
W_2^{-\nu }(x)W_1^{+\gamma }(y)Z_{1}^{\delta }(z)
\right)
\nonumber  \\
&+&
\left.
\left.
\left(
(x_2-y_2) \cos\xi \sin\xi
\right)
\left(
W_1^{-\nu }(x)W_2^{+\gamma }(y)Z_{2}^{\delta }(z)+
W_2^{-\nu }(x)W_1^{+\gamma }(y)Z_{2}^{\delta }(z)
\right)
\right]
\right\}
\mid_{x=y=z}.
\end{eqnarray}

The four gauge bosons interactions can be given in the comprehensive form:

\begin{eqnarray}
L_g^{4}=
&-&
g^2
\left(
2 g_{\mu \nu }g_{\gamma \delta }-g_{\mu \gamma }g_{\nu \delta }
-g_{\mu \delta }g_{\nu \gamma }
\right)
\left\{
\frac{1}{2}
\left[
\left(
y_1^2 \sin^2\xi + x_1^2 \cos^2\xi
\right)
\left(
W_1^{-\mu }W_1^{+\nu }Z_{1}^{\gamma }Z_{1}^{\delta }
\right)
\right.
\right.
\nonumber  \\
&+&
\left(
y_2^2 \sin^2\xi + x_2^2 \cos^2\xi
\right)
\left(
W_1^{-\mu }W_1^{+\nu }Z_{2}^{\gamma }Z_{2}^{\delta }
\right)
+
x_3^2
\left(
W_1^{-\mu }W_1^{+\nu }A^{\gamma }A^{\delta }
\right)
+
\left(
y_1^2 \cos^2\xi + x_1^2 \sin^2\xi
\right)
\left(
W_2^{-\mu }W_2^{+\nu }Z_{1}^{\gamma }Z_{1}^{\delta }
\right)
\nonumber  \\
&+&
\left(
y_2^2 \cos^2\xi + x_2^2 \sin^2\xi
\right)
\left(
W_2^{-\mu }W_2^{+\nu }Z_{2}^{\gamma }Z_{2}^{\delta }
\right)
+
x_3^2
\left(
W_2^{-\mu }W_2^{+\nu }A^{\gamma }A^{\delta }
\right) \nonumber  \\
&+&
\left(
(x_1^2-y_1^2 ) \cos\xi \sin\xi
\right)
\left(
W_1^{-\mu }W_2^{+\nu }Z_{1}^{\gamma }Z_{1}^{\delta }
+W_2^{-\mu }W_1^{+\nu }Z_{1}^{\gamma }Z_{1}^{\delta }
\right)
\nonumber  \\
&+&
\left.
\left(
(x_2^2-y_2^2 ) \cos\xi \sin\xi
\right)
\left(
W_1^{-\mu }W_2^{+\nu }Z_{2}^{\gamma }Z_{2}^{\delta }
+W_2^{-\mu }W_1^{+\nu }Z_{2}^{\gamma }Z_{2}^{\delta }
\right)
\right]
\nonumber  \\
&+&
\left(
y_1 y_2 \sin^2\xi + x_1 x_2 \cos^2\xi
\right)
\left(
W_1^{-\mu }W_1^{+\nu }Z_{1}^{\gamma }Z_{2}^{\delta }
\right)
+
\left(
y_1 y_3 \sin^2\xi + x_1 x_3 \cos^2\xi
\right)
\left(
W_1^{-\mu }W_1^{+\nu }Z_{1}^{\gamma }A^{\delta }
\right) \nonumber  \\
&+&
\left(
y_2 y_3 \sin^2\xi + x_2 x_3 \cos^2\xi
\right)
\left(
W_1^{-\mu }W_1^{+\nu }Z_{2}^{\gamma }A^{\delta }
\right)
+
\left(
y_1 y_2 \cos^2\xi + x_1 x_2 \sin^2\xi
\right)
\left(
W_2^{-\mu }W_2^{+\nu }Z_{1}^{\gamma }Z_{2}^{\delta }
\right) \nonumber  \\
&+&
\left(
y_1 y_3 \cos^2\xi + x_1 x_3 \sin^2\xi
\right)
\left(
W_2^{-\mu }W_2^{+\nu }Z_{1}^{\gamma }A^{\delta }
\right)
+
\left(
y_2 y_3 \cos^2\xi + x_2 x_3 \sin^2\xi
\right)
\left(
W_2^{-\mu }W_2^{+\nu }Z_{2}^{\gamma }A^{\delta }
\right)
\nonumber  \\
&+&
\left(
(x_1 x_2-y_1 y_2 ) \cos\xi \sin\xi
\right)
\left(
W_1^{-\mu }W_2^{+\nu }Z_{1}^{\gamma }Z_{2}^{\delta }
+W_2^{-\mu }W_1^{+\nu }Z_{1}^{\gamma }Z_{2}^{\delta }
\right) \nonumber  \\
&+&
\left(
(x_1 x_3-y_1 y_3 ) \cos\xi \sin\xi
\right)
\left(
W_1^{-\mu }W_2^{+\nu }Z_{1}^{\gamma }A^{\delta }
+W_2^{-\mu }W_1^{+\nu }Z_{1}^{\gamma }A^{\delta }
\right)
\nonumber  \\
&+&
\left(
(x_2 x_3-y_2 y_3 ) \cos\xi \sin\xi
\right)
\left(
W_1^{-\mu }W_2^{+\nu }Z_{2}^{\gamma }A^{\delta }
+W_2^{-\mu }W_1^{+\nu }Z_{2}^{\gamma }A^{\delta }
\right)
\nonumber  \\
&-&
\frac{1}{2}
\biggl[
\frac{1}{2}
\left(
\sin^4\xi+ \cos^4\xi
\right)
\left(
W_1^{-\mu }W_1^{-\nu }W_1^{+\gamma }W_1^{+\delta }
+W_2^{-\mu }W_2^{-\nu }W_2^{+\gamma }W_2^{+\delta }
\right)
\nonumber  \\
&+&
\left(
-\sin^3\xi \cos\xi +\cos^3\xi \sin\xi
\right)
\left(
W_1^{-\mu }W_1^{-\nu }W_1^{+\gamma }W_2^{+\delta }+
W_1^{-\mu }W_2^{-\nu }W_1^{+\gamma }W_1^{+\delta }
\right)
\nonumber  \\
&+&
\left(
-\cos^3\xi \sin\xi +\sin^3\xi \cos\xi
\right)
\left(
W_2^{-\mu }W_2^{-\nu }W_2^{+\gamma }W_1^{+\delta }+
W_2^{-\mu }W_1^{-\nu }W_2^{+\gamma }W_2^{+\delta }
\right)
\nonumber  \\
&+&
\sin^2\xi \cos^2\xi
\left(
W_2^{-\mu }W_2^{-\nu }W_1^{+\gamma }W_1^{+\delta }+
W_1^{-\mu }W_1^{-\nu }W_2^{+\gamma }W_2^{+\delta }
\right)
\biggr]
-2
\sin^2\xi \cos^2\xi \;
W_1^{-\mu }W_2^{-\nu }W_2^{+\gamma }W_1^{+\delta }
\biggr\}.
\end{eqnarray}

\subsubsection*{\bf 2.2.2.3 Fermion - scalar boson interactions.}

The fermion - Higgs particles interactions are given by Lagrangian
Eq.(\ref{yukava}) for quarks and Eq.(\ref{yuklep}) for leptons.
We have to express the weak fields by the physical ones and the Yukawa
couplings by fermion masses and CKM mixing matrices elements.
For quark - Higgs particles couplings we have

\begin{equation}
- \sum\limits_{i,j}\bar{Q}_{iL} \left( (h_q)_{ij}\phi+
(\tilde{h}_q)_{ij}\tilde{\phi}\right) Q_{jR}+h.c. =L^q_{mass}+
L_{quark-Higgs},
\end{equation}

where $L^q_{mass}$ is given by Eq.(\ref{mass}), and

\begin{eqnarray}
L_{quark-Higgs}(u,d) = &-& \bar{U}
\left[
P_L \left( M^u_{diag}B_0^{\ast}+U_R^{CKM}M^d_{diag}
U_L^{CKM\dagger}A_0 \right)
\right.\nonumber \\
&+&
\left.
P_R \left( M^u_{diag}B_0+U_L^{CKM}M^d_{diag}
U_R^{CKM\dagger}A_0^{\ast} \right)
\right]
U  \nonumber \\
&-& \bar{D}
\left[
P_L \left( M^d_{diag}B_0+U_R^{CKM\dagger}M^u_{diag}
U_L^{CKM}A_0^{\ast} \right)
\right.\nonumber \\
&+&
\left.
P_R \left(
M^d_{diag}B_0^{\ast}+U_L^{CKM\dagger}M^u_{diag}
U_R^{CKM}A_0 \right)
\right]
D
\nonumber \\
&-& \bar{U}
\left[
P_L
\left( M^u_{diag}U_L^{CKM}B^+-U_R^{CKM}M^d_{diag} A^+ \right)
\right.\nonumber \\
&+&
\left.
P_R
\left( M^u_{diag}U_R^{CKM}A^+-U_L^{CKM}M^d_{diag}
B^+ \right)
\right]
D  \nonumber \\
&-& \bar{D}
\left[
P_L
\left( U_R^{CKM\dagger}M^u_{diag}A^{-}-M^d_{diag}
U_L^{CKM\dagger}B^{-} \right)
\right.\nonumber \\
&+&
\left.
P_R \left(
U_L^{CKM\dagger}M^u_{diag}B^{-}-
M^d_{diag} U_R^{CKM\dagger}A^{-} \right)
\right]
U.
\end{eqnarray}

The $A_0,\;B_0,\;A^\pm $ and $B^\pm $ are combinations of physical
Higgs fields and Goldstone bosons:

\begin{eqnarray}
A_0 &=& \frac{\sqrt{2}}{\kappa_-^2}
\left(
\kappa_1 \phi^0_2 -\kappa_2 \phi^{0\ast}_1
\right) =
\frac{\kappa_+}{\kappa_-^2}
\left(
H^0_0b_0+H^0_1b_1+H^0_2b_2-iA^0_1
\right), \nonumber \\
B_0 &=& \frac{\sqrt{2}}{\kappa_-^2}
\left(
\kappa_1 \phi^0_1 -\kappa_2 \phi^{0\ast}_2
\right) \nonumber \\
&=&
\frac{1}{\kappa^2_-\kappa_+}
\left[
H^0_0
\left(
a_0 \kappa^2_- - 2\kappa_1 \kappa_2 b_0
\right)+
H^0_1
\left(
a_1 \kappa^2_- - 2\kappa_1 \kappa_2 b_1
\right)+
H^0_2
\left(
a_2 \kappa^2_- - 2\kappa_1 \kappa_2 b_2
\right)+
i\kappa_-^2\tilde{G}^0_1  -i2\kappa_1 \kappa_2 A^0_1
\right], \nonumber \\
A^\pm &=&
\frac{\sqrt{2}}{\kappa_-^2}
\left(
\kappa_1 \phi^\pm_1 +\kappa_2 \phi^{\pm}_2
\right) =
\left[
\frac{4\kappa^4_+ v^2_R}
{\kappa^4_-\left( 2\kappa^2_+ v^2_R +\kappa^4_-\right)}
\right]^{\frac{1}{2}}
H^\pm_2 -
\left[
\frac{2\kappa^2_+ }{\kappa^4_- +2\kappa^2_+ v^2_R }
\right]^{\frac{1}{2}}
G^\pm_R, \nonumber \\
B ^\pm &=&
\frac{\sqrt{2}}{\kappa_-^2}
\left(
\kappa_2 \phi^\pm_1 +\kappa_1 \phi^{\pm}_2
\right) =
\frac{4\kappa_1 \kappa_2 v_R}{\kappa^2_-\sqrt{2\kappa^2_+ v^2_R +\kappa^4_-}}
H^\pm_2-
\frac{2\sqrt{2}\kappa_1 \kappa_2}{\kappa_+
\sqrt{2\kappa^2_+ v^2_R +\kappa^4_-}}
G^\pm_R+
\frac{\sqrt{2}}{\kappa_+}
G^\pm_L.
\end{eqnarray}

For lepton - Higgs particle couplings there is

\begin{equation}
-\left\{ \bar{L}_L \left[ h_l\phi + \tilde{h}_l\tilde{\phi }
\right] L_R   +h.c. \right\}- \bar{L}^c_R \Sigma_L h_M
L_L-\bar{L}^c_L \Sigma_R h_M L_R=
L^\nu_{mass}+L^l_{mass}+L_{lepton - Higgs},
\end{equation}

where $L^\nu_{mass}\;(L^l_{mass})$ are given by Eq.(\ref{massnu}),
(Eq.(\ref{massl})) and

\begin{eqnarray}
L_{lepton - Higgs} = &-& \bar{\nu}^\prime_L \left(
h_l\phi_1^0+\tilde{h}_l{\phi_2^0}^{\ast} \right) \nu^\prime_R -
\bar{\nu}^\prime_L \left( h_l\phi_1^+ -\tilde{h}_l\phi_2^+ \right)
l^\prime_R - \bar{l}^\prime_L \left( h_l\phi_2^-
-\tilde{h}_l\phi_1^- \right) \nu^\prime_R  - \bar{l}^\prime_L
\left( h_l{\phi_2^0}+\tilde{h}_l{\phi_1^0}^{\ast} \right)
l^\prime_R \nonumber \\ &-&\delta_L^0 \bar{\nu}_R^{\prime
c}h_M\nu^\prime_L +
\frac{\delta_L^+}{\sqrt{2}}\left(\bar{\nu}_R^{\prime c}
h_Ml^\prime_L+ \bar{l}_R^{\prime c}h_M\nu^\prime_L \right)+
\delta_L^{++}\bar{l}_R^{\prime c}h_Ml^\prime_L \nonumber \\ & - &
\delta_R^0 \bar{\nu}_L^{\prime c}h_M\nu^\prime_R +
\frac{\delta_R^+}{\sqrt{2}}\left(\bar{\nu}_L^{\prime c}
h_Ml^\prime_R+\bar{l}_L^{\prime c}h_M\nu^\prime_R \right)+
\delta_R^{++}\bar{l}_L^{\prime c}h_Ml^\prime_R +h.c.\;.
\end{eqnarray}

Now in each term we can introduce the physical fields and the
physical parameters.\\ For two Majorana neutrinos - neutral Higgs
particle interaction we have:

\begin{eqnarray}
&&\bar{\nu}^\prime_L \left( h_l\phi_1^0+\tilde{h}_l{\phi_2^0}^{\ast} \right)
\nu^\prime_R +
h.c.= \hspace*{10.0cm} \nonumber \\
& \sum\limits_{a} & \bar{N}_a
\left\{
\left[
\left(
\Omega_L
\right)_{aa}m^N_a
B_0
+ \sum_{l}
\left(
K_L
\right)_{al}
\left(
K_R^{\ast}
\right)_{al}
m_lA_0^{\ast}
\right] P_R
+
\left[
m^N_a
\left(
\Omega_L
\right)_{aa}
B_0^{\ast} + \sum_{l}
\left(
K_L^{\ast}
\right)_{al}
\left(
K_R
\right)_{al}
m_lA_0
\right]
P_L
\right\}
N_a
\nonumber \\
&+&
\sum_{a>b,\;c} \bar{N}_a \left\{ \left[ \left( \left( \Omega_L
\right)_{ac}m^N_c \left( \Omega_R \right)_{cb} + \left( \Omega_L
\right)_{bc}m^N_c \left( \Omega_R \right)_{ca} \right) B_0
+ \left. \sum_{l} m_l \left( \left( K_L \right)_{al} \left( K_R^{\ast}
\right)_{bl} + \left( K_L \right)_{bl} \left( K_R^{\ast} \right)_{al}
\right) A_0^{\ast} \right] P_R  \right.\right.
\nonumber \\
&+&
\left.
\left[ \left( \left( \Omega_L \right)_{ac}m^N_c \left( \Omega_R
\right)_{cb} + \left( \Omega_L \right)_{bc}m^N_c \left( \Omega_R
\right)_{ca} \right) B_0^{\ast}
+  \sum_{l} m_l \left( \left( K_L^{\ast} \right)_{bl} \left(
K_R \right)_{al} + \left( K_L^{\ast} \right)_{al} \left( K_R \right)_{bl}
\right) A_0 \right] P_L \right\} N_b,
\end{eqnarray}

\begin{eqnarray}
\hspace*{-2.0cm} \delta_R^0\bar{\nu}_L^{\prime c}h_M\nu^\prime_R +
h.c. &=& \frac{1}{2v_R} \sum_{a} \bar{N}_a \left( \sum_{c} m_c^N
\left[ \left( \Omega_R \right)^2_{ca} P_R \delta_R^0+ \left(
\Omega_R^{\ast} \right)^2_{ca} P_L {\delta_R^0}^{\ast} \right]
\right) N_a \nonumber \\ &+& \frac{1}{v_R} \sum_{a>b} \bar{N}_a
\left( \sum_{c} m_c^N \left[ \left( \Omega_R \right)_{cb} \left(
\Omega_R \right)_{ca}P_R \delta_R^0+ \left( \Omega_R^{\ast}
\right)_{cb} \left( \Omega_R^{\ast} \right)_{ca} P_L
{\delta_R^0}^{\ast} \right] \right) N_b,
\end{eqnarray}

and

\begin{eqnarray}
\delta_L^0 \bar{\nu}_R^{\prime c}h_M\nu^\prime_L + h.c. =
\frac{1}{\sqrt{2}v_R} \sum_{a} \bar{N}_a
\left[
X_{aa}
\delta_L^0P_L
+
X^\ast_{aa}
{\delta_L^0}^{\ast}P_R
\right]
N_a
+ \frac{\sqrt{2}}{v_R} \sum_{a>b} \bar{N}_a
\left[
X_{ab}
\delta_L^0P_L
+
X^\ast_{ab}
{\delta_L^0}^{\ast}P_R
\right]
N_b ,
\end{eqnarray}

where the $6\times 6$ matrix X is given by

\begin{equation}
X=
\left(
K^\ast_L W^l K^T_R
\right)
\left( M_\nu \right)_{diag}
\left(
K_R W^l K^+_L
\right)=
X^T .
\end{equation}

Majorana neutrinos interact with charged leptons and charged Higgs particles
in the way:

\begin{eqnarray}
&&\bar{\nu}^\prime_L \left( h_l\phi_1^+ -\tilde{h}_l\phi_2^+ \right)
l^\prime_R + \bar{l}^\prime_L
\left( h_l\phi_2^- -\tilde{h}_l\phi_1^- \right) \nu^\prime_R + h.c.
=  \hspace*{12.0cm} \nonumber \\
&&\sum_{a,l}\bar{N}_a \left\{
\biggl[
\sum_{b} \left( \Omega_L \right)_{ab}
m_b^N \left( K_R \right)_{bl} A^+ -\left( K_L \right)_{al} m_l B^+
\biggr]
P_R
+
\biggl[
m^N_a \left( K_L \right)_{al} B^+ -\left( K_R
\right)_{al}m_l A^+
\biggr]
P_L \right\} l_l  \nonumber \\
&+&\sum_{a,l}\bar{l}_l \left\{
\biggl[
\sum_{b} \left( K_R^{\dagger}
\right)_{lb} m_b^N\left( \Omega_L \right)_{ba} A^{-}
-m_l \left( K_L^{\dagger}
\right)_{la} B^{-}
\biggr]
P_L
+
\left[
\left( K_L^{\dagger} \right)_{la} m_a^N
B^{-} - m_l \left(
K_R^{\dagger} \right)_{la} A^{-}
\right]
P_R \right\} N_a , \nonumber
\end{eqnarray}

\begin{equation}
\end{equation}

\begin{eqnarray}
\hspace*{-4.5cm} \frac{\delta_R^+}{\sqrt{2}}\left(
\bar{\nu}_L^{\prime c} h_Ml^\prime_R+ \bar{l}_L^{\prime
c}h_M\nu^\prime_R \right) + h.c. = &&\frac{1}{v_R} \sum_{a,l}
\left\{ \bar{N}_a \left[ \sum_{b} \left( \Omega_R^ {\ast}
\right)_{ab} m_b^N \left( K_R \right)_{bl} \right] P_Rl_l
\delta_R^+ \right.   \nonumber \\ &+& \left. \bar{l}_l \left[
\sum_{b} \left( K_R^{\dagger} \right)_{lb} m_b^N \left(
\Omega_R^{\ast} \right)_{ba} \right] P_LN_a \delta_R^- \right\},
\end{eqnarray}

and

\begin{eqnarray}
\hspace*{-2.0cm}
\frac{\delta_L^+}{\sqrt{2}}\left(\bar{\nu}_R^{^\prime c}
h_Ml^\prime_L+\bar{l}_R^{\prime c}h_M\nu^\prime_L
\right) + h.c. =
\frac{1}{v_R}
\sum_{a,l}
\left\{
\delta_L^+
\left[
\bar{N}_a
\left(
X K_L
\right)_{al}
P_Ll_l
\right]
+
\delta_L^-
\left[
\bar{l}_l
\left(
 K^+_L X^\ast
\right)_{la}
P_RN_a
\right]
\right\}.
\end{eqnarray}

Finally, two charged leptons interact with neutral Higgs particles:

\begin{eqnarray}
\hspace*{-2.0cm}
&&\bar{l}^\prime_L \left( h_l{\phi_2^0}+\tilde{h}_l{\phi_1^0}^{\ast} \right)
l^\prime_R
+ h.c.=
\nonumber \\
& \sum_{l,k}& \bar{l}_l \left\{ \left[ \delta_{lk} m_l B_0^{\ast}
+ \sum_{a}
\left( K_L^{\ast} \right)_{al} \left( K_R \right)_{ak} m_a^N A_0 \right] P_R
 +  \left[ \delta_{lk} m_l B_0 + \sum_{a} \left( K_L \right)_{ak}
\left( K_R^{\ast} \right)_{al} m_a^N A_0^{\ast} \right] P_L \right\}
l_k,
\end{eqnarray}

and doubly charged Higgs particles

\begin{eqnarray}
&& \delta_R^{++}\bar{l}_L^{\prime c}h_Ml^\prime_R +h.c. =\hspace*{14.2cm}
\nonumber \\
&& \frac{1}{\sqrt{2}v_R}\sum\limits_{l,k}
\left\{
\delta_R^{++}
\left[
l^T_{l}C
\left(
K^T_R \left( M_\nu \right)_{diag} K_R
\right)_{lk}
P_R
l_{k}
\right]
+
\delta_R^{--}
\left[
\bar{l}_{l}
\left(
K^+_R \left( M_\nu \right)_{diag} K_R^\ast
\right)_{lk}
P_L
C\bar{l}^T_{k}
\right]
\right\},
\end{eqnarray}

\begin{eqnarray}
\hspace*{-2.0cm}
\delta_L^{++}\bar{l}_R^{\prime c}h_Ml^\prime_L +h.c. =
\frac{1}{\sqrt{2}v_R}\sum\limits_{l,k}
\left\{
\delta_L^{++}
\left[
l^T_{k}C
\left(
K_L^T X K_L^\ast
\right)_{kl}
P_L l_{l}
\right]+
\delta_L^{--}
\left[
\bar{l}_{k}
\left(
K_L^T X^\ast K_L^\ast
\right)_{kl}
P_R
C\bar{l}^T_{l}
\right]
\right\} .
\end{eqnarray}

%%%%%%%%%%%%%%%%%%%%%%%%%%%%%%%%%%%%%%%%%%%%%%%%%%%%%%%%%%%%%%

\subsubsection*{\bf 2.2.2.4 Gauge bosons - Higgs particles interactions.}

The interactions among gauge bosons $ W_i^\pm,\;Z_i,\;A
\;\;(i=1,2)$ and Higgs particles $H^0_i$, \\
$(i=0,1,2,3)\;\;A^0_k,\;A^\pm_k$,
$(i=1,2),\;\;\delta_{L,R}^{\pm\pm} $ together with Goldstone
bosons are given by the Lagrangian $L^{kin}_{Higgs}$
(Eq.(\ref{tr})). After SSB (Eq.(\ref{vev})) we obtain

\begin{equation}
L^{kin}_{Higgs}=L^{(2)}_{Higgs}+L^{gauge}_{Higgs}.
\end{equation}

The Lagrangian $L^{(2)}_{Higgs}$ contains four components:

\begin{equation}
L^{(2)}_{Higgs}=L_H^{kin}+L_{GB}^{kin}+L_{bilinear}+L_M.
\end{equation}

The $L_H^{kin}$ term is the kinetic energy for Higgs particles

\begin{eqnarray}
L_H^{kin} &=& \frac{1}{2} \sum_{i=0}^3
\left( \partial^\mu H^0_i \right)
\left( \partial_\mu H^0_i \right)+
\frac{1}{2} \sum_{i=0}^2
\left( \partial^\mu A^0_i \right)
\left( \partial_\mu A^0_i \right) \nonumber \\
&+&
\sum_{i=1}^2
\left( \partial^\mu H^+_i \right)
\left( \partial_\mu H^-_i \right)
+
\left( \partial^\mu \delta_L^{++} \right)
\left( \partial_\mu \delta_L^{--} \right)+
\left( \partial^\mu \delta_R^{++} \right)
\left( \partial_\mu \delta_R^{--} \right),
\end{eqnarray}

and

\begin{eqnarray}
\label{gb1}
L_{GB}^{kin}=
\left( \partial^\mu G^+_L \right)
\left( \partial_\mu G^-_L \right)+
\left( \partial^\mu G^+_R \right)
\left( \partial_\mu G^-_R \right)+
\frac{1}{2}
\left( \partial^\mu \tilde{G}^0_1 \right)
\left( \partial_\mu \tilde{G}^0_1 \right)+
\frac{1}{2}
\left( \partial^\mu \tilde{G}^0_2 \right)
\left( \partial_\mu \tilde{G}^0_2 \right)
\end{eqnarray}

is the kinetic energy for Goldstone bosons. $L_M$ is the mass
Lagrangian of the Gauge bosons (Eq.(\ref{lm})). $L_{bilinear} $
contains bilinear terms between Gauge and Higgs particles and is
given by

\begin{eqnarray}
\label{bili}
L_{bilinear} &=& \frac{ig}{2\sqrt{2}}
\left[
\kappa_1 \partial^\mu
\left(
\phi^0_1-\phi^{0\ast }_1
\right)-
\kappa_2 \partial^\mu
\left(
\phi^0_2-\phi^{0\ast }_2
\right)
\right]
\left[
W_{3L\mu}-W_{3R\mu}
\right] \nonumber \\
&+&
\frac{ig}{2}
\left\{
\left[
\kappa_1 \left( \partial^\mu \phi^-_2 \right)-
\kappa_2 \left( \partial^\mu \phi^-_1 \right)
\right]
W^+_{L\mu}+
\left[
\kappa_2 \left( \partial^\mu \phi^+_1 \right)-
\kappa_1 \left( \partial^\mu \phi^+_2 \right)
\right]
W^-_{L\mu}
\right\} \nonumber \\
&+&
\frac{ig}{2}
\left\{
\left[
\kappa_1 \left( \partial^\mu \phi^-_1 \right)-
\kappa_2 \left( \partial^\mu \phi^-_2 \right)
\right]
W^+_{R\mu}+
\left[
\kappa_2 \left( \partial^\mu \phi^+_2 \right)-
\kappa_1 \left( \partial^\mu \phi^+_1 \right)
\right]
W^-_{R\mu}
\right\} \nonumber \\
&+&
ig\frac{v_R}{\sqrt{2}}
\left\{
\left( \partial^\mu \delta^+_R \right) W^-_{R\mu}-
\left( \partial^\mu \delta^-_R \right) W^+_{R\mu}
\right\}+
i\frac{v_R}{\sqrt{2}}
\left\{
\partial^\mu
\left[
\left(  \delta^{0\ast}_R \right) -
\left(  \delta^{0}_R \right)
\right]
\left(
g W_{3R\mu}-g'B_\mu
\right)
\right\} .
\end{eqnarray}

The gauge fixing Lagrangian (see next Chapter) will be chosen
to cancel the $L_{bilinear}$ which complicates the
calculation of Higgs particle propagators.

The remaining part $L^{gauge}_{Higgs}$ describes the interactions among
gauge bosons and Higgs particles. The number of couplings is large.
There are 324 couplings among four bosons.
We do not give all the couplings here explicitly.
Here we describe the procedure by which all couplings
can be obtained. We divide the $L^{gauge}_{Higgs}$ terms into
triple and quadruple couplings

\begin{equation}
L^{gauge}_{Higgs}=L^{(III)}+L^{(IV)}.
\end{equation}

\paragraph*{\bf III A. Three bosons coupling with one Higgs particle.}

There are 21 different couplings between two gauge and one charged
Higgs (Goldstone) particles which are obtained from the Lagrangian

\begin{eqnarray}
\label{podwojne}
&& \frac{g^2}{2}
\left\{
\left[
\phi^-_1 \kappa_2 -\phi^-_2 \kappa_1
\right]
W^\mu_{3R} W^+_{L\mu}+
\left[
\phi^-_2 \kappa_2 -\phi^-_1 \kappa_1
\right]
W^\mu_{3L} W^+_{R\mu}
\right\} \nonumber \\
&+&
\sqrt{2} g v_R W^+_{R\mu} \delta^-_R
\left[
g' B^\mu- \frac{1}{2} g W^\mu_{3R}
\right]-
\frac{g^2v_R}{\sqrt{2}} W^+_{R\mu} W^{\mu +}\delta_R^{--} + h.c.\;.
\end{eqnarray}

There are 20 couplings with two gauge bosons and one neutral Higgs particle.
They are obtained from the Lagrangian

\begin{equation}
\frac{g^2}{4\sqrt{2}}
\left[
\phi^0_1 \kappa_1 +\phi^0_2 \kappa_2
\right]
\left(
W_{3L\mu}-W_{3R\mu}
\right)
\left(
W^\mu_{3L}-W^\mu_{3R}
\right)+
\frac{v_R}{\sqrt{2}} \delta^0_R
\left(
g W_{3R\mu}-g'B_\mu
\right)
\left(
g W^\mu_{3R}-g'B^\mu
\right)
+ h.c.
\end{equation}

and

\begin{eqnarray}
&& \frac{g^2}{2\sqrt{2}}
\left[
\phi^0_1 \kappa_1 +\phi^0_2 \kappa_2
\right]
\left(
W^+_{L\mu}W^{-\mu}_{L}+W^+_{R\mu}W^{-\mu}_{R}
\right)-
\frac{g^2}{\sqrt{2}}
\left[
\phi^0_1 \kappa_2 +\phi^{0\ast}_2 \kappa_1
\right]
W^{-\mu}_{L}W^{+\mu}_{R} \nonumber \\
&+&
\frac{g^2 v_R}{\sqrt{2}} \delta^0_R W^+_{R\mu}W^{-\mu}_{R} +h.c.\;.
\end{eqnarray}

It is worth to stress that the combinations $W_{3L\mu}-W_{3R\mu}$
and $g W_{3R\mu}-g'B_\mu $ do not contain the photon field. As it should be,
the photon does not couple to chargeless particles. To understand the absence
of other couplings it is worth to explore the charge conjugation symmetry C.
The Gauge - Higgs particle part of the Lagrangian is C symmetric. We set
the P and C eigenvalues $J^{PC}=1^{--}$ to gauge bosons, and $0^{++}$ or
$0^{+-}$ to $H^0_i$ and $A^0_i\;\left( G^0_i \right)$ scalar Higgs particles
(Goldstone bosons) respectively. Then it is easy to understand that
the couplings of the type $A^0_i Z_k Z_j,\;\tilde{G}^0_i Z_k Z_j$
and also $\tilde{G}^0_i W^+_k W^-_k$ are absent as they are not symmetric
under C transformation.

\paragraph*{\bf III B. Couplings of one gauge boson and two
Higgs (Goldstone) bosons.}

There are 33 couplings of one gauge boson and two charged Higgs particles
which are given by the Lagrangian

\begin{eqnarray}
&& \frac{ig}{2}
\left[
\left( \partial^\mu \phi^+_1 \right)\phi^-_1-
\left( \partial^\mu \phi^-_2 \right)\phi^+_2
\right]
\left(
W_{3L\mu}+W_{3R\mu}
\right)-
ig'
\left[
\left( \partial^\mu \delta^-_R \right) \delta^+_R+
\left( \partial^\mu \delta^-_L \right) \delta^+_L
\right] B_\mu \nonumber \\
&-&
i
\left[
\left( \partial^\mu \delta^{--}_R \right) \delta^{++}_R
\right]
\left(
gW_{3R\mu}+g'B_\mu
\right)-
i
\left[
\left( \partial^\mu \delta^{--}_L \right) \delta^{++}_L
\right]
\left(
gW_{3L\mu}+g'B_\mu
\right) \nonumber \\
&-&
ig
\left[
\left( \partial^\mu \delta^+_R \right) \delta^{--}_R-
\left( \partial^\mu \delta^{--}_R \right) \delta^+_R
\right] W^+_{R\mu} -
ig
\left[
\left( \partial^\mu \delta^+_L \right) \delta^{--}_L-
\left( \partial^\mu \delta^{--}_L \right) \delta^+_L
\right] W^+_{L\mu}
+h.c.\;.
\end{eqnarray}

The other 38 couplings describe the interactions of one gauge boson with one
charged and one neutral Higgs (Goldstone) particles.
They follow from the Lagrangian

\begin{eqnarray}
&& \frac{ig}{\sqrt{2}}
\left\{
\left( \partial^\mu \phi^-_2 \right) \phi_1^{0\ast}-
\left( \partial^\mu \phi^-_1 \right) \phi_2^{0}+
\left( \partial^\mu \phi^0_2 \right) \phi_1^{-}-
\left( \partial^\mu \phi^{0\ast}_1 \right) \phi_2^{-}-
\sqrt{2}
\left[
\left( \partial^\mu \delta^-_L \right) \delta^{0}_L-
\left( \partial^\mu \delta^0_L \right) \delta^{-}_L
\right]
W^+_{L\mu}
\right\} \nonumber \\
&+&
\frac{ig}{\sqrt{2}}
\left\{
\left( \partial^\mu \phi^-_1 \right) \phi_1^{0}-
\left( \partial^\mu \phi^-_2 \right) \phi_2^{0\ast}+
\left( \partial^\mu \phi^{0\ast}_2 \right) \phi_2^{-}-
\left( \partial^\mu \phi^{0}_1 \right) \phi_1^{-}-
\sqrt{2}
\left[
\left( \partial^\mu \delta^-_R \right) \delta^{0}_R-
\left( \partial^\mu \delta^0_R \right) \delta^{-}_R
\right]
W^+_{R\mu}
\right\} + h.c.\;.
\end{eqnarray}

There are 20 couplings among one gauge boson and two neutral Higgs
(Goldstone) particles, given by the Lagrangian

\begin{eqnarray}
&& \frac{ig}{2}
\left[
\left( \partial^\mu \phi^0_1 \right) \phi_1^{0\ast}-
\left( \partial^\mu \phi^0_2 \right) \phi_2^{0\ast}
\right]
\left(
W_{3L\mu}-W_{3R\mu}
\right)+
i
\left[
\left( \partial^\mu \delta^{0\ast}_R \right) \delta^{0}_R
\right]
\left(
gW_{3R\mu}-g'B_\mu
\right) \nonumber \\
&+&
i
\left[
\left( \partial^\mu \delta^{0\ast}_L \right) \delta^{0}_L
\right]
\left(
gW_{3L\mu}-g'B_\mu
\right) +h.c.\;.
\end{eqnarray}

Here photon also does not couple to neutral particles. From C symmetry
there are no couplings of the type $H^0_iH^0_jZ_k$ or $\tilde{G}^0_1A^0_1
Z_i$.

\paragraph*{\bf IV. Couplings between two gauge bosons and two Higgs
(Goldstone) particles.}

In this case we divide all four bosons couplings into three classes with
two charged scalars, one charged and one neutral and two neutral
Higgs (Goldstone) particles. All 117 couplings among two gauge
bosons and two charged scalar particles are described by
the Lagrangian

\begin{eqnarray}
&& \frac{g^2}{4} \left[ \phi^+_2\phi^-_2+\phi^+_1\phi^-_1 \right]
\left( W_{3L\mu}+W_{3R\mu} \right) \left( W^\mu_{3L}+W^\mu_{3R}
\right)+ g'^2 \left[ \delta^+_L \delta^-_L+\delta^+_R \delta^-_R
\right] B_\mu B^\mu \nonumber \\ &+& \frac{g^2}{2} \left[
\phi^+_2\phi^-_2+\phi^+_1\phi^-_1 \right] \left(
W^-_{L\mu}W^{+\mu}_{L}+W^-_{R\mu}W^{+\mu}_{R} \right)+ 2g^2
\delta^+_R  \delta^-_R W^-_{R\mu}W^{+\mu}_{R}+ 2g^2 \delta^+_L
\delta^-_L W^-_{L\mu}W^{+\mu}_{L} \nonumber \\ &+& \left[
\delta^{++}_R \delta^{--}_R \right] \left[ \left( g W_{3R\mu}+g'
B_\mu \right) \left( g W^\mu_{3R}+g' B^\mu \right)+ g^2
W^-_{R\mu}W^{+\mu}_{R} \right] \nonumber \\ &+& \left[
\delta^{++}_L \delta^{--}_L \right] \left[ \left( g W_{3L\mu}+g'
B_\mu \right) \left( g W^\mu_{3L}+g' B^\mu \right)+ g^2
W^-_{L\mu}W^{+\mu}_{L} \right],
\end{eqnarray}

and

\begin{eqnarray}
&-& g^2
\left[
\phi^+_2\phi^+_1
\right]
W^-_{L\mu}W^{-\mu}_{R}-
g
\left[
\delta^{+}_R \delta^{--}_R
\right]
\left(
2 g' B^\mu +g W^\mu_{3R}
\right)
W^{+}_{R\mu} \nonumber \\
&-&
g
\left[
\delta^{+}_L \delta^{--}_L
\right]
\left(
2 g' B^\mu +g W^\mu_{3L}
\right)
W^{+}_{L\mu} +h.c.\;.
\end{eqnarray}

The 132 couplings among two gauge bosons and two scalars, one
neutral and one charged, are given by the Lagrangian

\begin{eqnarray}
&& \frac{g^2}{\sqrt{2}}
\left[
\phi^-_1 \phi^0_2 -\phi^-_2 \phi^{0\ast}_1
\right]
W^\mu_{3R} W^+_{L\mu}+
\frac{g^2}{\sqrt{2}}
\left[
\phi^-_2 \phi^{0\ast}_2 -\phi^-_1 \phi^{0}_1
\right]
W^\mu_{3L} W^+_{R\mu} \nonumber \\
&+&
g
\left[
\delta^{-}_R \delta^{0}_R
\right]
\left(
2 g' B^\mu - g W^\mu_{3R}
\right)
W^+_{R\mu}+
g
\left[
\delta^{-}_L \delta^{0}_L
\right]
\left(
2 g' B^\mu - g W^\mu_{3L}
\right)
W^+_{L\mu} \nonumber \\
&-&
g^2
\left[
\delta^{--}_R \delta^{0}_R
\right]
W^{+\mu}_{R}
W^{+}_{R\mu}-
g^2
\left[
\delta^{--}_L \delta^{0}_L
\right]
W^{+\mu}_{L}
W^{+}_{L\mu} +h.c.\;.
\end{eqnarray}

Finally, there are 75 couplings among two gauge bosons and two neutral
scalars. They are described by the Lagrangian

\begin{eqnarray}
&& \frac{g^2}{4}
\left[
\phi^0_1 \phi^{0\ast}_1 +\phi^0_2 \phi^{0\ast}_2
\right]
\left(
W_{3L\mu}-W_{3R\mu}
\right)
\left(
W^\mu_{3L}-W^\mu_{3R}
\right) +
\left[
\delta^{0}_R \delta^{0\ast}_R
\right]
\left(
g W_{3R\mu}-g'B_\mu
\right)
\left(
g W^\mu_{3R}-g'B^\mu
\right) \nonumber \\
&+&
\left[
\delta^{0}_L \delta^{0\ast}_L
\right]
\left(
g W_{3L\mu}-g'B_\mu
\right)
\left(
g W^\mu_{3L}-g'B^\mu
\right) +
g^2
\left[
\delta^{0}_L \delta^{0\ast}_L
\right]
W^{+}_{L\mu} W^{-\mu}_{L} \nonumber \\
&+&
g^2
\left[
\delta^{0}_R \delta^{0\ast}_R
\right]
W^{+}_{R\mu} W^{-\mu}_{R}
- g^2
\left[
\phi^{0}_1 \phi^{0\ast}_2
\right]
W^{-}_{L\mu} W^{+\mu}_{R}
- g^2
\left[
\phi^{0\ast}_1 \phi^{0}_2
\right]
W^{+}_{L\mu} W^{-\mu}_{R} .
\end{eqnarray}

From the structure of this Lagrangian it is easy to see that the
photon does not take part in this interaction. From charge conjugation
invariance it follows that only
couplings among two gauge particles and two scalars with the same
C parities appear. For this reason all couplings of the type
$H^0_i A^0_j Z_k Z_l $ or $ H^0_i \tilde{G}^0_j W^+_k W^-_k $ vanish.

\subsubsection*{\bf 2.2.2.5 Interaction among scalar bosons.}

From the Higgs potential Eq.(\ref{potencjal}) we can find all couplings
among Higgs and Goldstone bosons. The number of different couplings
is very large. However, in the one loop approximation for processes with
light external
fermions and gauge bosons only, the knowledge of scalar particle interactions
is useless. So we will not present them here. This means that we
also do not renormalize the parameters in the Higgs potential.

\part*{ III Quantization of the left-right symmetric \hspace*{1cm} model.}

The canonical quantization of any gauge theory is not straightforward.
For such theories the Lagrangian is singular and canonical conjugate momenta
do not exist. Then, it is impossible to define the Hamiltonian
of the system. The second problem is connected with the fact that gauge
fields, described by four vectors, have only two degrees of freedom.
Their fields must satisfy additional conditions which disagree with
the commutation relations. Finally, gauge particle equations of motion are
not reversible which causes problems with the definition of their
propagators. All these problems have been solved many years ago,
first for abelian gauge theories. In the fifties, the so called
gauge fixing condition was introduced \cite{gupta} and later the auxiliary
hermitian scalar fields B(x) were invented \cite{nakamishi}.
For non-abelian gauge theories the same procedure as for quantum
electrodynamics can not be applied. Feynman has shown \cite{feynman}
in 1963 that the physical S-matrix is not unitary.
Contrary to the abelian case, contributions from the unphysical
longitudinal and scalar modes of gauge bosons to intermediate states
do not cancel out. Feynman \cite{feynman} and later De Witt \cite{dewitt}
found
that the violation of unitarity can be eliminated by introducing
the missing contribution to loops from some massless scalar fermions.
A clear-cut explanation why such fictitious particles have to appear
was given by Faddeev and Popov \cite{faddeev} from the viewpoint
of path-integral formalism. Since then these fermions with strange
statistics are called Faddeev-Popov ghosts. The very important point
in proving the consistency of the theory (for quantization and also
for renormalization) are relations among Green functions
which follow from gauge invariance of the theory. Such relations are known
as Ward-Takahashi \cite{ward} for abelian theory and Slavnov-Taylor
\cite{slavnov} identities in non abelian case. Later it was shown
\cite{becchi} that the full Lagrangian of gauge theory, after adding
the gauge fixing and Faddeev-Popov terms, still possesses some global
symmetry. This symmetry is a remnant of local gauge invariance
and is known as BRST symmetry \cite{becchi}.
In the same time it was proved that all Slavnov-Taylor
identities can be found from the BRST invariance of the full Lagrangian
\cite{brandt}. There are two different popular approaches to constructing the
quantum theory of gauge fields. The first, very fashionable approach
bases on the path integral formalism. Almost all textbooks on the
non-abelian gauge theory are based on this approach \cite{weinberg}.
The second one is well known from QED, the canonical quantization
formalism. In the context of non-abelian gauge theories it was introduced
by Kugo and Ojima \cite{kugo} and is known as operator formalism
\cite{ojima}. Here  this approach is adopted. The left-right model is
the gauge theory with spontaneous symmetry breaking. To define the gauge
fixing Lagrangian we have to remember that as remnant of SSB some
bilinear terms appear (Eq.(\ref{podwojne})).
In the next Chapter the gauge fixing Lagrangian is introduced. The
ghost particles, BRST transformations and
the Faddeev - Popov Lagrangian of the theory are considered in Chapter 3.2.

\section*{ 3.1 The gauge fixing Lagrangian.}

It was mentioned in section 2.2.2.4 that bilinear terms with gauge
and Higgs particles in $L_{bilinear}$ (Eq.(\ref{bili})) complicate
the Higgs particles propagators. They can be canceled by an
appropriate choice of the gauge fixing Lagrangian $(L_{GF})$.
Using the equations (\ref{fiz1}) and (\ref{fiz2}) for gauge bosons
and Eq.(\ref{rfns}) for Higgs and Goldstone particles the
$L_{bilinear}$ can be expressed in the form

\begin{eqnarray}
\label{bilinear1}
L_{bilinear}&=&
 Z_{1\mu}
\left\{
-\frac{g \kappa_+}{2 \cos{\Theta_W}}
\left(
\cos{\phi}+\sqrt{\cos {2\Theta_W}}\sin{\phi}
\right)
\partial^\mu \tilde{G}^0_1
+
\cot{\Theta_W}
\left(
-v_R g'\sin{\phi}
\right)
\partial^\mu \tilde{G}^0_2
\right\} \nonumber \\
&+&
Z_{2\mu}
\left\{
-\frac{g \kappa_+}{2 \cos{\Theta_W}}
\left(
\sin{\phi}-\sqrt{\cos {2\Theta_W}}\cos{\phi}
\right)
\partial^\mu \tilde{G}^0_1+
\cot{\Theta_W}
\left(
v_R g'\cos{\phi}
\right)
\partial^\mu \tilde{G}^0_2
\right\} \nonumber \\
&+&
igW^+_{1\mu}
\left\{
\left(
\frac{\kappa_+}{2} \cos{\xi} + \frac{\kappa_1 \kappa_2}{\kappa_+} \sin{\xi}
\right)
\partial^\mu {G}^-_L +
\left(
\frac{\sqrt{\kappa^4_- + 2v_R^2 \kappa^2_+}}{2 \kappa_+}\sin{\xi}
\right)
\partial^\mu {G}^-_R
\right\}
+h.c. \nonumber \\
&+&
igW^+_{2\mu}
\left\{
\left(
\frac{\kappa_+}{2} \sin{\xi} - \frac{\kappa_1 \kappa_2}{\kappa_+} \cos{\xi}
\right)
\partial^\mu {G}^-_L +
\left(
-\frac{\sqrt{\kappa^4_- + 2v_R^2 \kappa^2_+}}{2 \kappa_+}\cos{\xi}
\right)
\partial^\mu {G}^-_R
\right\}
+h.c. \;.\nonumber \\
\end{eqnarray}

Now we can introduce the gauge fixing Lagrangian

\begin{eqnarray}
\label{gf1}
L_{GF}=&-& \frac{1}{\xi_{W_1}} \mid \partial^\mu W^+_{1\mu}+
i\xi_{W_1} M_{W_1} G^+_1 \mid^2
- \frac{1}{\xi_{W_2}} \mid \partial^\mu W^+_{2\mu}+
i\xi_{W_2} M_{W_2} G^+_2 \mid^2
\nonumber \\
&-& \frac{1}{2\xi_{Z_1}}
\left(
\partial^\mu Z_{1\mu}-\xi_{Z_1} M_{Z_1} G^0_{1}
\right)^2
- \frac{1}{2\xi_{Z_2}}
\left(
\partial^\mu Z_{2\mu}-\xi_{Z_2} M_{Z_2} G^0_{2}
\right)^2
- \frac{1}{2\xi_{A}}
\left(
\partial^\mu A_\mu
\right)^2.
\end{eqnarray}

It is easy to check that the following terms which are present
 in Eq.(\ref{gf1})

\begin{eqnarray}
&&iM_{W_1}
\left[
\left(
\partial^\mu W^+_{1\mu}
\right)
G^-_1
\right]
+
iM_{W_2}
\left[
\left(
\partial^\mu W^+_{2\mu}
\right)
G^-_2
\right]\;+h.c.
\nonumber \\
&+&
M_{Z_1}
\left[
\left( \partial^\mu Z_{1\mu} \right)
G^0_{1}
\right]
+
M_{Z_2}
\left[
\left( \partial^\mu Z_{2\mu} \right)
G^0_{2}
\right]
\end{eqnarray}

cancel $L_{bilinear}$ in Eq.(\ref{bilinear1}) if the following
Goldstone particles transformations are made

\begin{equation}
\left(
\begin{array}{c}
{G}_{1}^{\pm} \\
\\
{G}_{2}^{\pm}
\end{array}
\right)
\equiv \left(
\begin{array}{cc}
\cos \psi_c & -\sin \psi_c \\
&  \\
\sin \psi_c & \cos \psi_c
\end{array}
\right) \left(
\begin{array}{c}
{G}_{R}^{\pm} \\
\\
{G}_{L}^{\pm}
\end{array}
\right),
\end{equation}

\begin{eqnarray}
\cos {\psi_c} &=& \frac{g}{2 M_{W_1} \kappa_+}
\sqrt{\kappa_-^2 +2 v_R^2 \kappa_+^2}\; \sin{\xi }, \nonumber \\
\sin {\psi_c} &=& -\frac{g}{2 M_{W_1} \kappa_+}
\left(
\kappa_+^2 \cos{\xi } + 2 \kappa_1 \kappa_2 \sin{\xi }
\right),
\end{eqnarray}

and analogously in the neutral sector

\begin{equation}
\left(
\begin{array}{c}
{G}^{o}_{1} \\
\\
{G}^{o}_{2}
\end{array}
\right)  \equiv \left(
\begin{array}{cc}
\cos \psi_n & \sin \psi_n \\
&  \\
\sin \psi_n & -\cos \psi_n
\end{array}
\right) \left(
\begin{array}{c}
{\tilde {G}}^{o}_{1} \\
\\
{\tilde {G}}^{o}_{2}
\end{array}
\right) ,
\end{equation}

\begin{eqnarray}
\cos{\psi_n} &=& -\frac{g \kappa_+ }{2 M_{Z_1} \cos{\Theta_W}}
\left(
\cos{\phi } +\sqrt{\cos{2\Theta_W}} \sin{\phi }
\right),
\nonumber \\
\sin{\psi_n} &=& -\frac{1}{M_{Z_1}} \cot{\Theta_W} v_R g' \sin{\phi }.
\end{eqnarray}

Adding $L_{GF}$ (Eq.(\ref{gf1})), $L_{bilinear}$
(Eq.(\ref{bilinear1})) and $L_{GB}^{kin}$ (Eq.(\ref{gb1})) one
gets

\begin{eqnarray}
\label{suma3}
L_{GF}+L_{bilinear}+L_{GB}^{kin} &=&
-
\frac{1}{\xi_{W_1}} \mid \partial^\mu W^+_{1\mu} \mid^2
-\frac{1}{\xi_{W_2}} \mid \partial^\mu W^+_{2\mu} \mid^2
-\frac{1}{2\xi_{Z_1}} \left( \partial^\mu Z_{1\mu} \right)^2
-\frac{1}{2\xi_{Z_2}} \left( \partial^\mu Z_{2\mu} \right)^2
\nonumber \\
&+&
\left( \partial^\mu G^+_1 \right) \left( \partial_\mu G^-_1 \right)
-\xi_{W_1} M^2_{W_1}G^+_1G^-_1
+
\left( \partial^\mu G^+_2 \right) \left( \partial_\mu G^-_2 \right)
-\xi_{W_2} M^2_{W_2}G^+_2G^-_2
\nonumber \\
&+&
\frac{1}{2}
\left( \partial^\mu G^0_1 \right) \left( \partial_\mu G^0_1 \right)
-
\frac{1}{2}
\xi_{Z_1} M^2_{Z_1} \left( G^0_1 \right)^2
+
\frac{1}{2}
\left( \partial^\mu G^0_2 \right) \left( \partial_\mu G^0_2 \right)
\nonumber \\
&-&
\frac{1}{2}
\xi_{Z_2} M^2_{Z_2} \left( G^0_2 \right)^2.
\end{eqnarray}

The gauge boson part from Eq.(\ref{suma3}) together with Eq.(\ref{l2})
and Eq.(\ref{lm}) give the kinetic energy Lagrangian for the physical
gauge bosons

\begin{eqnarray}
\label{gbprop}
&-&
\frac{1}{2} F^+_{1\mu \nu} F_1^{-\mu \nu}+M^2_{W_1} W^+_{1\mu} W_{1}^{-\mu}
-\frac{1}{\xi_{W_1}} \mid \partial^\mu W^+_{1\mu } \mid^2
\nonumber \\
&-&
\frac{1}{2} F^+_{2\mu \nu} F_2^{-\mu \nu}+M^2_{W_2} W^+_{2\mu} W_{2}^{-\mu}
-\frac{1}{\xi_{W_2}} \mid \partial^\mu W^+_{2\mu } \mid^2
\nonumber \\
&-&
\frac{1}{4} F_{1\mu \nu} F_1^{\mu \nu}+\frac{1}{2} M^2_{Z_1}
Z_{1\mu} Z_{1}^{\mu}
-\frac{1}{2\xi_{Z_1}} \left( \partial^\mu Z_{1\mu } \right)^2
\nonumber \\
&-&
\frac{1}{4} F_{2\mu \nu} F_2^{\mu \nu}+\frac{1}{2} M^2_{Z_2}
Z_{2\mu} Z_{2}^{\mu}
-\frac{1}{2\xi_{Z_2}} \left( \partial^\mu Z_{2\mu } \right)^2
\nonumber \\
&-&
\frac{1}{4} F_{\mu \nu} F^{\mu \nu}
-\frac{1}{2\xi_{A}} \left( \partial^\mu A_{\mu } \right)^2.
\end{eqnarray}

From Eq.(\ref{gbprop}) the propagators for gauge bosons are
obtained

\begin{equation}
i\Delta ^{\mu\nu}(p)=\frac{i}{p^2-M_{i}^{2}+i\varepsilon}
\left[
-g^{\mu\nu}+(1-\xi_i)\frac{p^{\mu}p^{\nu}}{p^{2}-\xi_i M_{i}^{2}}
\right]\;\;,
\end{equation}

where

\begin{equation}
M_{i}=\left\{
\begin{array}{ccl}
M_{W_{1,2}} & for & W^{\pm}_{1,2} \\
M_{Z_{1,2}} & for & Z_{1,2} \\
0 & for & A
\end{array}\;\;,\;\;\;
\right.
\xi_{i}=\left\{
\begin{array}{ccl}
\xi_{W_{1,2}} & for & W^{\pm}_{1,2} \\
\xi_{Z_{1,2}} & for & Z_{1,2} \\
\xi_A & for & A
\end{array}\;\;.
\right.
\end{equation}

The second part of Eq.(\ref{suma3}) gives the Goldstone bosons
propagators

\begin{equation}
i\Delta (p)=\frac{i}{p^2-\xi_i M_{i}^{2}+i\varepsilon},\;\;\;\; where \;\;\;\;
M_{i},\;(\xi_i)=\left\{
\begin{array}{ccl}
M_{Z_{1,2}},\;\;\left( \xi_{Z_{1,2}} \right) & for & {G}^{0}_{1,2} \\
M_{W_{1,2}},\;\;\left( \xi_{W_{1,2}} \right) & for & {G}^{\pm}_{1,2}
\end{array}\;\;\;.
\right.
\end{equation}

\section*{ 3.2 The ghost particles, BRST transformations and
Faddeev - Popov Lagrangian. }

To avoid the violation of unitarity ghost particles must be
introduced to the model. This is achieved by imposing
infinitesimal gauge transformations for gauge and Higgs fields
where instead of the gauge parameter we put the product
$\lambda\;c$. The $\lambda$ parameter is a Grassman number which
commutes with all gauge bosons as well as with Higgs particles but
anticommutes with the ghost fields "c"

\begin{equation}
\lambda^2=0,\;\;\;\{\lambda , c\}=0.
\end{equation}

The BRST transformations for gauge bosons are as follows:

\begin{eqnarray}
\label{brsgauge}
\delta^{BRST} W^i_{L,R \mu} &=& \lambda
\left(
\partial_\mu c^i_{L,R} + \sum_{e,f=1}^3 g \varepsilon^{ief} W^e_{L,R \mu}
c^f_{L,R}
\right),
\nonumber \\
\delta^{BRST} B_\mu &=& \lambda \left( \partial_\mu c^4 \right).
\end{eqnarray}

From Eq.(\ref{brsgauge}) one obtains (we define $\tilde{\delta}^{BRST}
\equiv \frac{\delta^{BRST}}{\lambda}$):

\begin{eqnarray}
\label{brsgauge1}
\tilde{\delta}^{BRST}W_{L,R}^{3\mu}&=&\partial^{\mu}c_{L,R}^3+
g(W_{L,R}^{1\mu}c_{L,R}^2 -W_{L,R}^{2\mu}c_{L,R}^1),
\nonumber \\
\tilde{\delta}^{BRST}W_{L,R}^{1,2\mu}&=&\partial^{\mu}c_{L,R}^{1,2}+
g(W_{L,R}^{2,3\mu} c_{L,R}^{3,1} -W_{L,R}^{3,1\mu}c_{L,R}^{2,3}),
\nonumber \\
\tilde{\delta}^{BRST}B^{\mu}&=& \partial^{\mu}c^4.
\end{eqnarray}

The ghost particles $c^1_{L,R}$, $c^2_{L,R}$, $c^3_{L,R}$, $c^4$
anticommute with each other and commute with all remaining fields.
According to Eqs.(\ref{phigauge},\ref{deltagauge}) there are the following
BRST transformations for scalar fields:

\begin{eqnarray}
\label{sc1brs}
\tilde{\delta}^{BRST}\phi&=&-ig_{R}c^a_{R}\phi\frac{\tau_{a}}{2}+
ig_{L}c^a_{L}\frac{\tau_{a}}{2}\phi \;\;\; a=1,2,3, \nonumber \\
\tilde{\delta}^{BRST}\Delta_{R,L}&=&
\frac{ig_{R}}{2}c^a_{R,L}(\tau_{a}\Delta_{R,L} +
\Delta_{R,L}\tau_{a}) + ig^{\prime}c^{4}\Delta_{R,L}, \;\;\;
a=1,2,3,
\end{eqnarray}

explicitly

\begin{eqnarray}
\label{sc2brs}
\tilde{\delta}^{BRST}\phi^{0}_{1}&=&
-\frac{ig}{2} (\sqrt{2}c^{-}_{R}\phi^{+}_{1} +
c^3_{R}\phi^{0}_{1} - \sqrt{2}c^{+}_{L}\phi^{-}_{2} - c^3_{L}\phi^{0}_{1}),
\nonumber \\
\tilde{\delta}^{BRST}\phi^{+}_{1}&=&-\frac{ig}{2} (\sqrt{2}c^{+}_{R}\phi^{0}_{1} -
c^3_{R}\phi^{+}_{1} - \sqrt{2}c^{+}_{L}\phi^{0}_{2} - c^3_{L}\phi^{+}_{1}) ,
\nonumber \\
\tilde{\delta}^{BRST}\phi^{-}_{2}&=&-\frac{ig}{2} (\sqrt{2}c^{-}_{R}\phi^{0}_{2} +
c^3_{R}\phi^{-}_{2} - \sqrt{2}c^{-}_{L}\phi^{0}_{1} + c^3_{L}\phi^{-}_{2})  ,
\nonumber \\
\tilde{\delta}^{BRST}\phi^{0}_{2} &=&-\frac{ig}{2} (\sqrt{2}c^{+}_{R}\phi^{-}_{2} -
c^3_{R}\phi^{0}_{2} - \sqrt{2}c^{-}_{L}\phi^{+}_{1} + c^3_{L}\phi^{0}_{2}),
\nonumber \\
\tilde{\delta}^{BRST}\delta^{+}_{R,L}&=&+ig(\delta^{0}_{R,L}c^{+}_{R,L} -
\delta^{++}_{R,L}c^{-}_{R,L}) + ig^{\prime}c^{4}\delta^{+}_{R,L},
\nonumber \\
\tilde{\delta}^{BRST}\delta^{++}_{R,L} &=& -ig(\delta^{+}_{R,L}c^{+}_{R,L}
-\delta^{++}_{R,L}c^3_{R,L}) + ig^{\prime}c^{4}\delta^{++}_{R,L},
\nonumber \\
\tilde{\delta}^{BRST}\delta^{o}_{R,L}&=&+ig\delta^{+}_{R,L}c^{-}_{R,L} +
ig^{\prime}c^{4}\delta^{o}_{R,L} - ig\delta^{o}_{R,L}c^3_{R,L},
\end{eqnarray}

where

\begin{equation}
c_{L,R}^{\pm}=\frac{1}{\sqrt{2}}(c_{L,R}^{1}\mp ic_{L,R}^{2}).
\end{equation}

Let us now rewrite the gauge fixing Lagrangian form Eq.(\ref{gf1})
using auxiliary fields "E" (from now we will not distinguish among
$\xi_{Z_{1,2}}$, $\xi_{W_{1,2}}$ and  $\xi_{A}$ except a case of
renormalization
of two point functions for unphysical particles):

\begin{eqnarray}
\label{gf22}
L_{GF} &=&
E^-_1
\left(
\partial^\mu W^+_{1\mu }+ C^+_{1E}
\right)
+
E^+_1
\left(
\partial^\mu W^-_{1\mu }+ C^-_{1E}
\right)
+\xi E^+_1 E^-_1
\nonumber \\
&+&
E^-_2
\left(
\partial^\mu W^+_{2\mu }+ C^+_{2E}
\right)
+
E^+_2
\left(
\partial^\mu W^-_{2\mu }+ C^-_{2E}
\right)
+\xi E^+_2 E^-_2
\nonumber \\
&+&
E_{Z_1}
\left(
\partial^\mu Z_{1\mu }+ C^{Z_1}_{E}
\right)
+
\frac{\xi}{2}  E^2_{Z_1}+
E_{Z_2}
\left(
\partial^\mu Z_{2\mu }+ C^{Z_2}_{E}
\right)
+
\frac{\xi}{2}  E^2_{Z_2} +
E_A
\left(
\partial^\mu A_{\mu }
\right)
+
\frac{\xi}{2}  E^2_{A} ,
\end{eqnarray}

where

\begin{eqnarray}
\label{gf33}
C^+_{1E} &=& i \xi M_{W_1} G^+_1,
\nonumber \\
C^+_{2E} &=& i \xi M_{W_2} G^+_2,
\nonumber \\
C^{Z_1}_E &=& - \xi M_{Z_1} G^0_1,
\nonumber \\
C^{Z_2}_E &=& - \xi M_{Z_2} G^0_2.
\end{eqnarray}

The Faddeev - Popov Lagrangian ($L_{FP}$) is constructed
by requiring that the sum
$L_{GF}+L_{FP} $ is BRST invariant.
So, first we have to find the BRST transformation for $L_{GF}$
(for all auxiliary fields $\tilde{\delta}^{BRST}
E=0$ is imposed, where $E=E^\pm_{1,2},E_{Z_{1,2}},E_A$).
To do this one has to know:

\begin{eqnarray}
\label{brs11}
\tilde{\delta}^{BRST}
\left(
\partial^\mu W^+_{1\mu }+ C^+_{1E}
\right)
&\equiv &
D^+_1,
\nonumber \\
\tilde{\delta}^{BRST}
\left(
\partial^\mu W^-_{1\mu }+ C^-_{1E}
\right)
&\equiv &
D^-_1,
\nonumber \\
\tilde{\delta}^{BRST}
\left(
\partial^\mu W^+_{2\mu }+ C^+_{2E}
\right)
&\equiv &
D^+_2,
\nonumber \\
\tilde{\delta}^{BRST}
\left(
\partial^\mu W^-_{2\mu }+ C^-_{2E}
\right)
&\equiv &
D^-_2,
\nonumber \\
\tilde{\delta}^{BRST}
\left(
\partial^\mu Z_{1\mu }+ C^{Z_1}_E
\right)
&\equiv &
D_{Z_1},
\nonumber \\
\tilde{\delta}^{BRST}
\left(
\partial^\mu Z_{2\mu }+ C^{Z_2}_E
\right)
&\equiv &
D_{Z_1},
\nonumber \\
\tilde{\delta}^{BRST}
\partial^\mu A_{\mu }
&\equiv &
D_{A}.
\end{eqnarray}

To calculate BRST transformations in Eq.(\ref{brs11})
the $\tilde{\delta}^{BRST} W^{\pm}_{1,2\mu}$, $\tilde{\delta}^{BRST} Z_{1,2\mu}$,
$\tilde{\delta}^{BRST} A_{\mu}$, $\tilde{\delta}^{BRST} G^{\pm}_{1,2}$ and
$\tilde{\delta}^{BRST} G^{0}_{1,2}$ are necessary.
The BRST transformations for the physical gauge particles are obtained
from Eq.(\ref{brsgauge1}) using Eqs.(\ref{fiz1},\ref{fiz2}) for gauge
bosons as well as for ghost particles, i.e.:

\begin{equation}
\left(
\begin{array}{c}
c^{\pm}_{L} \\
c^{\pm}_{R}
\end{array}
\right)=\left(
\begin{array}{cc}
cos\xi & sin\xi \\
-sin\xi & cos\xi
\end{array}
\right)\left(
\begin{array}{c}
c^{\pm}_{1} \\
c^{\pm}_{2}
\end{array}
\right),
\end{equation}

\begin{equation}
\left(
\begin{array}{c}
c^{3}_{L} \\
c^{3}_{R} \\
c^{4}
\end{array}
\right)= \left(
\begin{array}{ccc}
x_{1} & x_{2} & x_{3} \\
y_{1} & y_{2} & y_{3} \\
v_{1} & v_{2} & v_{3}
\end{array}
\right)\left(
\begin{array}{c}
c_{1} \\
c_{2} \\
c_{0}
\end{array}
\right).
\end{equation}

The explicit form of the BRST transformations for gauge particles is presented
below:

\begin{eqnarray}
\tilde{\delta}^{BRST} W^\pm_{1\mu } &=&
\partial_\mu c^\pm_1 \pm ie
\left[
W^\pm_{1\mu }
\left(
c_1\; f(s_\xi ,c_\xi ) + c_2\; g(s_\xi ,c_\xi ) + c_0
\right)+
W^\pm_{2\mu }
\left(
c_1\; h(s_\phi ,c_\phi ) + c_2\; h(-c_\phi ,s_\phi )
\right)
\right.
\nonumber \\
&-&
\left.
Z_{1\mu }
\left(
c^\pm_1\; f(s_\xi ,c_\xi ) + c^\pm_2\; h(s_\phi ,c_\phi )
\right)
-
Z_{2\mu }
\left(
c^\pm_1\; g(s_\xi ,c_\xi ) + c^\pm_2\; h(-c_\phi ,s_\phi )
\right)
-A_\mu c^\pm_1
\right],
\nonumber \\
\end{eqnarray}

\begin{eqnarray}
\tilde{\delta}^{BRST} W^\pm_{2\mu } &=&
\partial_\mu c^\pm_2 \pm ie
\left[
W^\pm_{1\mu }
\left(
c_1\; h(s_\phi ,c_\phi ) + c_2\; h(-c_\phi ,s_\phi )
\right)
+
W^\pm_{2\mu }
\left(
c_1\; f(s_\xi ,c_\xi ) + c_2\; g(s_\xi ,c_\xi ) + c_0
\right)
\right.
\nonumber \\
&-&
\left.
Z_{1\mu }
\left(
c^\pm_1\; h(s_\phi ,c_\phi )  + c^\pm_2\; f(s_\xi ,c_\xi )
\right)-
Z_{2\mu }
\left(
c^\pm_1\; h(-c_\phi ,s_\phi ) + c^\pm_2\;  g(s_\xi ,c_\xi )
\right)
-A_\mu c^\pm_2
\right],
\nonumber \\
\end{eqnarray}

\begin{eqnarray}
\tilde{\delta}^{BRST} Z_{1\mu } &=&
\partial_\mu c_1 - ie
\left[
W^+_{1\mu }
\left(
c^-_1\; f_1(s_\xi ,c_\xi ) + c^-_2\; h_1(s_\phi ,c_\phi )
\right)-
W^-_{1\mu }
\left(
c^+_1\; f_1(s_\xi ,c_\xi ) + c^+_2\; h_1(s_\phi ,c_\phi )
\right)
\right.
\nonumber \\
&+&
\left.
W^+_{2\mu }
\left(
c^-_1\; h_1(s_\phi ,c_\phi ) + c^-_2\; f_1(c_\xi ,s_\xi )
\right)-
W^-_{2\mu }
\left(
c^+_1\; h_1(s_\phi ,c_\phi ) + c^+_2\; f_1(c_\xi ,s_\xi )
\right)
\right],
\nonumber \\
\end{eqnarray}

\begin{eqnarray}
\tilde{\delta}^{BRST} Z_{2\mu } &=&
\partial_\mu c_2 - ie
\left[
W^+_{1\mu }
\left(
c^-_1\; g(s_\xi ,c_\xi ) + c^-_2\; h(-c_\phi ,s_\phi )
\right)-
W^-_{1\mu }
\left(
c^+_1\; g(s_\xi ,c_\xi ) + c^+_2\; h(-c_\phi ,s_\phi )
\right)
\right.
\nonumber \\
&+&
\left.
W^+_{2\mu }
\left(
c^-_1\; h(-c_\phi ,s_\phi ) + c^-_2\; g(c_\xi ,s_\xi )
\right)-
W^-_{2\mu }
\left(
c^+_1\; h(-c_\phi ,s_\phi ) + c^+_2\; g(c_\xi ,s_\xi )
\right)
\right],
\nonumber \\
\end{eqnarray}

and

\begin{eqnarray}
\tilde{\delta}^{BRST} A_{\mu } =
\partial_\mu c_0 -ie
\left(
W^+_{1\mu} c^-_1 - W^-_{1\mu} c^+_1 + W^+_{2\mu} c^-_2 - W^-_{2\mu} c^+_2
\right),
\end{eqnarray}

with the following definitions of the $f(s_\xi ,c_\xi )$,
$f_1(s_\xi ,c_\xi )$, $g(s_\xi ,c_\xi )$, $g_2(s_\xi ,c_\xi )$,
$h(s_\phi ,c_\phi )$, $h_1(s_\phi ,c_\phi )$,
$h(-c_\phi ,c_\phi )$ functions:

\begin{eqnarray}
f(s_\xi ,c_\xi ) &=&
\cos^2{\xi} \cot{\Theta_W} \cos{\phi} - \sin^2{\xi} \tan{\Theta_W} \cos{\phi}
-
\sin^2{\xi} \frac{\sqrt{\cos{2\Theta_W}}}{\sin{\Theta_W}\cos{\Theta_W}}
\sin{\phi},
\nonumber \\
f_1(s_\xi ,c_\xi ) &=&
\cos^2{\xi} \cot{\Theta_W} \cos{\phi} + \sin^2{\xi} \tan{\Theta_W} \cos{\phi}
+
\sin^2{\xi} \frac{\sqrt{\cos{2\Theta_W}}}{\sin{\Theta_W}\cos{\Theta_W}}
\sin{\phi},
\nonumber \\
g(s_\xi ,c_\xi ) &=&
\cos^2{\xi} \cot{\Theta_W} \sin{\phi} - \sin^2{\xi} \tan{\Theta_W} \sin{\phi}
+
\sin^2{\xi} \frac{\sqrt{\cos{2\Theta_W}}}{\sin{\Theta_W}\cos{\Theta_W}}
\cos{\phi},
\nonumber \\
h(s_\phi ,c_\phi ) &=&
\sin{\xi} \cos{\xi}
\left(
\cot{\Theta_W} \cos{\phi} + \tan{\Theta_W} \cos{\phi}
+
\frac{\sqrt{\cos{\Theta_W}}}{\sin{\Theta_W}\cos{\Theta_W}}
\sin{\phi}
\right),
\nonumber \\
h_1(s_\phi ,c_\phi ) &=&
\sin{\xi} \cos{\xi}
\left(
\cot{\Theta_W} \cos{\phi} - \tan{\Theta_W} \cos{\phi}
-
\frac{\sqrt{\cos{\Theta_W}}}{\sin{\Theta_W}\cos{\Theta_W}}
\sin{\phi}
\right),
\nonumber \\
h(-c_\phi ,s_\phi ) &=&
\sin{\xi} \cos{\xi}
\left(
\cot{\Theta_W} \sin{\phi} + \tan{\Theta_W} \sin{\phi}
-
\frac{\sqrt{\cos{\Theta_W}}}{\sin{\Theta_W}\cos{\Theta_W}}
\cos{\phi}
\right).
\nonumber \\
\end{eqnarray}

The BRST transformations for Goldstone particles are obtained from
Eq.(\ref{sc2brs}) :

\begin{eqnarray}
\tilde{\delta}^{BRST} \tilde{G}^0_1 &=&
\frac{e}{\sqrt{2} \kappa_+}
\left\{
\kappa_2
\biggl[
\left(
c_1\;\tilde{h}(s_\phi , c_\phi ) + c_2\;\tilde{h}(-c_\phi , s_\phi )
\right)
\left(
\phi_2^{0\ast}+\phi_2^{0}
\right)
\right.
\nonumber \\
&-&
\frac{1}{\sqrt{2}\sin{\Theta_W}}
\left(
\cos{\xi }
\left(
c^+_1 \phi^-_1+c^-_1 \phi^+_1
\right)+
\sin{\xi }
\left(
c^+_2 \phi^-_1+c^-_2 \phi^+_1
\right)
\right)
\nonumber \\
&+&
\frac{1}{\sqrt{2}\sin{\Theta_W}}
\left(
-\sin{\xi }
\left(
c^-_1 \phi^+_2+c^+_1 \phi^-_2
\right)+
\cos{\xi }
\left(
c^-_2 \phi^+_2+c^+_2 \phi^-_2
\right)
\right)
\biggr]
\nonumber \\
&-&
\kappa_1
\biggl[
\left(
-c_1\;\tilde{h}(s_\phi , c_\phi ) + c_2\;\tilde{h}(-c_\phi , s_\phi )
\right)
\left(
\phi_1^{0\ast}+\phi_1^{0}
\right)
\nonumber \\
&-&
\frac{1}{\sqrt{2}\sin{\Theta_W}}
\left(
\cos{\xi }
\left(
c^-_1 \phi^+_2+c^+_1 \phi^-_2
\right)+
\sin{\xi }
\left(
c^-_2 \phi^+_2+c^+_2 \phi^-_2
\right)
\right)
\nonumber \\
&+&
\left.
\frac{1}{\sqrt{2}\sin{\Theta_W}}
\left(
-\sin{\xi }
\left(
c^-_1 \phi^+_1+c^+_1 \phi^-_1
\right)+
\cos{\xi }
\left(
c^-_2 \phi^+_1+c^+_2 \phi^-_1
\right)
\right)
\biggr]
\right\},
\end{eqnarray}

\begin{eqnarray}
\tilde{\delta}^{BRST} \tilde{G}^0_2 = &-&
\frac{e}{\sqrt{2}}
\left\{
\frac{\sin{\xi}}{\sin{\Theta_W}}
\left(
c^+_1 \delta^-_R + c^-_1 \delta^+_R
\right)
-\frac{\cos{\xi}}{\sin{\Theta_W}}
\left(
c^+_2 \delta^-_R + c^-_2 \delta^+_R
\right)
\right.
\nonumber \\
&-&
\left.
c_1
\left(
\frac{\cot{\Theta_W}}{\sqrt{\cos{2\Theta_W}}}\sin{\phi}
\right)
\left(
\delta_R^{0\ast} + \delta_R^{0}
\right)+
c_2
\left(
\frac{\cot{\Theta_W}}{\sqrt{\cos{2\Theta_W}}}\cos{\phi}
\right)
\left(
\delta_R^{0\ast} + \delta_R^{0}
\right)
\right\},
\end{eqnarray}

and

\begin{eqnarray}
\tilde{\delta}^{BRST} G^0_1 &=&
\cos{\psi_n} \tilde{\delta}^{BRST} \tilde{G}^0_1 +
\sin{\psi_n} \tilde{\delta}^{BRST} \tilde{G}^0_2,
\nonumber \\
\tilde{\delta}^{BRST} G^0_2 &=&
\sin{\psi_n} \tilde{\delta}^{BRST} \tilde{G}^0_1 -
\cos{\psi_n} \tilde{\delta}^{BRST} \tilde{G}^0_2,
\end{eqnarray}

\begin{eqnarray}
\tilde{\delta}^{BRST} G^+_1 &=&
ie
\left\{
\left(
\cos{\psi_c}\; a_{1R} - \sin{\psi_c}\; a_{1L}
\right)
\left[
\sqrt{\cos{2\Theta_W}}
\left(
-c_1\;\tilde{h}(-c_\phi , s_\phi ) +
c_2\;\tilde{h}(s_\phi , c_\phi ) + c_0
\right)
\phi^+_1
\right.
\right.
\nonumber \\
&+&
\left.
\frac{1}{\sqrt{2}\sin{\Theta_W}}
\left(
\cos{\xi }\; c^+_1 + \sin{\xi }\; c^+_2
\right)
\phi^0_2
-
\frac{1}{\sqrt{2}\sin{\Theta_W}}
\left(
-\sin{\xi }\; c^+_1 + \cos{\xi }\; c^+_2
\right)
\phi^0_1
\right]
\nonumber \\
&+&
\left(
\cos{\psi_c}\; a_{2R} - \sin{\psi_c}\; a_{2L}
\right)
\left[
\sqrt{\cos{2\Theta_W}}
\left(
-c_1\;\tilde{h}(-c_\phi , s_\phi ) +
c_2\;\tilde{h}(s_\phi , c_\phi ) + c_0
\right)
\phi^+_2
\right.
\nonumber \\
&-&
\left.
\frac{1}{\sqrt{2}\sin{\Theta_W}}
\left(
\cos{\xi }\; c^+_1 + \sin{\xi }\; c^+_2
\right)
\phi^{0\ast}_1
+
\frac{1}{\sqrt{2}\sin{\Theta_W}}
\left(
-\sin{\xi }\; c^+_1 + \cos{\xi }\; c^+_2
\right)
\phi^{0\ast}_2
\right]
\nonumber \\
&+&
\left(
\cos{\psi_c}\; a_{RR}
\right)
\left[
\left(
-\frac{\sin{\xi}}{\sin{\Theta_W}} c^+_1  +
\frac{\cos{\xi}}{\sin{\Theta_W}} c^+_2
\right)
\delta^0_R
-
\left(
-\frac{\sin{\xi}}{\sin{\Theta_W}} c^-_1  +
\frac{\cos{\xi}}{\sin{\Theta_W}} c^-_2
\right)
\delta^{++}_R
\right.
\nonumber \\
&+&
\left(
-\tan{\Theta_W} \cos{\phi} +
\frac{\sin{\Theta_W}}{\cos{\Theta_W}\sqrt{\cos{2\Theta_W}}} \sin{\phi}
\right)
c_1 \delta^{+}_R
\nonumber \\
&+&
\left.
\left.
\left(
-\tan{\Theta_W} \sin{\phi} -
\frac{\sin{\Theta_W}}{\cos{\Theta_W}\sqrt{\cos{2\Theta_W}}} \cos{\phi}
\right)
c_2 \delta^{+}_R  +c_0 \delta^{+}_R
\right]
\right\},
\end{eqnarray}

\begin{eqnarray}
\tilde{\delta}^{BRST} G^+_2 &=&
ie
\left\{
\left(
\sin{\psi_c}\; a_{1R} + \cos{\psi_c}\; a_{1L}
\right)
\left[
\sqrt{\cos{2\Theta_W}}
\left(
-c_1\;\tilde{h}(-c_\phi , s_\phi ) +
c_2\;\tilde{h}(s_\phi , c_\phi ) + c_0
\right)
\phi^+_1
\right.
\right.
\nonumber \\
&+&
\left.
\frac{1}{\sqrt{2}\sin{\Theta_W}}
\left(
\cos{\xi }\; c^+_1 + \sin{\xi }\; c^+_2
\right)
\phi^0_2
-
\frac{1}{\sqrt{2}\sin{\Theta_W}}
\left(
-\sin{\xi }\; c^+_1 + \cos{\xi }\; c^+_2
\right)
\phi^0_1
\right]
\nonumber \\
&+&
\left(
\sin{\psi_c}\; a_{2R} + \cos{\psi_c}\; a_{2L}
\right)
\left[
\sqrt{\cos{2\Theta_W}}
\left(
-c_1\;\tilde{h}(-c_\phi , s_\phi ) +
c_2\;\tilde{h}(s_\phi , c_\phi ) + c_0
\right)
\phi^+_2
\right.
\nonumber \\
&-&
\left.
\frac{1}{\sqrt{2}\sin{\Theta_W}}
\left(
\cos{\xi }\; c^+_1 + \sin{\xi }\; c^+_2
\right)
\phi^{0\ast}_1
+
\frac{1}{\sqrt{2}\sin{\Theta_W}}
\left(
-\sin{\xi }\; c^+_1 + \cos{\xi }\; c^+_2
\right)
\phi^{0\ast}_2
\right]
\nonumber \\
&+&
\left(
\sin{\psi_c}\; a_{RR}
\right)
\left[
\left(
-\frac{\sin{\xi}}{\sin{\Theta_W}} c^+_1  +
\frac{\cos{\xi}}{\sin{\Theta_W}} c^+_2
\right)
\delta^0_R
-
\left(
-\frac{\sin{\xi}}{\sin{\Theta_W}} c^-_1  +
\frac{\cos{\xi}}{\sin{\Theta_W}} c^-_2
\right)
\delta^{++}_R
\right.
\nonumber \\
&+&
\left(
-\tan{\Theta_W} \cos{\phi} +
\frac{\sin{\Theta_W}}{\cos{\Theta_W}\sqrt{\cos{2\Theta_W}}} \sin{\phi}
\right)
c_1 \delta^{+}_R
\nonumber \\
&+&
\left.
\left.
\left(
-\tan{\Theta_W} \sin{\phi} -
\frac{\sin{\Theta_W}}{\cos{\Theta_W}\sqrt{\cos{2\Theta_W}}} \cos{\phi}
\right)
c_2 \delta^{+}_R  +c_0 \delta^{+}_R
\right]
\right\},
\end{eqnarray}

where

\begin{eqnarray}
\tilde{h}(-c_\phi , s_\phi ) =
\frac{h(-c_\phi , s_\phi )}{2\sin{\xi }\sin{\xi }},\;\;\;
\tilde{h}(s_\phi , c_\phi ) =
\frac{h(s_\phi , c_\phi )}{2\sin{\xi }\sin{\xi }}.
\end{eqnarray}

Introducing "bar" fields for each ghost particle with the BRST
transformations:

\begin{eqnarray}
\tilde{\delta}^{BRST} \bar{c}^\pm_1 &=& \pm i E^\pm_1 ,
\nonumber \\
\tilde{\delta}^{BRST} \bar{c}^\pm_2 &=& \pm i E^\pm_2 ,
\nonumber \\
\tilde{\delta}^{BRST} \bar{c}_1 &=&  i E_{Z_1} ,
\nonumber \\
\tilde{\delta}^{BRST} \bar{c}_2 &=&  i E_{Z_2} ,
\nonumber \\
\tilde{\delta}^{BRST} \bar{c}_0 &=&  i E_{A} ,
\end{eqnarray}

we can construct the Faddeev - Popov Lagrangian in the form

\begin{eqnarray}
\label{fp11}
L_{FP} &=&
i \bar{c}^+_1 D^-_1 + i \bar{c}^-_1 D^+_1 +
i \bar{c}^+_2 D^-_2
+
i \bar{c}^-_2 D^+_2 + i \bar{c}_1 D_{Z_1} + i \bar{c}_2 D_{Z_2} +
i \bar{c}_0 D_{A} .
\end{eqnarray}

If we define BRST transformations for ghost particles in the following
way:

\begin{eqnarray}
\delta^{BRST} c^1_{L,R} &=&  - \lambda g c^2_{L,R} c^3_{L,R},
\nonumber \\
\delta^{BRST} c^2_{L,R} &=&  - \lambda g c^3_{L,R} c^1_{L,R},
\nonumber \\
\delta^{BRST} c^3_{L,R} &=&  - \lambda g c^1_{L,R} c^2_{L,R},
\nonumber \\
\delta^{BRST} c^4 &=& 0,
\end{eqnarray}

then it is easy to check that

\begin{eqnarray}
\delta^{BRST}
\left[
L_{GF} + L_{FP}
\right]
=0.
\end{eqnarray}

Since the rest of the Lagrangian is gauge invariant, it is also
BRST invariant ( the BRST is infinitesimal gauge transformation).
The kinetic energy Lagrangian for ghost particles is obtained
after the BRST transformations for all fields in Eq.(\ref{fp11})

\begin{eqnarray}
\label{ghostkin}
L^{kin}_{ghost} =
&-& i \left( \partial_\mu \bar{c}^+_1 \right)
\left( \partial^\mu c^-_1 \right)
+ i \xi M^2_{W_1}  \bar{c}^+_1 c^-_1
- i \left( \partial_\mu \bar{c}^+_2 \right)
\left( \partial^\mu c^-_2 \right)
+ i \xi M^2_{W_2}  \bar{c}^+_2 c^-_2 + h.c.
\nonumber \\
&-& i \left( \partial_\mu \bar{c}_1 \right)
\left( \partial^\mu c_1 \right)
+ i \xi M^2_{Z_1}  \bar{c}_1 c_1
- i \left( \partial_\mu \bar{c}_2 \right)
\left( \partial^\mu c_2 \right)
+i \xi M^2_{Z_2}  \bar{c}_1 c_2
-i \left( \partial_\mu \bar{c}_A \right)
\left( \partial^\mu c_A \right).
\end{eqnarray}

According to Eq.(\ref{ghostkin}) the masses of ghost particles are as
follows:

\[
(M_{c^{\pm}_{1,2}})^{2}=\xi (M_{W^{\pm}_{1,2}})^{2} ,
\]
\[
(M_{c_{1,2}})^{2}=\xi (M_{Z_{1,2}})^{2} ,
\]
\begin{equation}
M_{c_{0}}=0,
\end{equation}

and the propagators are (it is worth to stress that a -1 factor
appears in the numerator)

\begin{equation}
i\Delta_i (p)=\frac{-1}{p^2-\xi M_{i}^{2}+i\varepsilon},\;\;\;
where\;\;\; M_{i}=\left\{
\begin{array}{ccl}
M_{W_{1}} & for & c^{\pm}_{1}, \\ M_{W_{2}} & for & c^{\pm}_{2},
\\ M_{Z_{1}} & for & c_{1}, \\ M_{Z_{2}} & for & c_{2}, \\ 0 & for &
c_{0}.
\end{array}
\right.
\end{equation}

For further use we write down the ghost particles - gauge bosons
interactions obtained from Eq.(\ref{fp11})

\begin{eqnarray}
L^{GB-GH}_{FP} &=&
g
\biggl\{
\left(
x_1 \left( \partial^\mu\bar{c}_1\right)
+ x_2 \left( \partial^\mu\bar{c}_2\right)
+ x_3 \left( \partial^\mu\bar{c}_0 \right)
\right)
\nonumber \\
& \times &
\left[
\cos^2{\xi }\;  W^-_{1\mu }\; c^+_1 +
\cos{\xi } \sin{\xi }\;  W^-_{1\mu }\; c^+_2 +
\cos{\xi } \sin{\xi }\;  W^-_{2\mu }\; c^+_1 +
\sin^2{\xi }\;  W^-_{2\mu }\; c^+_2
\right]
\nonumber \\
& + &
\left(
y_1 \left( \partial^\mu\bar{c}_1 \right)
+ y_2 \left( \partial^\mu\bar{c}_2 \right)
+ y_3 \left( \partial^\mu\bar{c}_0 \right)
\right)
\nonumber \\
& \times &
\left[
\sin^2{\xi }\;  W^-_{1\mu }\; c^+_1
-\cos{\xi } \sin{\xi }\;  W^-_{1\mu }\; c^+_2
-\cos{\xi } \sin{\xi }\;  W^-_{2\mu }\; c^+_1 +
\cos^2{\xi } \; W^-_{2\mu }\; c^+_2
\right]
\nonumber \\
& + &
\left[
\cos^2{\xi } \left( \partial^\mu\bar{c}^-_1 \right)
W^+_{1\mu }  +
\cos{\xi } \sin{\xi } \left( \partial^\mu\bar{c}^-_2 \right)
W^+_{1\mu }
\right.
\nonumber \\
&+&
\left.
\cos{\xi } \sin{\xi } \left( \partial^\mu\bar{c}^-_1 \right)
W^+_{2\mu } +
\sin^2{\xi } \left( \partial^\mu\bar{c}^-_2 \right)
W^+_{2\mu }
\right] \left(
x_1 c_1  + x_2 c_2 + x_3 c_0 \right)
\nonumber \\
%& \times &
%\left(
%x_1 c_1  + x_2 c_2 + x_3 c_0
%\right)
%\nonumber \\
& + &
\left[
\sin^2{\xi } \left( \partial^\mu\bar{c}^-_1  \right)
W^+_{1\mu }
-\cos{\xi } \sin{\xi } \left( \partial^\mu\bar{c}^-_2 \right)
W^+_{1\mu }
\right.
\nonumber \\
&-&
\left.
\cos{\xi } \sin{\xi } \left( \partial^\mu\bar{c}^-_1 \right)
W^+_{2\mu } +
\cos^2{\xi } \left( \partial^\mu\bar{c}^-_2 \right)
W^+_{2\mu }
\right] \left(
y_1 c_1  + y_2 c_2 + y_3 c_0
\right)
\nonumber \\
%& \times &
%\left(
%y_1 c_1  + y_2 c_2 + y_3 c_0
%\right)
%\nonumber \\
& - &
\left(
x_1 Z_{1\mu}  + x_2 Z_{2\mu} + x_3 A_\mu
\right)
\nonumber \\
& \times &
\left[
\cos^2{\xi } \left( \partial^\mu\bar{c}^-_1 \right)
c^+_1 +
\cos{\xi } \sin{\xi } \left( \partial^\mu\bar{c}^-_1 \right)
c^+_2 +
\cos{\xi } \sin{\xi }  \left( \partial^\mu\bar{c}^-_2 \right)
 c^+_1 +
\sin^2{\xi } \left( \partial^\mu\bar{c}^-_2 \right)
c^+_2
\right]
\nonumber \\
& - &
\left(
y_1 Z_{1\mu}  + y_2 Z_{2\mu} + y_3 A_\mu
\right)
\nonumber \\
& \times &
\left[
\sin^2{\xi } \left( \partial^\mu\bar{c}^-_1 \right)
c^+_1
-\cos{\xi } \sin{\xi } \left( \partial^\mu\bar{c}^-_1 \right)
c^+_2
-\cos{\xi } \sin{\xi } \left( \partial^\mu \bar{c}^-_2 \right)
c^+_1 +
\cos^2{\xi } \left( \partial^\mu\bar{c}^-_2 \right)
c^+_2
\right]
\biggr\}
\nonumber \\
&+& h.c. \;.
\end{eqnarray}

In our approach the ghost fields are hermitian, so we have:

\begin{eqnarray}
\label{ghos1}
\left(
c_i^\pm
\right)^+  = c_i^\mp ,\;\;\;
\left(
\bar{c}_i^\pm
\right)^+  = \bar{c}_i^\mp ,\;\;\; i=1,2
\end{eqnarray}

for charged ghost, and

\begin{eqnarray}
\left(
c_i
\right)^+  = c_i ,\;\;\;
\left(
\bar{c}_i
\right)^+  = \bar{c}_i
\end{eqnarray}

for neutral ghost. The ghosts $c$ and $\bar{c}$ anticommute

\begin{eqnarray}
\label{ghos2}
\left\{ c,\bar{c} \right\} =0.
\end{eqnarray}

Using the relations Eqs.(\ref{ghos1}-\ref{ghos2}) one can prove the hermicity
of all terms in the Lagrangian with ghost particles.
\newpage

\part*{ IV On the mass - shell renormalization of the \\
\hspace*{1.2cm}left - right model.}

It has been proved first by 't Hooft that non abelian theories are
renormalizable \cite{hooft}. Further contributions have clarified
the structure of gauge theories \cite{slavnov}, \cite{lee},
\cite{hooft2}, \cite{fujikava}. In the following sections the full
(i.e. all Green functions are claimed to be finite) on - shell
renormalization scheme of the left - right symmetric model is
presented \cite{hooft1}, \cite{ross}, \cite{hioki}. First, the
definitions of renormalization constants for fields, masses,
couplings, mixing angles and mixing matrices are given. Then, in
Chapter 4.2.1, the renormalization of tadpoles is considered. In
Chapter 4.2.2 the renormalization conditions and counter terms for
two - point functions are presented. Finally in Chapter 4.2.3 the
renormalization of remaining parameters ($\cos\varphi ,\; \sin\xi
,\; \Omega_{L,R},\; K_{L,R}$) is performed.

\section*{ 4.1 Definitions of the renormalization constants.}

All quantities written below with (without) suffix 0 denote bare
(renormalized) ones.
Capital "IJ" (small
"ij") indices represent up (down) fermions in the left and right handed
doublets. For abbreviation the $i,j$ indices are used to denote the neutral
physical Higgses
($i=H^0,H_1^0,H_2^0,H_3^0,A_1^0,A_2^0$),
(e.g.$Z^{\frac{1}{2}}_{ij}$ in Eq.(\ref{zh})). Similarly, the single
(doubly) charged Higgses $H_{1,2}^{\pm}$ ($\delta^{\pm \pm}_{L,R})$
without '$\pm$' ('$\pm \pm $') prefixes are written
(e.g. $Z^{\frac{1}{2}}_{H_1^+H_2^+} \equiv Z^{\frac{1}{2}}_{H_1H_2}$,
$Z^{\frac{1}{2}}_{\delta_L^{++}\delta_R^{++}} \equiv
Z^{\frac{1}{2}}_{\delta_L \delta_R}$).

\subsubsection*{  \bf 4.1.1 Definitions of the renormalization constants
for fields.}

\begin{itemize}
\item[i)] Charged gauge bosons, charged Nambu-Goldstone bosons
and ghost particles:

\begin{equation}
\label{scb}
\left(
\begin{array}{c}
\stackrel{\circ}{W}_{1\mu}^{\pm} \\
\stackrel{\circ}{W}_{2\mu}^{\pm}
\end{array}
\right)= \left(
\begin{array}{cc}
Z_{W_{1}W_{1}}^{\frac{1}{2}} & Z_{W_{1}W_{2}}^{\frac{1}{2}} \\
Z_{W_{2}W_{1}}^{\frac{1}{2}} & Z_{W_{2}W_{2}}^{\frac{1}{2}}
\end{array}
\right) \left(
\begin{array}{c}
{W}_{1\mu}^{\pm} \\
{W}_{2\mu}^{\pm}
\end{array}
\right),
\end{equation}

\begin{equation}
\left(
\begin{array}{c}
\stackrel{\circ}{G}_{1}^{\pm} \\
\stackrel{\circ}{G}_{2}^{\pm}
\end{array}
\right)= \left(
\begin{array}{cc}
\left( Z_{G}^{\pm} \right)_{11} & \left( Z_{G}^{\pm} \right)_{12} \\
\left( Z_{G}^{\pm} \right)_{12}  & \left( Z_{G}^{\pm} \right)_{22}
\end{array}
\right) \left(
\begin{array}{c}
{G}_{1}^{\pm} \\
{G}_{2}^{\pm}
\end{array}
\right),
\end{equation}

\begin{equation}
\label{scd}
\left(
\begin{array}{c}
\stackrel{\circ}{c}_{1}^{\pm} \\
\stackrel{\circ}{c}_{2}^{\pm}
\end{array}
\right)= \left(
\begin{array}{cc}
\tilde{Z}_{W_1W_1}^{\frac{1}{2}} & \tilde{Z}_{W_1W_2}^{\frac{1}{2}} \\
\tilde{Z}_{W_2W_1}^{\frac{1}{2}} & \tilde{Z}_{W_2W_2}^{\frac{1}{2}}
\end{array}
\right) \left(
\begin{array}{c}
{c}_{1}^{\pm} \\
{c}_{2}^{\pm}
\end{array}
\right),
\end{equation}

\begin{equation}
\stackrel{\circ}{\bar{c}}_{1,2}^{\pm}={\bar{c}}_{1,2}^{\pm},
\end{equation}

where the renormalization constants from Eqs.(\ref{scb},\ref{scd})
are given by:

\begin{equation}
Z_{W_{i}W_{j}}^{\frac{1}{2}}= \delta_{ij} +
\delta Z_{W_{i}W_{j}}^{\frac{1}{2}},\;\;\;\;\;
\left( Z_{G}^{\pm} \right)_{ij}=\delta_{ij}+
\delta \left( Z_{G}^{\pm} \right)_{ij},\;\;\;\;\;
\tilde{Z}_{W_{i}W_{j}}^{\frac{1}{2}}= \delta_{ij} +
\delta \tilde{Z}_{W_{i}W_{j}}^{\frac{1}{2}}.
\end{equation}

\item[ii)] Neutral gauge bosons, neutral Nambu-Goldstone bosons
and ghost particles:

\begin{equation}
\label{gbmr}
\left(
\begin{array}{c}
\stackrel{\circ}{Z_{1\mu}} \\
\stackrel{\circ}{Z_{2\mu}} \\
\stackrel{\circ}{A_{\mu}}
\end{array}
\right)= \left(
\begin{array}{ccc}
Z_{Z_{1}Z_{1}}^{\frac{1}{2}} & Z_{Z_{1}Z_{2}}^{\frac{1}{2}} & Z_{Z_{1}A}^{%
\frac{1}{2}} \\
Z_{Z_{2}Z_{1}}^{\frac{1}{2}} & Z_{Z_{2}Z_{2}}^{\frac{1}{2}} & Z_{Z_{2}A}^{%
\frac{1}{2}} \\
Z_{AZ_{1}}^{\frac{1}{2}} & Z_{AZ_{2}}^{\frac{1}{2}} & Z_{AA}^{\frac{1}{2}}
\end{array}
\right) \left(
\begin{array}{c}
Z_{1\mu} \\
Z_{2\mu} \\
A_{\mu}
\end{array}
\right),
\end{equation}

\begin{equation}
\left(
\begin{array}{c}
\stackrel{\circ}{G}_{1}^{0} \\
\stackrel{\circ}{G}_{2}^{0}
\end{array}
\right)= \left(
\begin{array}{cc}
\left( Z_{G}^{0} \right)_{11} & \left( Z_{G}^{0} \right)_{12} \\
\left( Z_{G}^{0} \right)_{12}  & \left( Z_{G}^{0} \right)_{22}
\end{array}
\right) \left(
\begin{array}{c}
{G}_{1}^{0} \\
{G}_{2}^{0}
\end{array}
\right),
\end{equation}

\begin{equation}
\left(
\begin{array}{c}
\stackrel{\circ}{c_{1}} \\
\stackrel{\circ}{c_{2}} \\
\stackrel{\circ}{c_{0}}
\end{array}
\right)= \left(
\begin{array}{ccc}
\tilde{Z}_{Z_{1}Z_{1}}^{\frac{1}{2}} & \tilde{Z}_{Z_{1}Z_{2}}^{\frac{1}{2}}
& \tilde{Z}_{Z_{1}A}^{\frac{1}{2}} \\
\tilde{Z}_{Z_{2}Z_{1}}^{\frac{1}{2}} & \tilde{Z}_{Z_{2}Z_{2}}^{\frac{1}{2}}
& \tilde{Z}_{Z_{2}A}^{\frac{1}{2}} \\
\tilde{Z}_{AZ_{1}}^{\frac{1}{2}} & \tilde{Z}_{AZ_{2}}^{\frac{1}{2}} & \tilde{%
Z}_{AA}^{\frac{1}{2}}
\end{array}
\right) \left(
\begin{array}{c}
c_{1} \\
c_{2} \\
c_{0}
\end{array}
\right).
\end{equation}

Similarly as for charged particle there is:

\begin{eqnarray}
Z_{Z_iZ_j}^{\frac 12} &=& \delta_{ij}+\delta Z_{Z_iZ_j}^{\frac 12}, \\
\left( Z_{G}^{0} \right)_{ij}&=&\delta_{ij}+
\delta \left( Z_{G}^{0} \right)_{ij}, \\
\tilde{Z}_{Z_iZ_j}^{\frac 12} &=&
\delta_{ij}+\delta \tilde{Z}_{Z_iZ_j}^{\frac 12}, \\
Z_{Z_AZ_A}^{\frac 12} &=&1+\delta Z_{Z_AZ_A}^{\frac 12}.
\end{eqnarray}

For the $\bar{c}$ - type ghosts it is not necessary to introduce
renormalization constants (they occur always in pairs with $c$ -
type ghosts), and

\begin{eqnarray}
\stackrel{\circ}{\bar{c}}_{1} & = & \bar{c}_{1}, \\
\stackrel{\circ}{\bar{c}}_{2} & = & \bar{c}_{2}, \\
\stackrel{\circ}{\bar{c}}_{0} & = & \bar{c}_{0}.
\end{eqnarray}

\item[iii)] Fermions.

After SSB the left - right symmetric model is a chiral theory and left
and right handed parts of fermion fields interacts in a different way.
So the $\psi_L$ and $\psi_R$ fields must have independent renormalization
constants. To satisfy all on mass - shell renormalization conditions
one has to introduce the renormalization matrix.

\begin{eqnarray}
\stackrel{\circ}{\psi}_{LI} & = & \sum_{J} (Z_{L}^{\frac{1}{2}%
})_{IJ}\psi_{LJ}, \\
\stackrel{\circ}{\psi}_{Li} & = & \sum_{j} (Z_{L}^{\frac{1}{2}%
})_{ij}\psi_{Lj}, \\
\stackrel{\circ}{\psi}_{RI} & = & \sum_{J} (Z_{R}^{\frac{1}{2}%
})_{IJ}\psi_{RJ}, \\
\stackrel{\circ}{\psi}_{Ri} & = & \sum_{j} (Z_{R}^{\frac{1}{2}%
})_{ij}\psi_{Rj},
\end{eqnarray}

where

\begin{eqnarray}
(Z_{L,R}^{\frac{1}{2}})_{ij} &=&
\delta _{ij}+(\delta Z_{L,R}^{\frac{1}{2}})_{ij},\\
(Z_{L,R}^{\frac{1}{2}})_{IJ} &=&\delta _{IJ}+(\delta Z_{L,R}^{\frac{1}{2}})_{IJ},
\end{eqnarray}

and the sum is taken over up (IJ) and down (ij) quarks, or over
charged leptons (ij) and neutrinos (IJ). For convenience we will also
write $(Z_{L,R}^{l\frac{1}{2}})_{ij}$ and $(Z_{L,R}^{\nu\frac{1}{2}})_{IJ}$
for charged leptons and neutrinos respectively.
The matrices $Z_{L,R}$ can be complex and not unitary.
If neutrinos are Majorana particles then there are relations between left
and right handed fields, and the $Z_L$ and $Z_R$ are related by

\begin{eqnarray}
Z_L=Z_R^\ast .
\end{eqnarray}

\item[iv)] Higgs particles.

Since all charge-less Higgs particles mix between each other
(CP is not conserved),
the renormalization matrix must be introduced for them:

\begin{eqnarray}
\label{zh}
\stackrel{\circ}{H_i^0} &=& \sum_{j} Z^{\frac{1}{2}}_{ij}H_j^0\;,
\;\;\;\;\;\;\;\;\;
i,j=H^0_0,H_1^0,H_2^0,H_3^0,A_1^0,A_2^0,
\end{eqnarray}

and similarly for charged particles:

\begin{eqnarray}
\stackrel{\circ}{H_i} &=& \sum_{j}
Z^{\frac{1}{2}}_{H_iH_j}H_j\;, \;\;\;\;\;
i,j=1,2\;, \\
\left(
\begin{array}{c}
\stackrel{\circ}{\delta}_L^{\pm \pm} \\
\stackrel{\circ}{\delta}_R^{\pm \pm}
\end{array}
\right)&=& \left(
\begin{array}{cc}
Z^{\frac{1}{2}}_{\delta_L \delta_L} & Z^{\frac{1}{2}}_{\delta_L\delta_R} \\
Z^{\frac{1}{2}}_{\delta_R \delta_L} & Z^{\frac{1}{2}}_{\delta_R \delta_R}
\end{array}
\right) \left(
\begin{array}{c}
{\delta}_L^{\pm \pm} \\
{\delta}_R^{\pm \pm}
\end{array}
\right).
\end{eqnarray}

In the case of CP conservation the charge-less $H$ and $A$ particles
have different CP parities and do not mix.

\end{itemize}

\subsubsection*{  \bf 4.1.2 Definitions of the renormalization constants
for masses.}

The masses of all physical particles are renormalized and we define:

\hspace*{-1cm}
\begin{eqnarray}
\stackrel{\circ}{M}_{W_{i}}^{2} & = & M_{W_{i}}^{2} + \delta
M_{W_{i}}^{2} \equiv G_{W_{i}}^{2}M_{W_{i}}^{2},
\;\;\;\;\;\;\;\;\;\;\;\;\;\;\;\;\;\;i=1,2\;, \\
\stackrel{\circ}{M}_{Z_{i}}^{2} & = & M_{Z_{i}}^{2} + \delta
M_{Z_{i}}^{2} \equiv G_{Z_{i}}^{2}M_{Z_{i}}^{2},
\;\;\;\;\;\;\;\;\;\;\;\;\;\;\;\;\;\;\;\;\; i=1,2\;, \\
\stackrel{\circ}{M}_{i}^{2} & = & M_{i}^{2} + \delta M_{i}^{2}
\equiv G_{i}^{2}M_{i}^{2},
\;\;\;\;\;\;\;\;\;\;\;\;\;\;\;\;\;\;\;\;\;\;\;\;\;\;\;
i=H^0,H_1^0,H_2^0,H_3^0,A_1^0,A_2^0\;,\\
\stackrel{\circ}{M}_{H_i^\pm}^{2} & = & M_{H_i^\pm}^{2}+ \delta
M_{H_i^\pm}^{2} \equiv G_{H_i}^{2}M_{H_i}^{2},
\;\;\;\;\;\;\;\;\;\;\;\;\;\;\;\;\;\; i=1,2\;, \\
\stackrel{\circ}{M}_{\delta_L^{\pm\pm}}^{2} & = &
M_{\delta_L^{\pm\pm}}^{2}+ \delta M_{\delta_L^{\pm\pm}}^{2} \equiv
G_{\delta_L}^{2}M_{\delta_L}^{2}\;, \\
\stackrel{\circ}{M}_{\delta_R^{\pm\pm}}^{2} & = &
M_{\delta_R^{\pm\pm}}^{2}+ \delta M_{\delta_R^{\pm\pm}}^{2} \equiv
G_{\delta_R}^{2}M_{\delta_R}^{2}\;, \\ \stackrel{\circ}{m}_{f}^{2}
& = & m_{f}^{2} + \delta m_{f}^{2} \equiv G_{f}^{2}m_{f}^{2},
\;\;\;\;\;\;\;\;\;\;\;\;\;\;\;\;\;\;\;\;\;\;\;\;\;\;\;\;
f=\mbox{\rm leptons, quarks}.
\end{eqnarray}

\subsubsection*{  \bf 4.1.3 Definitions of the renormalization constants
for couplings, mixing angles and mixing matrices.}

There are more free parameters in the left - light symmetric model
than in the
SM ($\cos\varphi , \sin\xi , \Omega_{L,R}, K_{L,R}$). Thus, more
renormalization constants must be introduced. The definitions of the
renormalization constants for remaining parameters are as follows:

\begin{eqnarray}
& \stackrel{\circ}{e} & = Ye \equiv (1+ \delta Y)e, \\
& \stackrel{\circ}{\cos\varphi} & = G_{\varphi}\cos\varphi \equiv
(1+\delta G_\varphi)\cos\varphi , \\
& \stackrel{\circ}{\sin\Theta_W} & = G_{\Theta_W}\sin\Theta_W \equiv
(1+\delta G_{\Theta_W})\sin\Theta_W ,\\
& \stackrel{\circ}{\sin\xi} & = G_{\xi}\sin\xi \equiv (1+\delta
G_\xi)\sin\xi ,\\
& \stackrel{\circ}{\Omega}_{L,R} & = \Omega_{L,R} +\delta \Omega_{L,R}, \\
& \stackrel{\circ}{K}_{L,R} & = K_{L,R} +\delta K_{L,R}, \\
& \stackrel{\circ}{U}_{L,R}^{CKM} & = U_{L,R}^{CKM} +\delta U_{L,R}^{CKM}.
\end{eqnarray}

\section*{ 4.2 On mass - shell renormalization conditions and counter terms}

\subsection*{  4.2.1 Renormalization of tadpoles.}

Lorentz invariance prevents tadpoles for fields other than scalar ones.
To get the renormalization conditions for the tadpoles,
to each order of perturbation the requirement
that the tadpole terms in effective action disappear is imposed.
Thus, if the diagram

\begin{center}
\begin{figure}
\epsfig{file=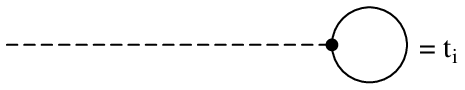,width=5cm}
\end{figure}
\end{center}

denotes tadpole $t_i$ ($i=H^0,H_1^0,H_2^0,H_3^0,A_1^0,A_2^0$) and the
diagram

\begin{center}
\begin{figure}
\epsfig{file=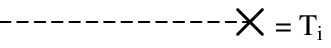,width=5cm}
\end{figure}
\end{center}

is an appropriate counter term $T_i$ the renormalization conditions
$t_i+T_i=0$
give renormalization constants $T_i$.

\subsection*{  4.2.2 Renormalization conditions and counter terms for
two-point functions.}

\subsubsection*{ \bf 4.2.2.1 Gauge bosons.}

There are two kind of counter terms of two - point functions for
gauge bosons. One is coming from kinetic energy

\begin{equation}
-\frac{1}{4} F^{\mu \nu } F_{\mu \nu },
\end{equation}

and the second is given by the mass term, e.q. for Z bosons

\begin{equation}
M^2_{Z_i} Z_i^\mu Z_{i\mu}.
\end{equation}

In the momentum space these two terms give contributions:

\begin{eqnarray}
kinetic\;energy &\longrightarrow &-ik^2
\left(
g^{\mu \nu }-\frac{k^\mu k^\nu }{k^2}
\right) ,
\nonumber \\
mass\; term &\longrightarrow &i g^{\mu \nu }.
\end{eqnarray}

It is better to use two orthogonal tensors:

\begin{eqnarray}
T^{\mu \nu}=g^{\mu \nu } -\frac{k^\mu k^\nu }{k^2},\;\;\;\;
L^{\mu \nu}=\frac{k^\mu k^\nu }{k^2}.
\end{eqnarray}

Then of course

\begin{eqnarray}
T^{\mu \nu} L_{\nu \rho}=0,\;\;\;\;T^{\mu \nu} +L^{\mu \nu}=g^{\mu \nu }.
\end{eqnarray}

Since the gauge fixing parameter is free it is not necessary to
renormalize the gauge fixing term, e.g.

\begin{eqnarray}
-\frac{1}{\xi } \left( \partial^\mu Z_{i\mu} \right)^2
\end{eqnarray}

for Z bosons.

In all the following pictures in this section the diagram

\begin{center}
\begin{figure}
\epsfig{file=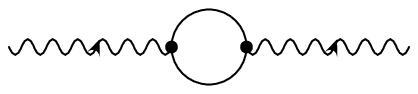,width=5cm}
\end{figure}
\end{center}
\noindent
represents n-loop self-energy contribution, whereas

\begin{center}
\begin{figure}
\epsfig{file=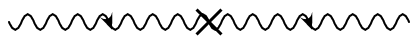,width=5cm}
\end{figure}
\end{center}
\noindent
is the appropriate counter term. The letter '$k$' stands for the external
particle momentum.
%For the scalar particles and fermions only external lines are
%changed.

For two charged gauge bosons we have two diagonal propagators and two mixed.

\begin{figure}
\epsfig{file=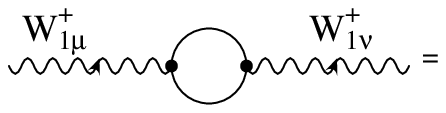,width=5cm}
\end{figure}
\begin{picture}(10,10)(30,25)
\put(175,65)
{\large
$ T^{\mu \nu }
A^{W_1}(k^2)+L^{\mu \nu }
B^{W_1}(k^2),$}
%\end{minipage}
\end{picture}

%\begin{eqnarray*}
%&&\left( g_{\mu \nu }-\frac{k_\mu
%k_\nu }{k^2}\right) A^{W_1}(k^2)+\frac{k_\mu k_\nu }{k^2}B^{W_1}(k^2),
%\end{eqnarray*}

\begin{figure}
\epsfig{file=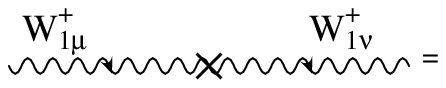,width=5cm}
\end{figure}

%\begin{picture}(10,10)(30,25)
%\put(220,60)

\begin{eqnarray}
&& T^{\mu \nu }
\left[ \left( M_{W_1}^2+\delta M_{W_1}^2 \right)
\left( Z^{\frac{1}{2}}_{W_1W_1} \right)^2+ \right. \nonumber \\
&& \nonumber \\
&& \left( M_{W_2}^2+\delta M_{W_2}^2
\right) \left( Z^{\frac{1}{2}}_{W_2W_1} \right)^2 -k^2 \left. \left( \left(
Z^{\frac{1}{2}}_{W_1W_1} \right)^2+ \left( Z^{\frac{1}{2}}_{W_2W_1}
\right)^2 \right) \right]+ \nonumber  \\
&&   \nonumber \\
&&
L^{\mu \nu }
\left[
\left( M^2_{W_1}+\delta M^2_{W_1} \right) \left( Z^{\frac{1}{2}}_{W_1W_1}
\right)^2 +\left( M^2_{W_2}+\delta M^2_{W_2} \right) \left( Z^{\frac{1}{2}%
}_{W_2W_1} \right)^2 \right].
\end{eqnarray}

\begin{figure}
\epsfig{file=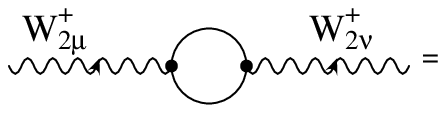,width=5cm}
\end{figure}

\begin{picture}(10,10)(30,25)
\put(175,65)
{\large $ T^{\mu \nu }
A^{W_2}(k^2)+L^{\mu \nu }B^{W_2}(k^2),$}
\end{picture}

\begin{figure}
\epsfig{file=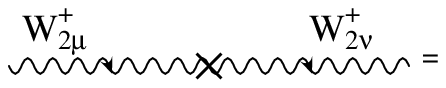,width=5cm}
\end{figure}

\begin{eqnarray}
&& T^{\mu \nu }
\left[ \left( M_{W_2}^2+\delta M_{W_2}^2\right)
\left( Z_{W_2W_2}^{\frac 12}\right) ^2+\right.
\nonumber \\
&&
\left. \left( M_{W_1}^2+\delta
M_{W_1}^2 \right) \left( Z_{W_1W_2}^{\frac 12} \right)^2 -k^2 \left( \left(
Z_{W_2W_2}^{\frac 12} \right) ^2+ \left( Z_{W_1W_2}^{\frac 12} \right) ^2
\right) \right] +\nonumber  \\
&&
L^{\mu \nu } \left[
\left( M_{W_2}^2+\delta M_{W_2}^2 \right) \left( Z_{W_2W_2}^{\frac 12}
\right) ^2 +\left( M_{W_1}^2+\delta M_{W_1}^2 \right) \left( Z_{W_1W_2}^{%
\frac{1}{2}} \right) ^2 \right] .
\end{eqnarray}

\newpage

\begin{figure}
\epsfig{file=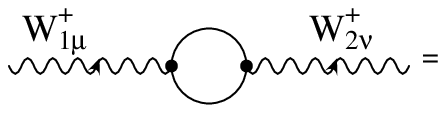,width=5cm}
\end{figure}

\begin{picture}(10,10)(30,25)
\put(175,65)
{\large $T^{\mu \nu }
A^{W_1W_2}(k^2)+L^{\mu \nu }
B^{W_1W_2}(k^2),$}
\end{picture}

\begin{figure}
\epsfig{file=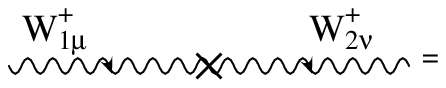,width=5cm}
\end{figure}

\begin{eqnarray}
&&T^{\mu \nu }
\left[ \left( M_{W_1}^2+\delta M_{W_1}^2\right)
Z_{W_1W_1}^{\frac 12}Z_{W_1W_2}^{\frac 12}+\right.
\nonumber \\
&& \left( M_{W_2}^2+\delta M_{W_2}^2\right)
Z_{W_2W_2}^{\frac 12}Z_{W_2W_1}^{\frac 12}-k^2\left( Z_{W_1W_1}^{\frac
12}Z_{W_1W_2}^{\frac 12}+\right. \left. \left. Z_{W_2W_2}^{\frac
12}Z_{W_2W_1}^{\frac 12}\right) \right] +
\nonumber \\
&& L^{\mu \nu }\left[ \left(
M_{W_1}^2+\delta M_{W_1}^2\right) Z_{W_1W_1}^{\frac 12}Z_{W_1W_2}^{\frac
12}+\right. \left. \left( M_{W_2}^2+\delta M_{W_2}^2\right)
Z_{W_2W_2}^{\frac 12}Z_{W_2W_1}^{\frac 12}\right] .
\end{eqnarray}

\begin{eqnarray}
\mbox{\rm Renormalization conditions} && \mbox{\rm Renormalization
constants} \nonumber \\
&& \nonumber \\
\left.
\begin{array}{lcc}
A^{W_{1}}(M^2_{W_1}) & = & 0 \\
A^{W^{\prime}_{1}}(M^2_{W_1}) & = & 0 \\
A^{W_{2}}(M^2_{W_2}) & = & 0 \\
A^{W^{\prime}_{2}}(M^2_{W_2}) & = & 0 \\
A^{W_{1}W_{2}}(M^2_{W_1}) & = & 0 \\
A^{W_{1}W_{2}}(M^2_{W_2}) & = & 0
\end{array}
\right\}
&  \;\;\;\;\; \Rightarrow \;\;\;\;\;\;\; &
\begin{array}{l}
Z^{\frac{1}{2}}_{W_1W_1},Z^{\frac{1}{2}}_{W_2W_2},\delta M^2_{W_1} \\
Z^{\frac{1}{2}}_{W_1W_2},Z^{\frac{1}{2}}_{W_2W_1},\delta M^2_{W_2} .
\end{array}
\end{eqnarray}

These equations allow us to calculate six renormalization
constants. Prime superscript  in the renormalization conditions
above means  differentiation with respect to $k^2$ e.g.

\begin{eqnarray}
A^{W^{\prime}_{1}}(M^2_{W_1}) =
\frac{\partial A^{W_{1}}(k^2)}{\partial k^2}\mid_{k^2=M^2_{W_1}}.
\end{eqnarray}

For neutral gauge bosons the situation is more complicated.
There are three diagonal propagators and three mixed.

\newpage

\begin{figure}
\epsfig{file=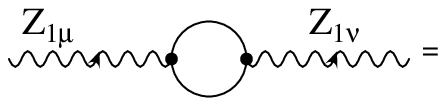,width=5cm}
\end{figure}

\begin{picture}(10,10)(30,25)
\put(175,65)
{\large $ T^{\mu \nu }
 A^{Z_1}(k^2)+L^{\mu \nu }B^{Z_1}(k^2),
$}
\end{picture}

\begin{figure}
\epsfig{file=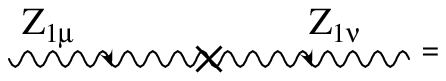,width=5cm}
\end{figure}

\begin{eqnarray}
\label{ctz1}
&&T^{\mu \nu }
 \left[ \left( M_{Z_1}^2+\delta M_{Z_1}^2\right)
\left( Z_{Z_1Z_1}^{\frac 12}\right) ^2+\right.
\nonumber \\
&& \left. \left( M^2_{Z_2}+\delta M^2_{Z_2}\right) \left( Z^{%
\frac{1}{2}}_{Z_2Z_1}\right) ^2- k^2\left( \left( Z^{\frac{1}{2}%
}_{Z_1Z_1}\right) ^2+ \left( Z^{\frac{1}{2}}_{Z_2Z_1}\right) ^2+ \left( Z^{%
\frac{1}{2}}_{AZ_1}\right) ^2\right) \right] +
\nonumber \\
&&
L^{\mu \nu } \left[ \left( M^2_{Z_1}+ \delta M^2_{Z_1}\right)
\left( Z^{\frac{1}{2}}_{Z_1Z_1}\right) ^2+\left( M^2_{Z_2}+ \delta
M^2_{Z_2}\right) \left( Z^{\frac{1}{2}}_{Z_2Z_1}\right) ^2 \right] .
\end{eqnarray}

\begin{figure}
\epsfig{file=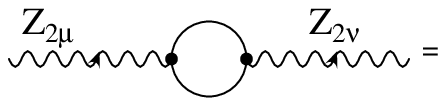,width=5cm}
\end{figure}

\begin{picture}(10,10)(30,25)
\put(175,65)
{\large $T^{\mu \nu }
 A^{Z_2}( k^2)+L^{\mu \nu }B^{Z_2}(k^2),
$}
\end{picture}

\begin{figure}
\epsfig{file=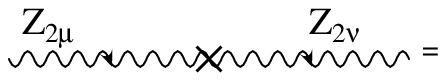,width=5cm}
\end{figure}

\begin{eqnarray}
\label{ctz2}
&&T^{\mu \nu }
 \left[ \left( M_{Z_2}^2+\delta M_{Z_2}^2\right)
\left( Z_{Z_2Z_2}^{\frac 12}\right) ^2+\right.
\nonumber \\
&&
\left. \left( M^2_{Z_1}+\delta M^2_{Z_1}\right) \left( Z^{%
\frac{1}{2}}_{Z_1Z_2}\right) ^2- k^2\left( \left( Z^{\frac{1}{2}%
}_{Z_2Z_2}\right) ^2+ \left( Z^{\frac{1}{2}}_{Z_1Z_2}\right) ^2+ \left( Z^{%
\frac{1}{2}}_{AZ_2}\right) ^2\right) \right] +
\nonumber \\
&&
L^{\mu \nu } \left[ \left( M^2_{Z_2}+ \delta M^2_{Z_2}\right)
\left( Z^{\frac{1}{2}}_{Z_2Z_2}\right) ^2+\left( M^2_{Z_1}+ \delta
M^2_{Z_1}\right) \left( Z^{\frac{1}{2}}_{Z_1Z_2}\right) ^2 \right] .
\end{eqnarray}

\newpage

\begin{figure}
\epsfig{file=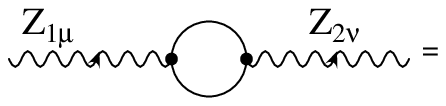,width=5cm}
\end{figure}

\begin{picture}(10,10)(30,25)
\put(175,65)
{\large $T^{\mu \nu }
 A^{Z_1Z_2}(k^2)+L^{\mu \nu }B^{Z_1Z_2}(k^2),
$}
\end{picture}

\begin{figure}
\epsfig{file=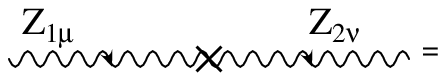,width=5cm}
\end{figure}

\begin{eqnarray}
\label{ctz1z2}
&&
T^{\mu \nu } \left[ \left( M_{Z_1}^2+\delta
M_{Z_1}^2\right) Z_{Z_1Z_1}^{\frac 12}Z_{Z_1Z_2}^{\frac 12}+\right.
\nonumber \\
&&
\left. \left( M_{Z_2}^2+\delta M_{Z_2}^2\right)
Z_{Z_2Z_2}^{\frac 12}Z_{Z_2Z_1}^{\frac 12}-k^2\left( Z_{Z_1Z_1}^{\frac
12}Z_{Z_1Z_2}^{\frac 12}+Z_{Z_2Z_2}^{\frac 12}Z_{Z_2Z_1}^{\frac
12}+Z_{AZ_1}^{\frac 12}Z_{AZ_2}^{\frac 12}\right) \right]+
\nonumber \\
&&
L^{\mu \nu }\left[ \left( M_{Z_1}^2+\delta M_{Z_1}^2\right)
Z_{Z_1Z_1}^{\frac 12}Z_{Z_1Z_2}^{\frac 12}+\left( M_{Z_2}^2+\delta
M_{Z_2}^2\right) Z_{Z_2Z_2}^{\frac 12}Z_{Z_2Z_1}^{\frac 12}\right] .
\end{eqnarray}

\begin{figure}
\epsfig{file=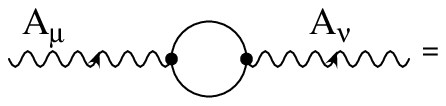,width=5cm}
\end{figure}

\begin{picture}(10,10)(30,25)
\put(175,65)
{\large $
T^{\mu \nu } A^A(k^2)+L^{\mu \nu }B^A(k^2),
$}
\end{picture}

\begin{figure}
\epsfig{file=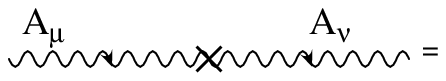,width=5cm}
\end{figure}

\begin{eqnarray}
\label{cta}
&&T^{\mu \nu }
 \left[ \left( M_{Z_1}^2+\delta M_{Z_1}^2\right) \left(
Z_{Z_1A}^{\frac 12}\right) ^2+\right.
\nonumber \\
&&
\left. \left( M^2_{Z_2}+ \delta M^2_{Z_2}\right) \left(
Z^{\frac{1}{2}}_{Z_2A}\right) ^2- k^2\left( \left( Z^{\frac{1}{2}%
}_{Z_1A}\right) ^2+ \left( Z^{\frac{1}{2}}_{Z_2A}\right) ^2+ \left( Z^{\frac{%
1}{2}}_{AA}\right) ^2\right) \right]+
\nonumber \\
&&
L^{\mu \nu } \left[ \left( M^2_{Z_1}+ \delta M^2_{Z_1}\right)
\left( Z^{\frac{1}{2}}_{Z_1A}\right) ^2 +\left( M^2_{Z_2}+ \delta
M^2_{Z_2}\right) \left( Z^{\frac{1}{2}}_{Z_2A}\right) ^2 \right] .
\end{eqnarray}

\newpage

\begin{figure}
\epsfig{file=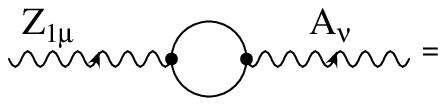,width=5cm}
\end{figure}

\begin{picture}(10,10)(30,25)
\put(175,65)
{\large $T^{\mu \nu }
 A^{Z_1A}(k^2)+L^{\mu \nu }B^{Z_1A}(k^2),
$}
\end{picture}

\begin{figure}
\epsfig{file=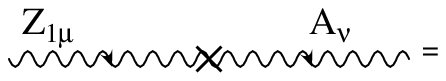,width=5cm}
\end{figure}

\begin{eqnarray}
\label{ctz1a}
&& T^{\mu \nu }
 \left[ \left( M_{Z_1}^2+\delta M_{Z_1}^2\right)
Z_{Z_1Z_1}^{\frac 12}Z_{Z_1A}^{\frac 12}+\right.
\nonumber \\
&&
\left. \left( M_{Z_2}^2+\delta M_{Z_2}^2\right)
Z_{Z_2Z_1}^{\frac 12}Z_{Z_2A}^{\frac 12}-k^2\left( Z_{Z_1Z_1}^{\frac
12}Z_{Z_1A}^{\frac 12}+Z_{Z_2Z_1}^{\frac 12}Z_{Z_2A}^{\frac
12}+Z_{AZ_1}^{\frac 12}Z_{AA}^{\frac 12}\right) \right] +
\nonumber \\
&&
L^{\mu \nu }\left[ \left( M_{Z_1}^2+\delta M_{Z_1}^2\right)
Z_{Z_1Z_1}^{\frac 12}Z_{Z_1A}^{\frac 12}+\left( M_{Z_2}^2+\delta
M_{Z_2}^2\right) Z_{Z_2Z_1}^{\frac 12}Z_{Z_2A}^{\frac 12}\right] .
\end{eqnarray}

\begin{figure}
\epsfig{file=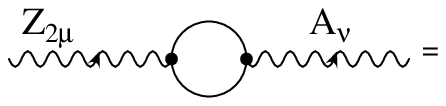,width=5cm}
\end{figure}

\begin{picture}(10,10)(30,25)
\put(175,65)
{\large $ T^{\mu \nu }
 A^{Z_2A}(k^2)+L^{\mu \nu }B^{Z_2A}(k^2),
$}
\end{picture}

\begin{figure}
\epsfig{file=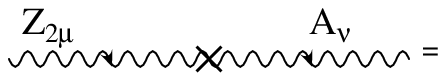,width=5cm}
\end{figure}

\begin{eqnarray}
\label{ctz2a}
&&T^{\mu \nu }
 \left[ \left( M_{Z_2}^2+\delta M_{Z_2}^2\right)
Z_{Z_2Z_2}^{\frac 12}Z_{Z_2A}^{\frac 12}+\right.
\nonumber \\
&&
\left. \left( M_{Z_1}^2+\delta M_{Z_1}^2\right)
Z_{Z_1Z_2}^{\frac 12}Z_{Z_1A}^{\frac 12}-k^2\left( Z_{Z_2Z_2}^{\frac
12}Z_{Z_2A}^{\frac 12}+Z_{Z_1Z_2}^{\frac 12}Z_{Z_1A}^{\frac
12}+Z_{AZ_2}^{\frac 12}Z_{AA}^{\frac 12}\right) \right] +
\nonumber \\
&&
L^{\mu \nu }\left[ \left( M_{Z_2}^2+\delta M_{Z_2}^2\right)
Z_{Z_2Z_2}^{\frac 12}Z_{Z_2A}^{\frac 12}+\left( M_{Z_1}^2+\delta
M_{Z_1}^2\right) Z_{Z_1Z_2}^{\frac 12}Z_{Z_1A}^{\frac 12}\right] .
\end{eqnarray}

There are 12 renormalization conditions but only 11 renormalization
constants. In fact not all 12 conditions are independent.
The BRST invariance gives relations among them, so only 11
are independent. It is worth to stress that to satisfy all 11
conditions, the renormalization matrix (Eq. (\ref{gbmr}))
does not have to be a symmetric one. From these 11 independent conditions
all constants can be determined.

\newpage

\begin{eqnarray}
\label{rcbosons}
\mbox{\rm Renormalization conditions}\;\;\;\; && \;\; \mbox{\rm Renormalization
constants} \nonumber \\
&& \nonumber \\
\left.
\begin{array}{ll}
A^{Z_{1}}(M^2_{Z_1})=0, \;\; & A^{Z^{\prime}_{1}}(M^2_{Z_1})=0, \\
A^{Z_{2}}(M^2_{Z_2})=0, \;\; & A^{Z^{\prime}_{2}}(M^2_{Z_2})=0, \\
A^{Z_1A}(M^2_{Z_1})=0,\;\; & A^{Z_1A}(0)=0, \\
A^{Z_2A}(M^2_{Z_2})=0,\;\; & A^{Z_2A}(0)=0, \\
A^{Z_1Z_2}(M^2_{Z_1})=0,\;\; & A^{Z_1Z_2}(M^2_{Z_2})=0, \\
A^{A}(0)=0, \;\;\;\;\;\; & A^{A^{\prime}}(0)=0
\end{array}
\right\}\ & \Rightarrow &
\begin{array}{l}
Z^{\frac{1}{2}}_{Z_1Z_1}, Z^{\frac{1}{2}}_{Z_1A},Z^{\frac{1}{2}}_{Z_2A},Z^{%
\frac{1}{2}}_{AZ_1}, Z^{\frac{1}{2}}_{AZ_2},Z^{\frac{1}{2}}_{AA}, \\
\delta M^2_{Z_1},\delta M^2_{Z_2},Z^{\frac{1}{2}}_{Z_1Z_2}, Z^{\frac{1}{2}%
}_{Z_2Z_1},Z^{\frac{1}{2}}_{Z_2Z_2}.
\end{array}
\end{eqnarray}

\subsubsection*{\bf 4.2.2.2 Scalar particles.}

We don't write the tadpole part for scalar bosons because
according to the discussion in section 4.2.1 it equals zero. For
neutral scalar particles the counter terms come from the following
part of the inverse propagators Lagrangian

\begin{eqnarray}
-
\sum^6_{i=1}
\frac{1}{2} H^0_{i}
\left(
\Box + M^2_{i}
\right)
H^0_{i}.
\end{eqnarray}

As all 6 neutral particles mix together we have six diagonal propagators
and 15 off diagonal ones.

\begin{figure}
\epsfig{file=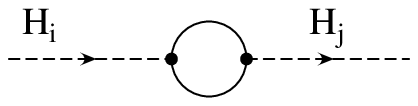,width=5cm}
\end{figure}

\begin{picture}(10,10)(30,25)
\put(175,65)
{\large $=\;F_{ij}(k^2), \;\; i,j=H^0,H_1^0,H_2^0,H_3^0,A_1^0,A_2^0
\;\; i\geq j,$}
\end{picture}

\begin{figure}
\epsfig{file=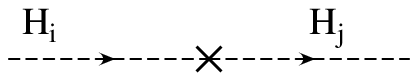,width=5cm}
\end{figure}

\begin{picture}(10,10)(30,25)
\put(177,60)
{\large $=\sum\limits_{k=1}^6 Z^{\frac{1}{2}}_{ki} Z^{\frac{1}{2}}_{kj}
\left(
k^2 -M^2_k - \delta M^2_k
\right). $}
\end{picture}

Altogether 42 renormalization conditions give us 42 renormalization
constants.

\begin{eqnarray}
\mbox{\rm Renormalization conditions} &&  \mbox{\rm Renormalization
constants} \nonumber \\
&& \nonumber \\
\left.
\begin{array}{lcl}
F_{ii}(M_{i}^{2}) & = & 0 \\
F^{\prime}_{ii}(M_{i}^{2}) & = & 0 \\
F_{ij}(M_{i}^{0}) & = & 0,\;\;i> j \\
F_{ij}(M_{j}^{2}) & = & 0,\;\;i> j
\end{array}
\right\} &  \Rightarrow &
\begin{array}{l}
Z^{\frac{1}{2}}_{ii}, Z^{\frac{1}{2}}_{ij}, Z^{\frac{1}{2}}_{ji},
\delta M_{i}^{2}.
\end{array}
\end{eqnarray}

\newpage

The following part of the inverse propagators Lagrangian gives
contributions to counter terms of the single charged scalar
particles

\begin{eqnarray}
-
\sum^2_{i=1} H^+_{i}
\left(
\Box + M^2_{H_i}
\right)
H^-_{i}.
\end{eqnarray}

\begin{figure}
\epsfig{file=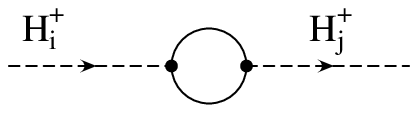,width=5cm}
\end{figure}

\begin{picture}(10,10)(30,25)
\put(175,65)
{\large $=F_{H_iH_j}(k^2),\;\;\; i,j=1,2,;\;\; i \geq j ,
$}
\end{picture}

\begin{figure}
\epsfig{file=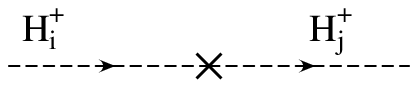,width=5cm}
\end{figure}

\begin{picture}(10,10)(30,25)
\put(177,60)
{\large $=
\sum\limits_{k=1}^2 Z^{\frac{1}{2}}_{H_kH_i} Z^{\frac{1}{2}}_{H_kH_j}
\left(
k^2 -M^2_{H_k} - \delta M^2_{H_k}
\right). $}
\end{picture}

In this case 6 renormalization constants is fixed:
\begin{eqnarray}
\mbox{\rm Renormalization conditions} &&  \mbox{\rm Renormalization
constants} \nonumber \\
&& \nonumber \\
\left.
\begin{array}{lcl}
F_{H_iH_i}(M_{H_i}^{2}) & = & 0 \\
F^{\prime}_{H_iH_i}(M_{H_i}^{2}) & = & 0 \\
F_{H_iH_j}(M_{H_i}^{2}) & = & 0,\;\;i> j \\
F_{H_iH_j}(M_{H_j}^{2}) & = & 0,\;\;i> j
\end{array}
\right\} &  \Rightarrow &
\begin{array}{l}
Z^{\frac{1}{2}}_{H_1H_1},Z^{\frac{1}{2}}_{H_2H_2}, Z^{\frac{1}{2}%
}_{H_1H_2},Z^{\frac{1}{2}}_{H_2H_1}, \delta M_{H_1}^{2},
\delta M_{H_2}^{2}.
\end{array}
\end{eqnarray}

Finally, 6 renormalization constants connected with doubly charged Higgses
can be found.

For doubly charged scalar particles the counter terms come from
the following part of the inverse propagators Lagrangian

\begin{eqnarray}
-
\delta^{++}_{L}
\left(
\Box + M_{\delta_L}
\right)
\delta^{--}_{L}
-
\delta^{++}_{R}
\left(
\Box + M_{\delta_R}
\right)
\delta^{--}_{R} .
\end{eqnarray}

\newpage

\begin{figure}
\epsfig{file=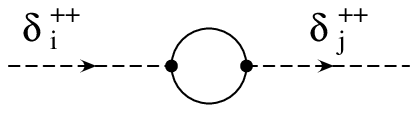,width=5cm}
\end{figure}

\begin{picture}(10,10)(30,25)
\put(175,65)
{\large $=F_{\delta_i\delta_j}(k^2),\;\;\; i,j=L,R,\;\;\;\; i \geq j ,$}
\end{picture}

\begin{figure}
\epsfig{file=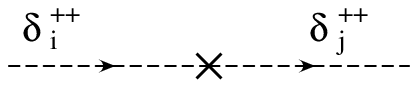,width=5cm}
\end{figure}

\begin{picture}(10,10)(30,25)
\put(177,60)
{\large $=\sum\limits_{k=1}^2 Z^{\frac{1}{2}}_{\delta_k \delta_i}
Z^{\frac{1}{2}}_{\delta_k \delta_j}
\left(
k^2 -M^2_{\delta_k} - \delta M^2_{\delta_k}
\right). $}
\end{picture}

\begin{eqnarray}
\mbox{\rm Renormalization conditions} &&  \mbox{\rm Renormalization
constants} \nonumber \\
&& \nonumber \\
\left.
\begin{array}{lcl}
F_{\delta _i}(M_{\delta _i}^{2}) & = & 0 \\
F^{\prime}_{\delta _i}(M_{\delta _i}^{2}) & = & 0 \\
F_{\delta _i\delta _j}(M_{\delta _i}^{2}) & = & 0,\;\;i>j \\
F_{\delta _i\delta _j}(M_{\delta _j}^{2}) & = & 0,\;\;i>j
\end{array}
\right\} & \Rightarrow  &
\begin{array}{l}
Z^{\frac{1}{2}}_{\delta _L\delta _L}, Z^{\frac{1}{2}}_{\delta
_L\delta _R}, Z^{\frac{1}{2}}_{\delta _R\delta_L},
Z^{\frac{1}{2}}_{\delta _R\delta _R},
\delta M_{\delta _L}^{2}, \delta M_{\delta _R}^{2}.
\end{array}
\end{eqnarray}

\subsubsection*{\bf 4.2.2.3 Fermions.}

The following part of the inverse propagators Lagrangian gives
contributions to the counter terms of fermions

\begin{eqnarray}
\sum_{i}\bar{\psi }_i
\left(
i\gamma^\mu \partial_\mu -m_i
\right)
\psi_i
+\sum_{I}\bar{\psi }_I
\left(
i\gamma^\mu \partial_\mu -m_I
\right)
\psi_I.
\end{eqnarray}

It is suitable to divide the renormalized inverse fermion
propagator matrix $K=\left( K_{ij} \right) $ into four parts

\begin{eqnarray}
K^{ren}=A_L P_L + A_R P_R +B_L \widehat{k} P_L + B_R \widehat{k} P_R,
\end{eqnarray}

where

\begin{eqnarray}
A_{L,R}=\left\{ \left( A_{L,R} \right)_{ij} \right\},\;\;\;
B_{L,R}=\left\{ \left( B_{L,R} \right)_{ij} \right\}
\end{eqnarray}

are independent matrices in flavor space separate for up - quarks
($A^{up}_{L,R},\;B^{up}_{L,R}$), down - quarks
($A^{down}_{L,R},\;B^{down}_{L,R}$), charged leptons
($A^{l}_{L,R},\;B^{l}_{L,R}$) and neutrinos
($A^{\nu}_{L,R},\;B^{\nu}_{L,R}$). These matrices for up and down
quarks and charged leptons are 3 dimensional. For neutrinos
$A^{\nu}_{L,R}$ and $B^{\nu}_{L,R}$ have $6\times 6 $ dimension.
So we have

\newpage

\begin{figure}
\epsfig{file=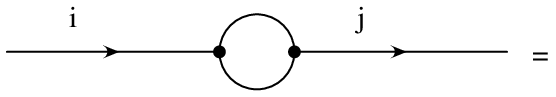,width=5cm}
\end{figure}

\begin{picture}(10,10)(30,25)
\put(175,65)
{\large $ A_L^{ij} P_L + A_R^{ij} P_R + B_L^{ij} \widehat{k} P_L
+ B_R^{ij} \widehat{k} P_R, $}
\end{picture}

and counter terms \vspace*{1.0cm}

\begin{figure}
\epsfig{file=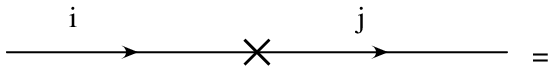,width=5cm}
\end{figure}

\begin{eqnarray}
\label{rcferm}
&-&
\left(
Z^+_L (m+\delta m) Z_R
\right)_{ij} P_R
-
\left(
Z^+_R (m+\delta m) Z_L
\right)_{ij} P_L \nonumber \\
&+&
\left(
Z^+_R Z_R
\right)_{ij}
\widehat{k} P_R
+
\left(
Z^+_L Z_L
\right)_{ij}
\widehat{k} P_L .
\end{eqnarray}

\noindent

\vspace*{1.0cm}

The on - mass shell renormalization conditions are the following
\vspace*{0.5cm} \newline
 for $\;i=j:$

\begin{equation}
\;\;\;\;\;\;\;\;\;\; \left\{
\begin{array}{l}
A_L^{ii}(m^2_i)=-m_i B_R^{ii}(m^2_i),\\
A_R^{ii}(m^2_i)=-m_i B_R^{ii}(m^2_i), \\
B_L^{ii}(m^2_i)=B_R^{ii}(m^2_i), \\
m_i
\left(
A_L^{ii\prime}(m^2_i) + A_R^{ii\prime}(m^2_i)
\right)
+
\frac{1}{2}
\left(
B_L^{ii}(m^2_i) + B_R^{ii}(m^2_i)
\right)
+m_i^2
\left(
B_L^{ii\prime}(m^2_i) + B_R^{ii\prime}(m^2_i)
\right)=1,
\end{array}
\right.
\end{equation}

\noindent

\hspace*{-0.5cm} and for $i>j:$

\begin{equation}
\hspace*{-7.8cm}
\label{ctferm}
\left\{
\begin{array}{l}
A_L^{ij}(m^2_j) +m_j B_R^{ij}(m^2_j)=0,  \\
A_R^{ij}(m^2_j) +m_j B_L^{ij}(m^2_j)=0,  \\
A_L^{ij}(m^2_i) +m_i B_R^{ij}(m^2_i)=0, \\
A_R^{ij}(m^2_i) +m_i B_L^{ij}(m^2_i)=0.
\end{array}
\right.
\end{equation}

For Dirac particles of N generations (quarks and charged leptons)
there are $4N^2$ independent on mass - shell conditions and
the same number of renormalization constants. So, for three generations
there are altogether $3\times 4 \times 3^2=108$ renormalization constants
which have to be determined. For Majorana particles there are
$2\times (2N)^2 +2N =78$ renormalization constants which must be found.

\subsubsection*{\bf 4.2.2.4 Unphysical particles.}

The unphysical particles do not appear on external lines, so we may
impose any renormalization conditions for these "particles". It is
most convenient to use the minimal subtraction scheme.
To express the Faddeev-Popov Lagrangian (Eq.(\ref{fp11})) by bare quantities
we define the renormalization constants for auxiliary fields in the
following way:

\begin{equation}
\left(
\begin{array}{c}
E_{1}^{\pm} \\
E_{2}^{\pm}
\end{array}
\right)=
\left(
Z_W^{\frac{1}{2} T}
\right)_{ij}
\left(
\begin{array}{c}
\stackrel{\circ}{E}_{1}^{\pm} \\
\stackrel{\circ}{E}_{2}^{\pm}
\end{array}
\right),
\end{equation}

\begin{equation}
\left(
\begin{array}{c}
E_{Z_1} \\
E_{Z_2} \\
E_{A} \\
\end{array}
\right)=
\left(
Z_Z^{\frac{1}{2}T}
\right)_{ij}
\left(
\begin{array}{c}
\stackrel{\circ}{E}_{Z_1} \\
\stackrel{\circ}{E}_{Z_2} \\
\stackrel{\circ}{E}_{A}
\end{array}
\right).
\end{equation}

Then the gauge fixing Lagrangian (Eq.\ref{gf22}) in bare fields
(omitting bilinear terms of auxiliary fields) reads

\begin{eqnarray}
\stackrel{\circ}{L}_{GF} &=&
\stackrel{\circ}{E}^-_{1}\partial^{\mu} \stackrel{\circ}{W}^+_{1\mu}
+ \stackrel{\circ}{E}^-_{2}\partial^{\mu} \stackrel{\circ}{W}^+_{2\mu} +
i \stackrel{\circ}{E}^-_{1}
\left(
\stackrel{\circ}{\xi }_{W_1} \stackrel{\circ}{M}_{W_1}
\stackrel{\circ}{G}^+_1 +
\stackrel{\circ}{\gamma}_{c1} \stackrel{\circ}{G}^+_2
\right)+
i \stackrel{\circ}{E}^-_{2}
\left(
\stackrel{\circ}{\xi }_{W_2} \stackrel{\circ}{M}_{W_2}
\stackrel{\circ}{G}^+_2 +
\stackrel{\circ}{\gamma}_{c2}
\stackrel{\circ}{G}^+_1
\right) + \;h.c. \nonumber \\
&+&
\stackrel{\circ}{E}_{Z_1}\partial^{\mu} \stackrel{\circ}{Z}_{1\mu} +
\stackrel{\circ}{E}_{Z_2}\partial^{\mu} \stackrel{\circ}{Z}_{2\mu} +
\stackrel{\circ}{E}_{A}\partial^{\mu} \stackrel{\circ}{A}_{\mu}
- \stackrel{\circ}{E}_{Z_1}
\left(
\stackrel{\circ}{\xi }_{Z_1} \stackrel{\circ}{M}_{Z_1}
\stackrel{\circ}{G}^0_1 +
\stackrel{\circ}{\gamma}_{n1} \stackrel{\circ}{G}^0_2
\right)
-
\stackrel{\circ}{E}_{Z_2}
\left(
\stackrel{\circ}{\xi }_{Z_2} \stackrel{\circ}{M}_{Z_1}
\stackrel{\circ}{G}^0_2 +
\stackrel{\circ}{\gamma}_{n2} \stackrel{\circ}{G}^0_1
\right) \nonumber \\
&-& E_A \beta_1 \stackrel{\circ}{G}^0_1
-E_A \beta_2 \stackrel{\circ}{G}^0_2,
\end{eqnarray}

where

\begin{eqnarray}
\stackrel{\circ}{\xi }_{W_1} &=&
\left(
\xi \frac{M_{W_1}}{\stackrel{\circ}{M}_{W_1}}
\left(
Z_W^{\frac{1}{2} T}
\right)_{11}
\left(
Z_G^{+ -1}
\right)_{11}+
\xi \frac{M_{W_2}}{\stackrel{\circ}{M}_{W_1}}
\left(
Z_W^{\frac{1}{2} T}
\right)_{21}
\left(
Z_G^{+ -1}
\right)_{21}
\right),   \nonumber \\
\stackrel{\circ}{\xi }_{W_2} &=&
\left(
\xi \frac{M_{W_1}}{\stackrel{\circ}{M}_{W_2}}
\left(
Z_W^{\frac{1}{2} T}
\right)_{12}
\left(
Z_G^{+ -1}
\right)_{12}+
\xi \frac{M_{W_2}}{\stackrel{\circ}{M}_{W_2}}
\left(
Z_W^{\frac{1}{2} T}
\right)_{22}
\left(
Z_G^{+ -1}
\right)_{22}
\right), \nonumber \\
\stackrel{\circ}{\xi }_{Z_1} &=&
\left(
\xi \frac{M_{Z_1}}{\stackrel{\circ}{M}_{Z_1}}
\left(
Z_Z^{\frac{1}{2} T}
\right)_{11}
\left(
Z_G^{0 -1}
\right)_{11}+
\xi_{Z_2} \frac{M_{Z_2}}{\stackrel{\circ}{M}_{Z_1}}
\left(
Z_Z^{\frac{1}{2} T}
\right)_{21}
\left(
Z_G^{0 -1}
\right)_{21}
\right),   \nonumber \\
\stackrel{\circ}{\xi }_{Z_2} &=&
\left(
\xi \frac{M_{Z_1}}{\stackrel{\circ}{M}_{Z_2}}
\left(
Z_Z^{\frac{1}{2} T}
\right)_{12}
\left(
Z_G^{0 -1}
\right)_{12}+
\xi \frac{M_{Z_2}}{\stackrel{\circ}{M}_{Z_2}}
\left(
Z_Z^{\frac{1}{2} T}
\right)_{22}
\left(
Z_G^{0 -1}
\right)_{22}
\right), \nonumber \\
\stackrel{\circ}{\gamma}_{n1} &=&
\xi M_{Z_1} \left( Z^{\frac{1}{2} T}_{Z}\right)_{11}
\left( Z^{0 -1}_{G}\right)_{12} +
\xi M_{Z_2} \left( Z^{\frac{1}{2} T}_{Z}\right)_{21}
\left( Z^{0 -1}_{G}\right)_{22} , \nonumber \\
\stackrel{\circ}{\gamma}_{n2} &=&
\xi M_{Z_1} \left( Z^{\frac{1}{2} T}_{Z}\right)_{12}
\left( Z^{0 -1}_{G}\right)_{11} +
\xi M_{Z_2} \left( Z^{\frac{1}{2} T}_{Z}\right)_{22}
\left( Z^{0 -1}_{G}\right)_{21} , \nonumber \\
\stackrel{\circ}{\gamma}_{c1} &=&
\xi M_{W_1} \left( Z^{\frac{1}{2} T}_{W}\right)_{11}
\left( Z^{+ -1}_{G}\right)_{12} +
\xi M_{W_2} \left( Z^{\frac{1}{2} T}_{W}\right)_{21}
\left( Z^{+ -1}_{G}\right)_{22} , \nonumber \\
\stackrel{\circ}{\gamma}_{c2} &=&
\xi M_{W_1} \left( Z^{\frac{1}{2} T}_{W}\right)_{12}
\left( Z^{+ -1}_{G}\right)_{11} +
\xi M_{W_2} \left( Z^{\frac{1}{2} T}_{W}\right)_{22}
\left( Z^{+ -1}_{G}\right)_{21} , \nonumber \\
\beta_{1} &=&
\xi M_{Z_1} \left( Z^{\frac{1}{2} T}_{Z}\right)_{13}
\left( Z^{0 -1}_{G}\right)_{11} +
\xi M_{Z_2} \left( Z^{\frac{1}{2} T}_{Z}\right)_{23}
\left( Z^{0 -1}_{G}\right)_{21} , \nonumber \\
\beta_{2} &=&
\xi M_{Z_1} \left( Z^{\frac{1}{2} T}_{Z}\right)_{13}
\left( Z^{0 -1}_{G}\right)_{12} +
\xi M_{Z_2} \left( Z^{\frac{1}{2} T}_{Z}\right)_{23}
\left( Z^{0 -1}_{G}\right)_{22} . \nonumber \\
\end{eqnarray}

The bare "D" quantities from Eq.(\ref{brs11}) read:

\begin{eqnarray}
\stackrel{\circ}{D}_{Z_1} &=& \tilde{\delta}^{BRST}
\left(
\partial^{\mu} \stackrel{\circ}{Z}_{1\mu} - \stackrel{\circ}{\xi }_{Z_1}
\stackrel{\circ}{M}_{Z_1} \stackrel{\circ}{G}^0_1
-\stackrel{\circ}{\gamma}_{n1} \stackrel{\circ}{G}^0_2
\right), \nonumber \\
\stackrel{\circ}{D}_{Z_2} &=& \tilde{\delta}^{BRST}
\left(
\partial^{\mu} \stackrel{\circ}{Z}_{2\mu} - \stackrel{\circ}{\xi }_{Z_2}
\stackrel{\circ}{M}_{Z_2} \stackrel{\circ}{G}^0_2
-\stackrel{\circ}{\gamma}_{n2} \stackrel{\circ}{G}^0_1
\right), \nonumber \\
\stackrel{\circ}{D}_{A} &=& \tilde{\delta}^{BRST}
\left(
\partial^{\mu} \stackrel{\circ}{A}_{\mu} - \beta_1
\stackrel{\circ}{G}^0_1 - \beta_2 \stackrel{\circ}{G}^0_2
\right), \nonumber \\
\stackrel{\circ}{D}^+_{1} &=& \tilde{\delta}^{BRST}
\left(
\partial^{\mu} \stackrel{\circ}{W}^+_{1\mu} +i \stackrel{\circ}{\xi }_{W_1}
\stackrel{\circ}{M}_{W_1} \stackrel{\circ}{G}^+_1
+i\stackrel{\circ}{\gamma}_{c1} \stackrel{\circ}{G}^+_2
\right), \nonumber \\
\stackrel{\circ}{D}^+_{2} &=& \tilde{\delta}^{BRST}
\left(
\partial^{\mu} \stackrel{\circ}{W}^+_{2\mu} +i \stackrel{\circ}{\xi }_{W_2}
\stackrel{\circ}{M}_{W_2} \stackrel{\circ}{G}^+_2
+i\stackrel{\circ}{\gamma}_{c2} \stackrel{\circ}{G}^+_1
\right) .
\end{eqnarray}

\paragraph*{\bf a. Goldstone - bosons.}

Three renormalization constants for charged Goldstone bosons are determined
from three inverse propagators

\begin{figure}
\epsfig{file=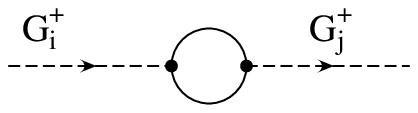,width=5cm}
\end{figure}

\begin{picture}(10,10)(30,25)
\put(175,65)
{\large $\;=\;F^\pm_{ij}(k^2),\;\;\; i \geq j ,
$}
\end{picture}

\begin{figure}
\epsfig{file=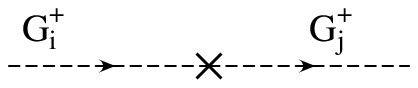,width=5cm}
\end{figure}

\begin{picture}(10,10)(30,25)
\put(175,60)
{\large $\;=$}
\end{picture}

\vspace*{-1cm}

\begin{eqnarray}
&&k^2
\left[
\left( Z^\pm_{G} \right)_{1i} \left( Z^\pm_{G} \right)_{1j}+
\left( Z^\pm_{G} \right)_{2i} \left( Z^\pm_{G} \right)_{2j}
\right] \nonumber \\
&-&
\left[
\stackrel{\circ}{\xi }_{W_1}
\left(
M^2_{W_1} + \delta M^2_{W_1}
\right)
\left( Z^\pm_{G} \right)_{1i} \left( Z^\pm_{G} \right)_{1j}+
\stackrel{\circ}{\xi }_{W_2}
\left(
M^2_{W_2} + \delta M^2_{W_2}
\right)
\left( Z^\pm_{G} \right)_{2i} \left( Z^\pm_{G} \right)_{2j}
\right]. \nonumber
\end{eqnarray}

\vspace*{1cm}

From any renormalization conditions  $\;\;\;\Rightarrow \;\;\;
\left( Z^\pm_{G} \right)_{11},\;
\left( Z^\pm_{G} \right)_{12},\;
\left( Z^\pm_{G} \right)_{22}.$

\vspace*{1cm}

For neutral Goldstone - bosons there is a similar situation:

\begin{figure}
\epsfig{file=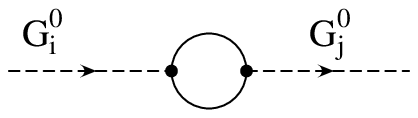,width=5cm}
\end{figure}

\begin{picture}(10,10)(30,25)
\put(175,65)
{\large $\;=\;F^0_{ij}(k^2),\;\;\; i \geq j ,
$}
\end{picture}

\begin{figure}
\epsfig{file=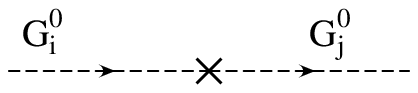,width=5cm}
\end{figure}

\begin{picture}(10,10)(30,25)
\put(175,60)
{\large $\;=$}
\end{picture}

\vspace*{-1cm}

\begin{eqnarray}
&&k^2
\left[
\left( Z^0_{G} \right)_{1i} \left( Z^0_{G} \right)_{1j}+
\left( Z^0_{G} \right)_{2i} \left( Z^0_{G} \right)_{2j}
\right] \nonumber \\
&-&
\left[
\stackrel{\circ}{\xi }_{Z_1}
\left(
M^2_{Z_1} + \delta M^2_{Z_1}
\right)
\left( Z^0_{G} \right)_{1i} \left( Z^0_{G} \right)_{1j}+
\stackrel{\circ}{\xi }_{Z_2}
\left(
M^2_{Z_2} + \delta M^2_{Z_2}
\right)
\left( Z^0_{G} \right)_{2i} \left( Z^0_{G} \right)_{2j}
\right]. \nonumber
\end{eqnarray}

\vspace*{1cm}

And from any renormalization conditions  $\;\;\;\Rightarrow \;\;\;
\left( Z^0_{G} \right)_{11},\;
\left( Z^0_{G} \right)_{12},\;
\left( Z^0_{G} \right)_{22}.$

\paragraph*{\bf b. Faddeev - Popov ghosts.}

For charged Faddeev - Popov ghosts there are four inverse propagators and
four counter terms to them:

\begin{figure}
\epsfig{file=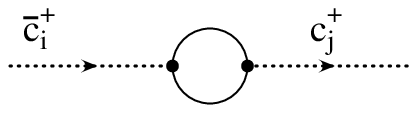,width=5cm}
\end{figure}

\begin{picture}(10,10)(30,25)
\put(175,65)
{\large $\;=\;\gamma^+_{ij}(k^2),\;\;\; i,j=1,2\; ,
$}
\end{picture}

\begin{figure}
\epsfig{file=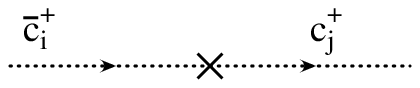,width=5cm}
\end{figure}

\begin{picture}(10,10)(30,25)
\put(175,60)
{\large $\;=\;i
\left(
-k^2 + \stackrel{\circ}{\xi }_{W_i}
\left( M^2_{W_i} + \delta M^2_{W_i}
\right)
\right)
\tilde{Z}^{\frac{1}{2}}_{W_i W_j}. $}
\end{picture}

From any renormalization condition, four renormalization constants
$\tilde{Z}^{\frac{1}{2}}_{W_i W_j} $  ($i,j=1,2$) will be determined.

\vspace*{1cm}

For neutral Faddeev - Popov ghosts there are nine inverse propagators
and the same number of counter terms:

\begin{figure}
\epsfig{file=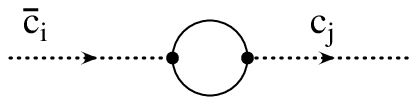,width=5cm}
\end{figure}

\begin{picture}(10,10)(30,25)
\put(175,65)
{\large $\;=\;\gamma^0_{ij}(k^2),\;\;\; i,j=Z_1,\;Z_2,\;A\; ,
$}
\end{picture}

\begin{figure}
\epsfig{file=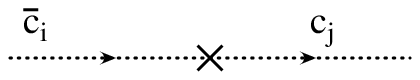,width=5cm}
\end{figure}

\begin{picture}(10,10)(30,25)
\put(175,60)
{\large $\;=\;i
\left(
-k^2 + \stackrel{\circ}{\xi_i }
\left( M^2_{i} + \delta M^2_{i}
\right)
\right)
\tilde{Z}^{\frac{1}{2}}_{i j}. $}
\end{picture}

The already determined renormalization constants make finite all other
two - point functions.

\subsection*{  4.2.3 Renormalization of
couplings, mixing angles and mixing matrices.}

As it was mentioned in section 4.1.3 there are more free parameters
in the left - right symmetric model than in SM. To calculate
the renormalization constants for them, additional renormalization conditions
must be imposed. In the next section the renormalization of charge
is performed. Subsequently, the renormalization of mixing angles
$\varphi $, $\Theta_W $, $\xi $ and mixing matrices is analyzed.

\subsubsection*{ \bf 4.2.3.1 Renormalization of charge.}

In order to calculate the renormalization constant for charge
($\delta Y$), the $A^\mu e^{-}e^{+}$ vertex is considered

\begin{center}
\begin{figure}
\epsfig{file=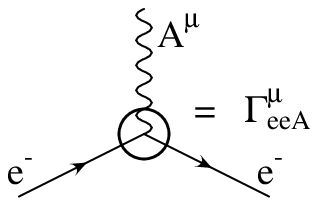,width=5cm}
\end{figure}
\end{center}

\hspace*{3.7cm} \mbox{\rm Renormalization condition} \hspace*{0.7cm}
\mbox{\rm Renormalization constant}
\begin{eqnarray}
\label{rcy}
\bar{u}(m_e)\Gamma^{\mu}_{eeA} u(m_e)\mid_{k^{2}=0}=0 \;\;\;\;\;
 \Rightarrow  \;\;\;\;\;\;\;\;\; \delta Y.
\end{eqnarray}

\vspace*{1.cm}

The vertex $\Gamma^{\mu}_{eeA}$ contains the counter term part and
the proper vertex of all the Feynman diagrams

\vspace*{1.cm}

\begin{equation}
\Gamma^{\mu}_{eeA}= ie\gamma^{\mu} \left( \Gamma_{CT}^{eeA}+\Gamma_{D}^{eeA}
\right).
\end{equation}

Putting the definitions of appropriate renormalization constants
from Chapter 4.1 to $L^{(l)}_{NC}$ (Eq.(\ref{lnc})), the following
form of $\Gamma^{\mu}_{eeA}$ in one loop approximation is obtained

\begin{eqnarray}
\label{gcteea}
\Gamma_{CT}^{eeA}& =&
-\left[
\delta Y+\delta Z_{AA}^{\frac{1}{2}}+
(\delta Z^l_L)_{ee} P_L+(\delta Z^l_R)_{ee} P_R
\right]
\nonumber \\
&+& Z_{Z_1A}^{\frac{1}{2}} \frac{1}{\sin 2\Theta_W } \left[ \left( \cos
\varphi (-1+2\sin^2\Theta_W ) -\sin\varphi \frac{\sin^2\Theta_W}
{\sqrt{\cos2\Theta_W }} \right) P_L \right.
\nonumber \\
&+&
\left.
\left( 2 \cos \varphi \sin^2\Theta_W +\sin\varphi
\frac{1-3\sin^2\Theta_W}{\sqrt{\cos2\Theta_W }} \right) P_R \right]  \nonumber \\
&+& Z_{Z_2A}^{\frac{1}{2}} \frac{1}{\sin2\Theta_W } \left[ \left( \sin\varphi
(-1+2\sin^2\Theta_W ) +\cos\varphi \frac{\sin^2\Theta_W}{\sqrt{\cos2\Theta_W }}
\right)  P_L \right. \nonumber \\
&+&
\left.
\left( 2\sin\varphi \sin^2\Theta_W +\cos\varphi
\frac{-1+3\sin^2\Theta_W}{\sqrt{\cos2\Theta_W }} \right) P_R \right],
\end{eqnarray}

where

\begin{equation}
(\delta Z^l_{L,R})_{ee}= (\delta Z_{L,R}^{l\frac{1}{2}})_{ee}+
(\delta Z_{L,R}^{l\frac{1}{2}})_{ee}^{\dagger}.
\end{equation}

\subsubsection*{ \bf 4.2.3.2 Renormalization of mixing angle in neutral
sector ($\phi $) and mixing angle $\Theta_W $.}

In order to calculate the $\delta G_\varphi $ and $\delta
G_{\Theta_W} $ renormalization constants an additional vertex -
$Z_1e^{-}e^{+}$ is exploited

\begin{center}
\begin{figure}
\epsfig{file=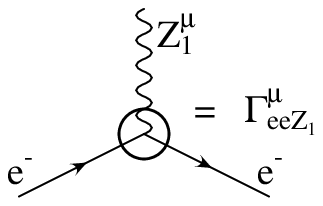,width=5cm}
\end{figure}
\end{center}

By considering renormalization conditions separately for vector
and axial vector parts, two renormalization constants can be
obtained

\begin{eqnarray}
\label{eez11}
\mbox{\rm Renormalization conditions}
\;\;\;\;\;\;\;\;\;\;\;\;\;\;\;\;\;\;\;\;\;\;\;\;\;\;\;\;\;\;\;\;\;\;\;\;\;\;\;
\;\;\;\;\;\;&&
\mbox{\rm Renormalization
constants} \nonumber \\
&& \nonumber \\
\bar{u}(m_e)\Gamma^{\mu}_{eeZ_1} u(m_e)\mid_{k^{2}=M^2_{Z_1}}=0 \;\; \left(
\begin{array}{l}
from\;\; the \;\;vector\;\; \\
and\;\; axial\;\;vector\;\; parts
\end{array}
\right) & \Rightarrow & \;\;\;\delta G_\varphi ,\delta G_{\Theta_W} .
\end{eqnarray}

Similarly as in previous two sections, $\Gamma^{\mu}_{eeZ_1}$ can
be split into two terms

\begin{equation}
\Gamma^{\mu}_{eeZ_1}= ie\gamma^{\mu} \left(
\Gamma_{CT}^{eeZ_1}+\Gamma_{D}^{eeZ_1}\right).
\end{equation}

By applying definitions from Chapter 4.1 to Eq.(\ref{lnc}), for
the one loop level the following form of counter term ensues

\begin{eqnarray}
\label{eez111}
\Gamma_{CT}^{eeZ_1} & = & \left[ \left( \delta Y+(\delta Z^l_L)_{ee}+ \delta
Z_{Z_{1}Z_{1}}^{\frac{1}{2}}+\delta G_\varphi+\delta G_{\Theta_W}
\frac{1}{\cos^2\Theta_W (-1+2\sin^2\Theta_W)}  \right) \right.
\nonumber \\
& \times & \left( \cos\varphi \frac{(-1+2\sin^2\Theta_W )}{\sin2\Theta_W }
\right)  \nonumber \\
& + & \left( \delta Y+(\delta Z^l_L)_{ee}+\delta Z_{Z_{1}Z_{1}}^{\frac{1}{2}}-
\delta G_\varphi \cot^2{\varphi } +\delta G_{\Theta_W}
\frac{(1-2\sin^4\Theta_W )}{(1-\sin^2\Theta_W)(1-2\sin^2\Theta_W)} \right)
\nonumber \\
&\times & \left. \left( \sin\varphi \frac{(-\sin^2\Theta_W )}{\sin2\Theta_W
\sqrt{\cos2\Theta_W }} \right) \right] P_L  \nonumber \\
& + & \left[ \left( \delta Y+(\delta Z^l_R)_{ee}+
\delta Z_{Z_{1}Z_{1}}^{\frac{1}{2}}+\delta G_\varphi+
\delta G_{\Theta_W} \frac{1}{\cos^2\Theta_W } \right.
\right) \times \left( \cos\varphi \frac{2\sin^2\Theta_W }{\sin2\Theta_W } \right)
\nonumber \\
& + & \left( \delta Y+(\delta Z^l_R)_{ee}+
\delta Z_{Z_{1}Z_{1}}^{\frac{1}{2}}-
\delta G_\varphi \cot^2{\phi}
\right.
\nonumber \\
&+&
\left.
\delta  G_{\Theta_W} \frac{-3\sin^2\Theta_W +6\sin^4\Theta_W
-6\sin^6\Theta_W +1}{(-1+3\sin^2\Theta_W)(1-\sin^2\Theta_W)(1-2\sin^2\Theta_W)}
\right)   \nonumber \\
& \times &
\left. \left( \sin\varphi
\frac{(1-3\sin^2\Theta_W )}{\sin2\Theta_W\sqrt{\cos2\Theta_W }}
\right) \right] P_R  \nonumber \\
& + & Z_{Z_{2}Z_{1}}^{\frac{1}{2}} \left[ \left( \cos\varphi
 \frac{\sin^2\Theta_W }{\sin2\Theta_W \sqrt{\cos2\Theta_W }}+ \sin\varphi
 \frac{(-1+2\sin^2\Theta_W )}{\sin2\Theta_W } \right. \right) P_L
 \nonumber \\
& + & \left. \left( \cos\varphi \frac{(-1+3\sin^2\Theta_W )}{\sin2\Theta_W
\sqrt{\cos2\Theta_W }}+ \sin\varphi \frac{2\sin^2\Theta_W } {\sin2\Theta_W }
\right) P_R
\right] - Z_{AZ_{1}}^{\frac{1}{2}}.
\end{eqnarray}

\subsubsection*{ \bf 4.2.3.3 Renormalization of mixing angle in charged
sector ($\xi $).}

The renormalization constant for mixing angle in charged sector
($\delta G_{\xi}$) can be determined from the $W_1\nu_i e$ vertex

\begin{center}
\begin{figure}
\epsfig{file=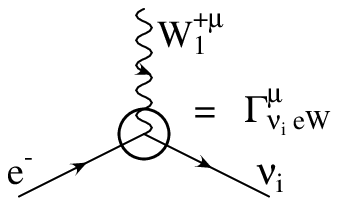,width=5cm}
\end{figure}
\end{center}

\hspace*{3.7cm} \mbox{\rm Renormalization condition} \hspace*{0.7cm}
\mbox{\rm Renormalization constant}

\begin{eqnarray}
\bar{u}(m_{\nu_i})\Gamma^{\mu}_{N_ieW_1} u(m_e)\mid_{k^{2}=M^2_{W_1}}=0
\;\;\;\;\;\;\;\;\;
 \Rightarrow  \;\;\;\;\;\;\;\;\; \delta G_{\xi}.
\end{eqnarray}

Similarly as for $A^\mu e^{-}e^{+}$ vertex,
$\Gamma^{\mu}_{\nu_ieW_1}$ is the sum of the counter term  and
Feynman diagrams part

\begin{equation}
\Gamma^{\mu}_{N_ieW_1}= ie\gamma^{\mu} \left( \Gamma_{CT}^{N_ieW_1}+
\Gamma_{D}^{N_ieW_1} \right).
\end{equation}

After setting to $L^{(l)}_{CC}$ (Eq.(\ref{lcc})) the definitions
of appropriate renormalization constants from Chapter 4.1, in one
loop approximation the counter term reads

\begin{eqnarray}
\Gamma_{CT}^{N_ieW_1}&=& \frac{\cos \xi}{\sqrt{2} \sin \Theta_W} \left\{
\left[ \delta Y+\delta Z^{\frac{1}{2}}_{W_1W_1}-\delta G_{\Theta_W}- \delta
G_{\xi} \frac{\sin^2 \xi}{\cos^2 \xi} \right] \left( K_L \right)_{Ie}
\right. \nonumber \\
&+&
\frac{1}{2} \sum_{A} \left( (\delta Z_L^{\nu\frac{1}{2}})_{IA}+
(\delta Z_L^{\nu\frac{1}{2}})^\dagger _{IA} \right) (K_{L})_{Ae}
 \nonumber \\
&+& \left. \frac{1}{2} \sum_{b} \left( (\delta Z_L^{l\frac{1}{2}})_{be}+
(\delta Z_L^{l\frac{1}{2}})^\dagger _{be} \right) (K_{L})_{Ib}
+ Z^{\frac{1}{2}}_{W_2W_1}\frac{\sin \xi}{\sqrt{2} \sin \Theta_W }
\left( K_L \right)_{Ie} \right\} P_L \nonumber  \\
&-& \frac{\sin \xi}{\sqrt{2} \sin \Theta_W} \left\{ \left[ \delta Y+
\delta Z^{\frac{1}{2}}_{W_1W_1}-\delta G_{\Theta_W}+ \delta G_{\xi}
\right] \left( K_R
\right)_{Ie}  \right. \nonumber \\
&+&
 \frac{1}{2} \sum_{A} \left( (\delta Z_R^{\nu\frac{1}{2}})_{IA}+
 (\delta Z_R^{\nu\frac{1}{2}})^\dagger _{IA} \right) (K_{R})_{Ae}
 \nonumber \\
&+& \frac{1}{2} \sum_{b} \left( (\delta Z_R^{l\frac{1}{2}})_{be}+(\delta
Z_R^{l\frac{1}{2}})^\dagger _{be} \right) (K_{R})_{Ib}
+ \left. Z^{\frac{1}{2}}_{W_2W_1}\frac{\cos \xi}{\sqrt{2} \sin \Theta_W}
\left( K_R \right)_{Ie} \right\} P_R.
\end{eqnarray}

\subsubsection*{ \bf 4.2.3.4 Renormalization of mixing matrices.}

The renormalization of the mixing matrices $U_{L,R}^{CKM}$ in the quark
sector and $K_{L,R}$ in the weak charged current Lagrangian
has already been analyzed \cite{sack},\cite{kniehl}.
In this section the renormalization of mixing matrices in the
weak neutral current sector ($\Omega_{L,R}$) is performed.

First, let us consider the left handed part of the unrenormalized
charged current interactions (for the right handed part all
formulas are analogous) in the form

\begin{equation}
\label{left}
{\left( \stackrel{\circ}{L}_{CC} \right)}_{left} \propto \;\;
\stackrel{\circ}{\bar{N}} \gamma^{\mu}P_L\stackrel{\circ}{K}
\stackrel{\circ}{l}.
\end{equation}

In terms of the renormalized quantities the right hand side of
Eq.(\ref{left}) reads

\begin{eqnarray}
\bar{N}_L Z^{\nu\frac{1}{2}\dagger }_{L} \gamma^{\mu}
\left( K_L +\delta K_L \right)
Z^{l\frac{1}{2}}_{L} l_L &=&
\bar{N}_L \left( 1+\frac{1}{2} \delta Z^{\nu\dagger }_{L} \right)
\gamma^{\mu} \left(
K_L +\delta K_L \right) \left( 1+ \frac{1}{2} \delta Z^l_{L} \right) l_L
\nonumber  \\
&=& \left( J^{\mu}_L\right)_{CC} +\left( \delta J^{\mu}_L\right)_{CC},
\end{eqnarray}

where

\begin{eqnarray}
\left( \delta J^{\mu}_L\right)_{CC}&=&
\bar{N}_L\gamma^{\mu} \left( \frac{1}{2}
\delta Z^{\nu\dagger }_{L} K_L+\delta K_L+
\frac{1}{2}K_L \delta Z^l_{L} \right) l_L
\nonumber \\
& =& \bar{N}_L\gamma^{\mu} \left( \frac{1}{2} \delta
Z^{\nu\dagger }_{L} K_L+\delta K_L+\frac{1}{2}K_L \delta Z^l_{L} \right) l_L+
\bar{N}_L\gamma^{\mu} \left( \frac{1}{4} \delta
Z^\nu_{L} K_L -\frac{1}{4} \delta Z^\nu_{L}K_L \right) l_L \nonumber \\
&+& \bar{N}_L\gamma^{\mu} \left( \frac{1}{4} K_L
\delta Z^{l\dagger }_{L} -\frac{1}{4}K_L \delta Z^{l\dagger }_{L} \right) l_L.
\end{eqnarray}

Using the formulas (satisfied both for the bare and renormalized quantities):

\begin{eqnarray}
&K_L&=V^{\nu\dagger }_L V^l_L, \\
&V^\nu_L V^{\nu\dagger }_L&=I, \\
&V^l_{L} V^{l\dagger }_{L}&=V^{l\dagger }_{L} V^l_{L}=I ,
\end{eqnarray}

in one loop approximation one gets

\begin{eqnarray}
\stackrel{\circ}{V}^{\nu\dagger }_{L} \stackrel{\circ}{V}^l_{L} & = & \left(
V^{\nu\dagger }_{L}+\delta V^{\nu\dagger }_{L} \right)
\left( V^l_{L}+\delta V^l_{L} \right)=
K_L+V^{\nu\dagger }_{L} \delta V^l_{L}+\delta
V^{\nu\dagger }_{L}V^l_{L}=
K_L+\delta V^{\nu\dagger }_{L}V^\nu_{L}K_L+ K_L V^{l\dagger }_{L}
\delta V^l_{L} \nonumber \\
& = &K_L+\delta K_L,
\end{eqnarray}

and thus

\begin{eqnarray}
\label{dcc}
\left( \delta J^{\mu}_L\right)_{CC} & =& \bar{N}_L\gamma^{\mu}
\left\{ \frac{1}{4}
\left( \delta Z^{\nu\dagger }_{L}+\delta Z^\nu_{L} \right) K_L
+ \frac{1}{4}K_L
\left(
\delta Z^l_{L}+\delta Z^{l\dagger }_{L} \right) \right\} l_L \nonumber \\
& + & \bar{N}_L\gamma^{\mu} \left[ \frac{1}{4} \left(
\delta Z^{\nu\dagger }_{L}-\delta Z^\nu_{L} \right)
+\delta V^{\nu\dagger }_{L}V^\nu_{L}
\right]
K_L l_L \nonumber \\
&+&
\bar{N}_L\gamma^{\mu} K_L \left[ \frac{1}{4}
\left( \delta Z^l_{L}-\delta Z^{l\dagger }_{L} \right) + V^{l\dagger }_{L}
\delta V^l_{L}
\right] l_L.
\end{eqnarray}

We may fix $\delta V^{\nu\dagger }_{L}$ and $\delta V^\nu_{L}$ by
requiring that the antihermitian part in Eq.(\ref{dcc})
(i.e. the terms in square brackets) vanishes:

\begin{equation}
\label{ul5}
\delta V^{\nu\dagger }_{L}= -\frac{1}{4} \left( \delta Z^{\nu\dagger }_{L}-
\delta Z^\nu_{L} \right)
V^{\nu\dagger }_{L},
\end{equation}

\begin{equation}  \label{ul2}
\delta V^\nu_{L}=
-\frac{1}{4} V^\nu_{L} \left( \delta Z^\nu_{L}-\delta Z^{\nu\dagger }_{L}
\right).
\end{equation}

The same procedure used for the right part of the neutral current Lagrangian
yields similar expressions for $\delta V^\nu_{R}$ and
$\delta V^{\nu\dagger }_{R}$:

\begin{equation}
\delta V^{\nu\dagger }_{R}= -\frac{1}{4} \left( \delta Z^{\nu\dagger }_{R}-
\delta Z^\nu_{R} \right)
V^{\nu\dagger }_{R},
\end{equation}

\begin{equation}
\delta V^\nu_{R}=
-\frac{1}{4} V^\nu_{R} \left( \delta Z^\nu_{R}-\delta Z^{\nu\dagger }_{R}
\right).
\end{equation}

Let us consider now the left handed part of the unrenormalized
neutral current interactions (for the right handed part all
formulas are analogous) in the form

\begin{equation}
\label{lnc5}
\left( \stackrel{\circ}{L}_{NC} \right)_{left}\propto \;\;
\stackrel{\circ}{\bar{N}} \gamma^{\mu}P_L\stackrel{\circ}{\Omega}_L
\stackrel{\circ}{N}.
\end{equation}

Rewriting the right side of Eq.(\ref{lnc5}) in terms of
renormalized quantities one gets

\begin{eqnarray}
\bar{N}_L Z^{\nu\frac{1}{2}\dagger }_{L} \gamma^{\mu} \left( \Omega_L +
\delta \Omega_L
\right) Z^{\nu\frac{1}{2}}_{L} N_L &=&
\bar{N}_L \left( 1+\frac{1}{2} \delta Z^{\nu\dagger }_{L} \right)
 \gamma^{\mu} \left(
\Omega_L +\delta \Omega_L \right) \left( 1
+ \frac{1}{2} \delta Z^\nu_{L} \right)
N_L \nonumber \\
& = &
\left( J^{\mu}_L\right)_{NC} +\left( \delta J^{\mu}_L\right)_{NC},
\end{eqnarray}

with the following expression for $\left( \delta
J^{\mu}_L\right)_{NC}$

\begin{eqnarray}
\left( \delta J^{\mu}_L\right)_{NC} &=& \bar{N}_L\gamma^{\mu}
\left( \frac{1}{2}
\delta Z^{\nu\dagger }_{L} \Omega_L+\delta \Omega_L+\frac{1}{2}
\Omega_L \delta Z^\nu_{L}
\right) N_L \nonumber \\
&=& \bar{N}_L\gamma^{\mu} \left( \frac{1}{2} \delta
Z^{\nu\dagger }_{L} \Omega_L+\delta \Omega_L+\frac{1}{2}\Omega_L
\delta Z^\nu_{L} \right)
N_L+   \bar{N}_L\gamma^{\mu} \left( \frac{1}{4} \delta
Z^\nu_{L} \Omega_L
-\frac{1}{4} \delta Z^\nu_{L}\Omega_L \right) N_L \nonumber \\
&+&
\bar{N}_L\gamma^{\mu} \left( \frac{1}{4} \Omega_L
\delta Z^{\nu\dagger }_{L} -\frac{1}{4}\Omega_L \delta Z^{\nu\dagger }_{L}
\right) N_L.
\end{eqnarray}

Applying the formula

\begin{equation}
\Omega_L=V^{\nu\dagger }_{L} V^\nu_{L}
\end{equation}

satisfied for the bare and renormalized quantities, at the one
loop level one obtains

\begin{eqnarray}
\stackrel{\circ}{V}^{\nu\dagger }_{L} \stackrel{\circ}{V}^\nu_{L}&=&
\left( V^{\nu\dagger }_{L}+\delta
V^{\nu\dagger }_{L} \right) \left( V^\nu_{L}+\delta V^\nu_{L} \right)=
\Omega_L+V^{\nu\dagger }_{L} \delta V^\nu_{L}+\delta V^{\nu\dagger }_{L}
V^\nu_{L}=
 \Omega_L+\Omega_L V^{\nu\dagger }_{L} \delta V^\nu_{L}+\delta
V^{\nu\dagger }_{L}V^\nu_{L}\Omega_L \nonumber \\
&=& \Omega_L+\delta \Omega_L
\end{eqnarray}

and further

\begin{eqnarray}  \label{neu}
\left( \delta J^{\mu}_L\right)_{NC} &=& \bar{N}_L\gamma^{\mu}
\left\{ \frac{1}{4}
\left( \delta Z^{\nu\dagger }_{L}+\delta Z^\nu_{L} \right) \Omega_L+
\frac{1}{4}\Omega_L
\left( \delta Z^\nu_{L}+\delta Z^{\nu\dagger }_{L} \right) \right\}
N_L \nonumber \\
&+& \bar{N}_L\gamma^{\mu} \left[ \frac{1}{4} \left(
\delta Z^{\nu\dagger }_{L}-\delta Z^\nu_{L} \right) \Omega_L+\delta
V^{\nu\dagger }_{L}V^\nu_{L}\Omega_L
\right] N_L \nonumber \\
&+&
\bar{N}_L\gamma^{\mu} \left[ \frac{1}{4}\Omega_L
\left( \delta Z^\nu_{L}-\delta Z^{\nu\dagger }_{L} \right) +
\Omega_L V^{\nu\dagger }_{L}\delta V^\nu_{L}
\right] N_L.
\end{eqnarray}

It is easy to see that after using the formulas from
Eq.(\ref{ul5}) and Eq.(\ref{ul2}), the antihermitian part in Eq.(\ref{neu})
(i.e. the terms written in square brackets) cancels,
and when the mixing matrices for the charged currents $K_L$, $K_R$ are
renormalized the matrices $\Omega_{L,R}$ are renormalized as well.

\newpage

\part*{ V Renormalization of the Z - two Majorana \\
\hspace*{0.7cm} neutrinos vertex
($Z_1 N_i N_j $) in the frame of \\
\hspace*{0.7cm} the left - right symmetric model.}

In the previous Chapter the  method to renormalize the
left - right
symmetric model was presented. Analyzing the procedure, it is easy to realize
that the expressions used for renormalization of the
left - right symmetric model
are much more complicated than for SM. First of all, there are more fields
(and what follows more masses) that have to be renormalized. More particles
in the model means more Feynman diagrams one has to take into account
when calculating processes. Furthermore, additional renormalization
conditions for new free parameters
($\phi $, $\xi $, $\Omega_{L,R}$,
$K_{L,R}$) must be imposed.

To have a good test whether the method developed
in Chapter IV  works a cancelation of the infinite part
(ultraviolet divergences) in the $Z_1 N_i N_j $ vertex
in the next section is verified.
In Chapter 5.1 the general renormalization scheme
for the $Z_1 N_i N_j $ vertex is presented. Then, in Chapter 5.2
the calculation of the necessary renormalization constants is performed.
Finally, in Chapter 5.3 the numerical calculations for the
$Z_1 N_i N_j $ vertex are described.

\section*{ 5.1 The general scheme - definitions of the counter terms.}

Having the definitions of renormalization constants and
appropriate renormalization conditions we can go to the
renormalization of the $Z_1 N_i N_j $ vertex

\begin{center}
\begin{figure}
\epsfig{file=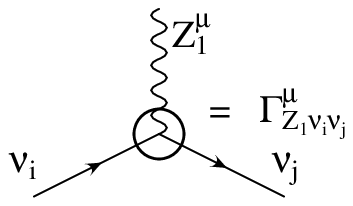,width=5cm}
\end{figure}
\end{center}

As usually, the full $Z_1 N_i N_j $ vertex is the sum of the
counter term part and proper vertex of all the Feynman diagrams

\begin{equation}
\label{gzij}
\Gamma^{\mu}_{Z_1N_iN_j}= ie\gamma^{\mu} \left(
\Gamma_{CT}^{Z_1N_iN_j} +\Gamma_{D}^{Z_1N_iN_j} \right).
\end{equation}

We can decompose the counter term part into left and right
components

\begin{equation}
\label{gzij1}
\Gamma_{CT}^{Z_1N_iN_j} \equiv \left( P_L(CT)_L+P_R(CT)_R \right).
\end{equation}

The following bare Lagrangian (see Eq.(\ref{lnc},\ref{imi}))
contributes to the $Z_1N_iN_j$ vertex

\begin{eqnarray}
\label{barlnij}
\stackrel{\circ}{L}_{Z_1\nu_i\nu_j}=
\frac{\stackrel{\circ}{e}}{2\stackrel{\circ}{\sin}\Theta_W
\stackrel{\circ}{\cos}\Theta_W}\sum_{a,b}
&& \left\{
\stackrel{\circ}{\bar{N}}_a\gamma^{\mu} P_L \left(
\stackrel{\circ}{A^{1\nu}_{L}}
(\stackrel{\circ}{\Omega}_L)_{ab}- \stackrel{\circ}{A^{1\nu}_{R}}
(\stackrel{\circ}{\Omega}_R)_{ba} \right) \stackrel{\circ}{N}_b
\stackrel{\circ}{Z}_{1\mu}+ \right. \nonumber \\
&&
\stackrel{\circ}{\bar{N}}_a\gamma^{\mu} P_R \left(
\stackrel{\circ}{A^{1\nu}_{R}}(\stackrel{\circ}{\Omega}_R)_{ab}-
\stackrel{\circ}{A^{1\nu}_{L}}
(\stackrel{\circ}{\Omega}_L)_{ba} \right) \stackrel{\circ}{N}_b
\stackrel{\circ}{Z}_{1\mu}+ \nonumber \\
&&
\stackrel{\circ}{\bar{N}}_a\gamma^{\mu} P_L \left(
\stackrel{\circ}{A^{2\nu}_{L}}(\stackrel{\circ}{\Omega}_L)_{ab}-
\stackrel{\circ}{A^{2\nu}_{R}}
(\stackrel{\circ}{\Omega}_R)_{ba} \right) \stackrel{\circ}{N}_b
\stackrel{\circ}{Z}_{2\mu}+ \nonumber \\
&&
\left.
\stackrel{\circ}{\bar{N}}_a\gamma^{\mu} P_R \left(
\stackrel{\circ}{A^{2\nu}_{R}}(\stackrel{\circ}{\Omega}_R)_{ab}-
\stackrel{\circ}{A^{2\nu}_{L}}
(\stackrel{\circ}{\Omega}_L)_{ba} \right) \stackrel{\circ}{N}_b
\stackrel{\circ}{Z}_{2\mu} \right\}.
\end{eqnarray}

To be consistent with the definitions of renormalization constants
from section 4.1.3, the couplings $A_{L,R}^{1,2\nu}$ should be taken
in the form:

\begin{eqnarray}
\stackrel{\circ}{A^{1\nu}_{L}} & = & \stackrel{\circ}{\cos}\phi-
\sqrt{1-\stackrel{\circ}{\cos}^2\phi} \;\;\;
\frac{\stackrel{\circ}{\sin}^2\Theta_W}
{\sqrt{1 -2\stackrel{\circ}{\sin}^2\Theta_W}}\;\;, \\
\stackrel{\circ}{A^{1\nu}_{R}} & = & -\sqrt{1-\stackrel{\circ}{\cos}^2\phi}
\;\;\; \frac{1-\stackrel{\circ}{\sin}^2\Theta_W} {\sqrt{1 -2
\stackrel{\circ}{\sin}^2\Theta_W}}\;\;, \\
\stackrel{\circ}{A^{2\nu}_{L}} & = & \sqrt{1-\stackrel{\circ}{\cos}^2\phi}\;+
\stackrel{\circ}{\cos}\phi \;\;\; \frac{\stackrel{\circ}{\sin}^2\Theta_W}
{\sqrt{1 -2\stackrel{\circ}{\sin}^2\Theta_W}}\;\;, \\
\stackrel{\circ}{A^{2\nu}_{R}} & = & \stackrel{\circ}{\cos}\phi \;\;\;
\frac{1-\stackrel{\circ}{\sin}^2\Theta_W}
{\sqrt{1 -2\stackrel{\circ}{\sin}^2\Theta_W}}\;\;.
\end{eqnarray}

After transformation to the renormalized quantities,
(including renormalization of the mixing matrices $\Omega_{L,R}$)
the following form of $(CT)_L$ and $(CT)_R$ is obtained
from Eq.(\ref{barlnij}):

\begin{eqnarray}
\label{ctl}
(CT)_L & = & \frac{1}{\sin2\Theta_W} \left\{ \left[ \delta Y-\delta
G_{\Theta_W}
\frac{1-2\sin^2\Theta_W}{\cos^2\Theta_W}+ \delta Z_{Z_{1}Z_{1}}^{\frac{1}{2}}+
\delta G_\varphi \right] \cos\varphi(\Omega_{L})_{ji} \right.
  \nonumber \\
& + & \left[ \delta Y+\delta G_{\Theta_W}
 \frac{1-2\sin^4\Theta_W}{\cos^2\Theta_W
(1-2\sin^2\Theta_W )}+ \delta Z_{Z_{1}Z_{1}}^{\frac{1}{2}}- \delta G_\varphi
\cot^2{\varphi} \right]  \nonumber \\
& \times & \sin\varphi \frac{-\sin^2\Theta_W}{\sqrt{\cos2\Theta_W}}
(\Omega_{L})_{ji}  \nonumber \\
& + & \left[ \delta Y+\delta G_{\Theta_W}
\frac{-1+4\sin^2\Theta_W-2\sin^4\Theta_W}
{\cos^2\Theta_W (1-2\sin^2\Theta_W )} +\delta Z_{Z_{1}Z_{1}}^{\frac{1}{2}}-
\delta G_\varphi \cot^2{\varphi} \right]  \nonumber \\
& \times  & \sin\varphi \frac{\cos^2\Theta_W}{\sqrt{\cos2\Theta_W}}
(\Omega_{R})_{ij}  \nonumber \\
& + & Z_{Z_{2}Z_{1}}^{\frac{1}{2}} \left[ \cos\varphi \frac{\sin^2\Theta_W}
{\sqrt{\cos2\Theta_W}}+ \sin\varphi \right] (\Omega_{L})_{ji}-
Z_{Z_{2}Z_{1}}^{\frac{1}{2}}
\cos\varphi \frac{\cos^2\Theta_W}{\sqrt{\cos2\Theta_W}}
(\Omega_{R})_{ij}  \nonumber \\
& + & \left[ \frac{1}{2} \sum_{A} \left(
(\delta Z_L^{\nu\frac{1}{2}})_{JA}+
(\delta Z_L^{\nu\frac{1}{2}})^\dagger_{JA} \right)
(\Omega_{L})_{Ai}+ \frac{1}{2}
\sum_{B} \left( (\delta Z_L^{\nu\frac{1}{2}})_{BI}+
(\delta Z_L^{\nu\frac{1}{2}})^\dagger_{BI} \right) (\Omega_{L})_{jB} \right]
\nonumber \\
&\times  &\left( \cos\varphi + \sin\varphi
\frac{-\sin^2\Theta_W}{\sqrt{\cos2\Theta_W}} \right)  \nonumber \\
& + & \left[ \frac{1}{2} \sum_{A} \left(
(\delta Z_L^{\nu\frac{1}{2}})_{JA}+
(\delta Z_L^{\nu\frac{1}{2}})^\dagger_{JA} \right) (\Omega_{R})_{iA}+
\frac{1}{2} \sum_{B} \left( (\delta Z_L^{\nu\frac{1}{2}})_{BI}+
(\delta Z_L^{\nu\frac{1}{2}})^\dagger_{BI} \right) (\Omega_{R})_{Bj} \right]
\nonumber \\
& \times  & \left. \left( \sin\varphi \frac{\cos^2\Theta_W}{\sqrt{\cos2\Theta_W}}
\right) \right\},  \nonumber
\end{eqnarray}

\begin{equation}
\end{equation}

\begin{eqnarray}
(CT)_R & = & -\frac{1}{\sin2\Theta_W} \left\{ \left[ \delta Y-\delta
G_{\Theta_W}
\frac{1-2\sin^2\Theta_W}{\cos^2\Theta_W}+ \delta Z_{Z_{1}Z_{1}}^{\frac{1}{2}}+
\delta G_\varphi \right] \cos\varphi(\Omega_{L})_{ij} \right.
 \nonumber \\
& + & \left[ \delta Y+\delta G_{\Theta_W}
\frac{1-2\sin^4\Theta_W}{\cos^2\Theta_W
(1-2\sin^2\Theta_W )}+ \delta Z_{Z_{1}Z_{1}}^{\frac{1}{2}}- \delta G_\varphi
\cot^2{\varphi} \right]  \nonumber \\
& \times  & \sin\varphi \frac{-\sin^2\Theta_W}{\sqrt{\cos2\Theta_W}}
(\Omega_{L})_{ij}  \nonumber \\
& + & \left[ \delta Y+\delta G_{\Theta_W}
\frac{-1+4\sin^2\Theta_W-2\sin^4\Theta_W}{\cos^2\Theta_W (1-2\sin^2\Theta_W )}
+\delta Z_{Z_{1}Z_{1}}^{\frac{1}{2}}-
\delta G_\varphi \cot^2{\varphi} \right]  \nonumber \\
& \times & \sin\varphi \frac{\cos^2\Theta_W}{\sqrt{\cos2\Theta_W}}
(\Omega_{R})_{ji}  \nonumber \\
& + & Z_{Z_{2}Z_{1}}^{\frac{1}{2}} \left[ \cos\varphi
\frac{\sin^2\Theta_W}{\sqrt{\cos2\Theta_W}}+
\sin\varphi \right] (\Omega_{L})_{ij}-
Z_{Z_{2}Z_{1}}^{\frac{1}{2}} \cos\varphi
\frac{\cos^2\Theta_W}{\sqrt{\cos2\Theta_W}}(\Omega_{R})_{ji}  \nonumber \\
& + & \left[ \frac{1}{2} \sum_{A} \left(
(\delta Z_R^{\nu\frac{1}{2}})_{JA}+
(\delta Z_R^{\nu\frac{1}{2}})^\dagger_{JA} \right) (\Omega_{L})_{iA}+
\frac{1}{2} \sum_{B} \left(
(\delta Z_R^{\nu\frac{1}{2}})_{BI}+
(\delta Z_R^{\nu\frac{1}{2}})^\dagger_{BI} \right) (\Omega_{L})_{Bj} \right]
\nonumber \\
&\times &  \left( \cos\varphi + \sin\varphi \frac{-\sin^2\Theta_W}
{\sqrt{\cos2\Theta_W}} \right)  \nonumber \\
& + & \left[ \frac{1}{2} \sum_{A} \left( (\delta Z_R^{\nu\frac{1}{2}
})_{JA}+(\delta Z_R^{\nu\frac{1}{2}})^\dagger_{JA} \right)
(\Omega_{R})_{Ai}+ \frac{1}{2}
\sum_{B} \left( (\delta Z_R^{\nu\frac{1}{2}})_{BI}+
(\delta Z_R^{\nu\frac{1}{2}})^\dagger_{BI} \right) (\Omega_{R})_{jB} \right]
\nonumber \\
& \times & \left. \left( \sin\varphi \frac{\cos^2\Theta_W}{\sqrt{\cos2\Theta_W}}
\right) \right\}.
\end{eqnarray}

There is a symmetry between the left and the right part
of the Lagrangian that allows us
to check the cancelation of the infinite part only for one
(left or right) component. In the section 5.3 the calculations for
the left - handed part of $Z_1N_iN_j$ vertex are described.

\section*{ 5.2 Calculation of the renormalization constants.}

In this section the calculations (in one loop approximation )
of the renormalization constants
indispensable for renormalization of the $Z_1N_iN_j$ vertex are presented.
All calculations are performed in the Feynman gauge ($\xi =1)$.

\subsection*{  5.2.1 Calculation of the renormalization constants for gauge
bosons.}

\begin{itemize}
\item[i)] The renormalization constant $Z^{\frac{1}{2}}_{Z_1Z_1}$.

The $Z^{\frac{1}{2}}_{Z_1Z_1}$ constant is obtained from renormalization
condition (see Eq.(\ref{rcbosons}))

\begin{equation}
\label{az1}
A^{Z_1\prime} \left( M^2_{Z_1} \right)=0.
\end{equation}

After taking the counter term from Eq.(\ref{ctz1}) and splitting
$A^{Z_1} \left( k^2 \right)$ into counter term part $A_{CT}^{Z_1}
\left( k^2 \right)$ and Feynman diagram component $A_{D}^{Z_1}
\left( k^2 \right)$, Eq.(\ref{az1}) leads to

\begin{equation}
\label{az1d}
\sum_n A_{Dn}^{Z_1\prime} \left( M^2_{Z_1} \right)-
2 Z^{\frac{1}{2}}_{Z_1Z_1}=0\Rightarrow Z^{\frac{1}{2}}_{Z_1Z_1}=
\frac{1}{2} \sum_n A_{Dn}^{Z_1\prime} \left( M^2_{Z_1} \right),
\end{equation}

where the sum in Eq.(\ref{az1d}) is over all Feynman diagrams that contribute
to the $Z_1$ - $Z_1$ Green function
(see Fig. 1).

\begin{figure}
\epsfig{file=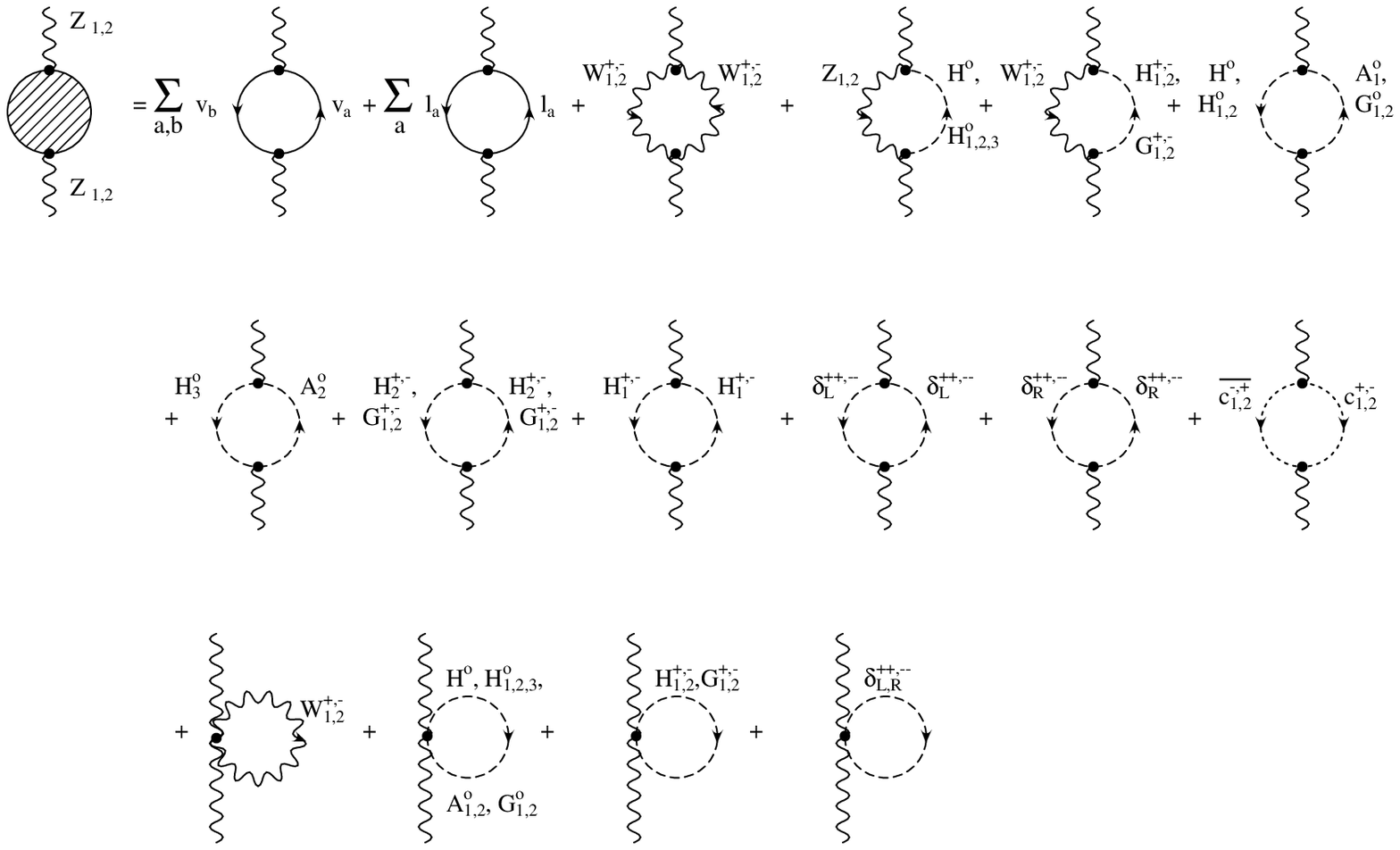,width=15cm}
\end{figure}

{\small Fig.1 \newline
The $Z_{1,2}-Z_{1,2}$ one loop Green functions.}

\vspace*{0.5cm}

\item[ii)] The renormalization constant $Z^{\frac{1}{2}}_{AA}$.

Taking the counter term from Eq.(\ref{cta}) and renormalization
condition ( Eq.(\ref{rcbosons}))

\begin{equation}
\label{aa}
A^{A\prime} ( 0 )=0,
\end{equation}

one gets the following expression for $Z^{\frac{1}{2}}_{AA}$

\begin{equation}
\label{aad}
Z^{\frac{1}{2}}_{AA}=
\frac{1}{2} \sum_n A_{Dn}^{A\prime} ( 0 ).
\end{equation}

Similarly like for $Z^{\frac{1}{2}}_{Z_1Z_1}$ the sum in
Eq.(\ref{aad}) is over all Feynman diagrams that contribute
to the $A$ - $A$ Green function (see Fig. 2).

\begin{figure}
\epsfig{file=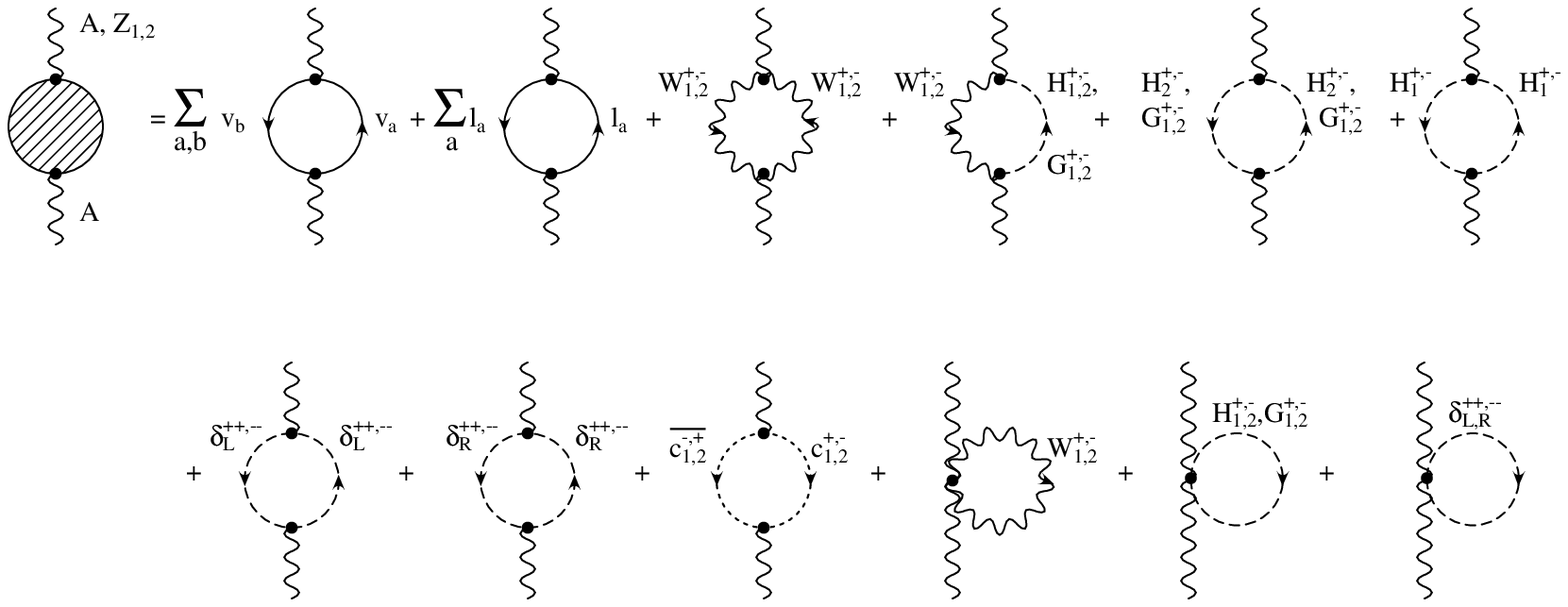,width=16cm}
\end{figure}

{\small Fig.2 \newline
The $A-A,Z_{1,2}$ one loop Green functions. The summation is
over the neutrino ($N_a,N_b$) and the charged lepton flavors ($l_a$).}

\vspace*{0.5cm}

\item[iii)] The renormalization constant $Z^{\frac{1}{2}}_{Z_1A}$.

From the counter term Eq.(\ref{ctz1a}) and renormalization
condition ( Eq.(\ref{rcbosons}))

\begin{equation}
\label{z1a}
A^{Z_1 A} ( 0 )=0,
\end{equation}

the following expression for $Z^{\frac{1}{2}}_{Z_1A}$ is obtained

\begin{equation}
\label{z1ad}
Z^{\frac{1}{2}}_{Z_1A}=
-\frac{1}{M^2_{Z_1}} \sum_n A_{Dn}^{Z_1A} ( 0 ).
\end{equation}

The diagrams contributing to $Z_1$ - $A$ Green function are presented
in Fig. 2.

\item[iv)] The renormalization constant $Z^{\frac{1}{2}}_{Z_2A}$.

Taking the counter term from Eq.(\ref{ctz2a}) and renormalization
condition ( Eq.(\ref{rcbosons}))

\begin{equation}
\label{z2a}
A^{Z_2 A} ( 0 )=0,
\end{equation}

one gets

\begin{equation}
\label{z2ad}
Z^{\frac{1}{2}}_{Z_2A}=
-\frac{1}{M^2_{Z_2}} \sum_n A_{Dn}^{Z_2A} ( 0 ).
\end{equation}

The diagrams contributing to $Z_2$ - $A$ Green function are the same
as for $Z_1$ - $A$  (see Fig. 2).

\item[v)] The renormalization constant $Z^{\frac{1}{2}}_{AZ_1}$.

From the counter term from Eq.(\ref{ctz1a}) and renormalization
condition ( Eq.(\ref{rcbosons}))

\begin{equation}
\label{z1a1}
A^{Z_1 A} \left( M^2_{Z_1} \right)=0,
\end{equation}

the following expression for $Z^{\frac{1}{2}}_{AZ_1}$ is obtained

\begin{equation}
\label{z1ad1}
Z^{\frac{1}{2}}_{AZ_1}=
\frac{1}{M^2_{Z_1}} \sum_n A_{Dn}^{Z_1A} \left( M^2_{Z_1} \right).
\end{equation}

The sum in Eq.(\ref{z1ad1}) is also over diagrams presented in Fig. 2.

\item[vi)] The renormalization constant $Z^{\frac{1}{2}}_{Z_2Z_1}$.

Taking the counter term from Eq.(\ref{ctz1z2}) and renormalization
condition ( Eq.(\ref{rcbosons}))

\begin{equation}
\label{z1z21}
A^{Z_1 Z_2} \left( M^2_{Z_1} \right)=0,
\end{equation}

one gets

\begin{equation}
\label{z1ad2}
Z^{\frac{1}{2}}_{Z_2Z_1}=
\frac{1}{M^2_{Z_1}-M^2_{Z_2}} \sum_n A_{Dn}^{Z_1Z_2} \left( M^2_{Z_1}
\right).
\end{equation}

The sum is over diagrams given in Fig. 1.

\end{itemize}

\subsection*{   5.2.2 Calculation of the renormalization constants
for fermions.}

First of all, one should notice that renormalization constants for
fermions occur in $\Gamma_{CT}^{eeA}$ (Eq.(\ref{gcteea})) and
$(CT)_L$ (Eq.(\ref{ctl})) only in combination

\begin{equation}
\label{comb}
\delta Z^{\frac{1}{2}}_{L,R}+\delta Z^{\frac{1}{2}\dagger }_{L,R}.
\end{equation}

Hence, it is sufficient to find the expression for all terms from
Eq.(\ref{comb})
not worrying about the form of particular elements. Furthermore,
for the infinite part of $A_{L}^{ij}$, $A_{R}^{ij}$, $B_{L}^{ij}$,
$B_{R}^{ij}$ the following relations take place:

\begin{eqnarray}
A_{L}^{ij} (m_j^2)=A_{L}^{ij} (m_i^2) \equiv A_{L}^{ij}, \nonumber \\
A_{R}^{ij} (m_j^2)=A_{R}^{ij} (m_i^2) \equiv A_{R}^{ij}, \nonumber \\
B_{L}^{ij} (m_j^2)=B_{L}^{ij} (m_i^2) \equiv B_{L}^{ij}, \nonumber \\
B_{R}^{ij} (m_j^2)=B_{R}^{ij} (m_i^2) \equiv B_{R}^{ij}.
\end{eqnarray}

Using renormalization condition from Eq.(\ref{ctferm}) and counter term
given in Eq.(\ref{rcferm}) one obtains:

\begin{eqnarray}
\left[
\left( \delta Z^{\frac{1}{2}}_{L}\right)_{ij}+
\left( \delta Z^{\frac{1}{2}\dagger }_{L}\right)_{ij}
\right]_{infinite\;part}=
-2 B_L^{ij},
\\
\left[
\left( \delta Z^{\frac{1}{2}}_{R}\right)_{ij}+
\left( \delta Z^{\frac{1}{2}\dagger }_{R}\right)_{ij}
\right]_{infinite\;part}=
-2 B_R^{ij}.
\end{eqnarray}

Here indices $i$, $j$ mean the charged lepton and neutrino flavors
separately,
and $i=j$ or $i\neq j$. As usually index $D$ means the Feynman diagrams
contribution. The electron and neutrino self energy diagrams are
presented in Fig. 3 and Fig. 4 respectively.

\begin{figure}
\epsfig{file=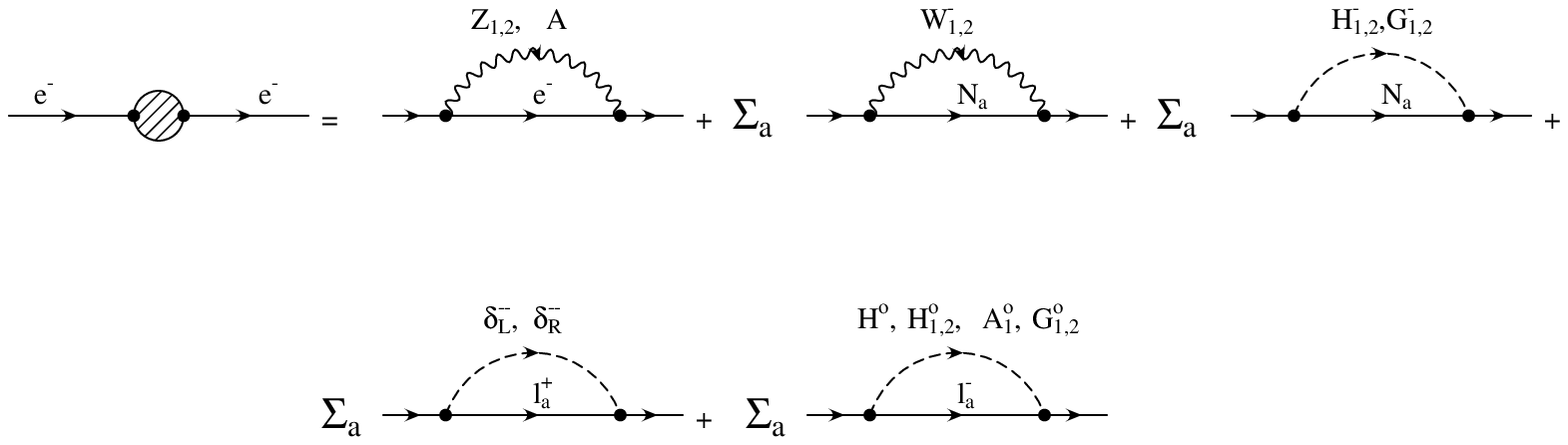,width=16cm}
\end{figure}

{\small Fig.3 \newline
The electron one-loop self energy diagrams. } \newline
\vspace*{0.5cm}

\begin{figure}
\epsfig{file=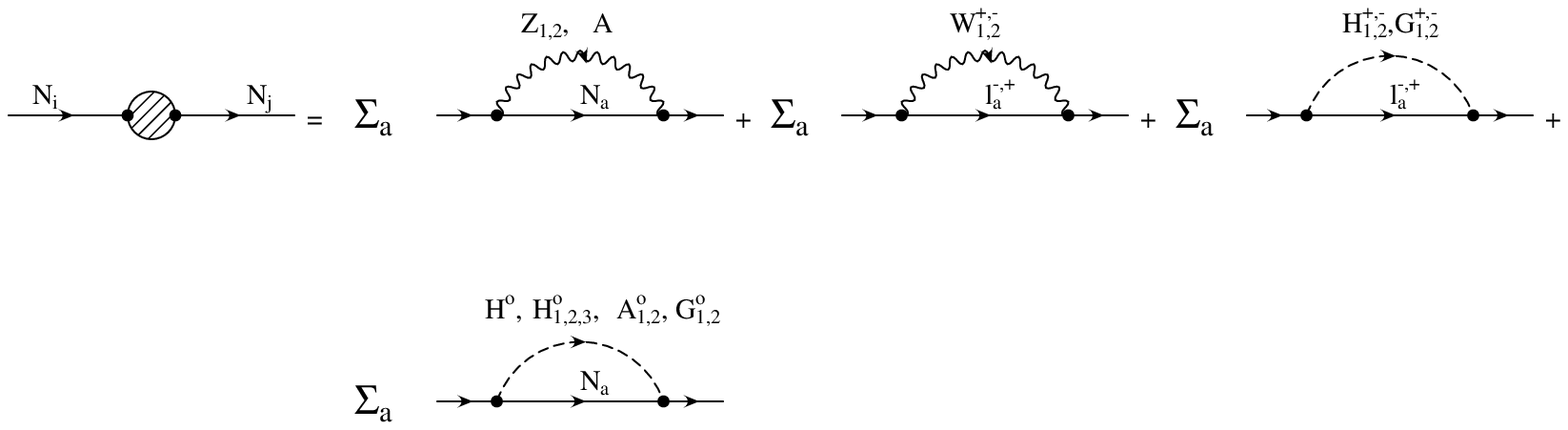,width=16cm}
\end{figure}

{\small Fig.4 \newline
The neutrino one-loop self energy diagrams. } \newline
\vspace*{0.5cm}

\subsection*{  5.2.3 Calculation of the renormalization constant
for charge.}

Having renormalization constants for gauge bosons and fermions,
the renormalization constant for charge (Y) can be easily calculated
from renormalization condition Eq.(\ref{rcy}). The $\Gamma_{CT}^{eeA}$
component of $\Gamma^\mu_{eeA}$ is given in Eq.(\ref{gcteea}).
The diagrams contributing to $\Gamma_{D}^{eeA}$ are shown in Fig. 5.

\begin{figure}
\epsfig{file=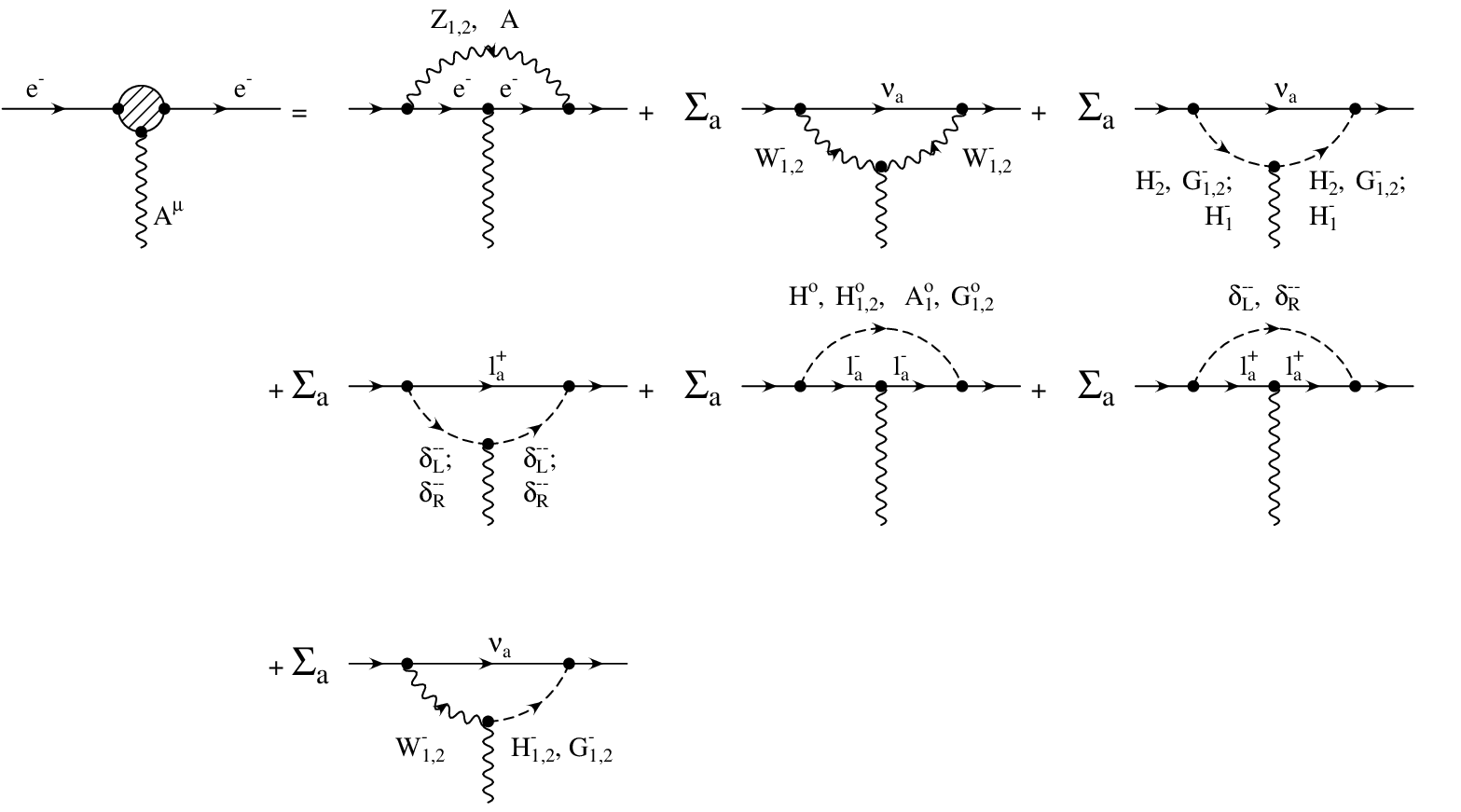,width=15cm}
\end{figure}

{\small Fig.5 \newline
The one-loop diagrams for the $Aee$ vertex.}

\subsection*{   5.2.4 Calculation of the renormalization constants
for mixing angles $\phi $ and $\Theta_W $.}

Using renormalization constants for gauge bosons, fermions and charge,
the renormalization constants $\delta_{G_{\phi}}$ and $\delta_{G_{\Theta_W}}$
are obtained from renormalization condition Eq.(\ref{eez11}).
The $\Gamma_{CT}^{eeZ_1}$ component of $\Gamma^\mu_{eeZ_1}$ is presented
in Eq.(\ref{eez111}). The diagrams contributing to $\Gamma_{D}^{eeZ_1}$
are shown in Fig. 6.

\begin{figure}
\epsfig{file=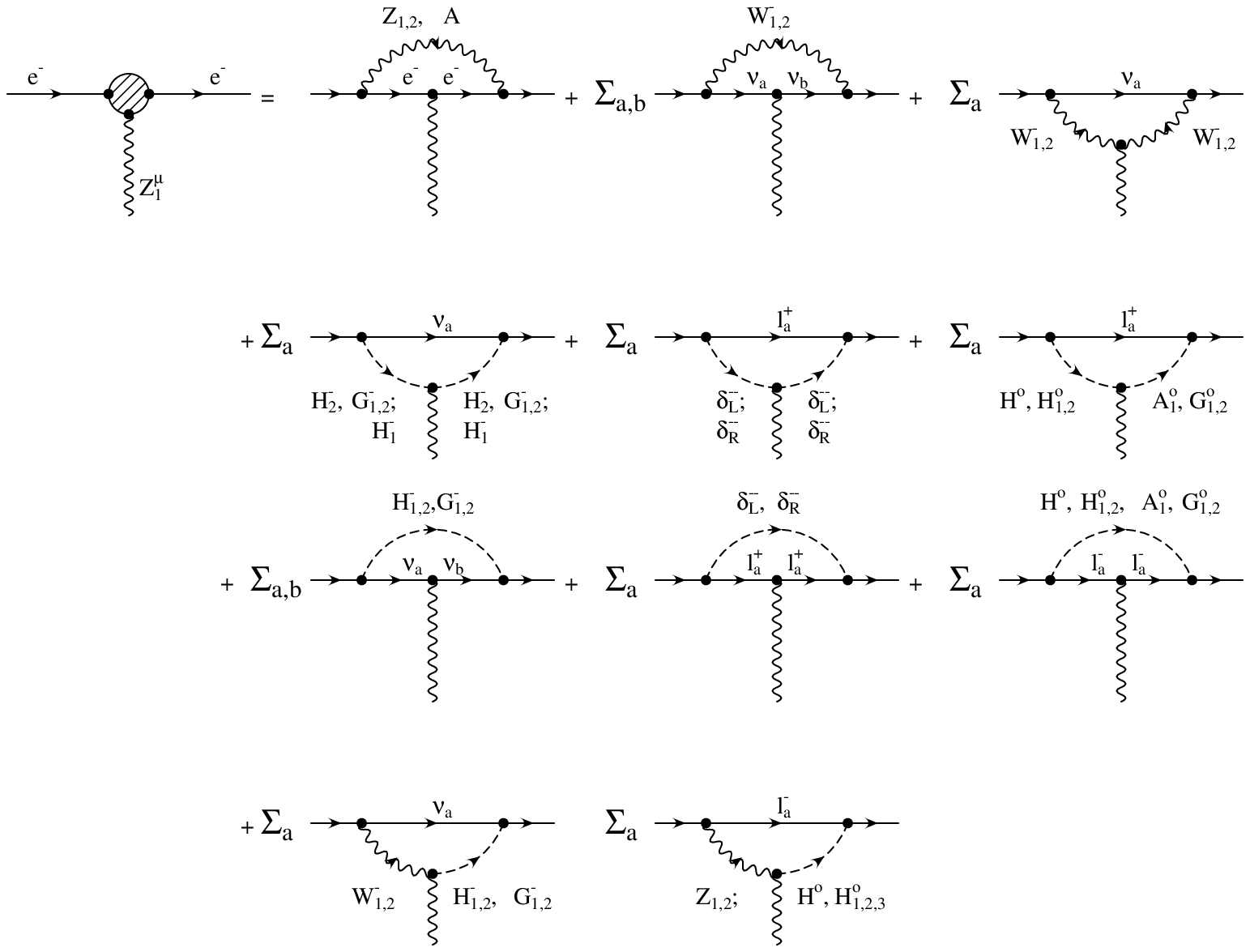,width=16cm}
\end{figure}

{\small Fig.6 \newline
The one-loop diagrams for the $Z_1ee$ vertex. }

\section*{ 5.3 Infinite part of the $Z_1 N_i N_j $ vertex and renormalization.}

%In the previous section the calculation of all renormalization constants
%necessary for renormalization of the $Z_1N_iN_j$ vertex has been performed.
%Now we can proceed to the main part of the work, i.e. to the check of the
%infinite part's cancellation of the $Z_1N_iN_j$ vertex.
%To prove it, one must show that the infinite part of $\Gamma^\mu_{Z_1N_iN_j}$
%(see Eq.(\ref{gzij})) equals zero.
As it was already mentioned in Chapter 5.1, the symmetry between
the left and the right part of the Lagrangian allows us to check
the cancellation of the infinite part only for one (left or right)
component of the $\Gamma^\mu_{Z_1N_iN_j}$. For further
consideration the left part has been chosen. Therefore, the
following equation had to be verified (see
Eq.(\ref{gzij},\ref{gzij1}))

\newpage

\begin{eqnarray}
\label{gctl1}
\left[
\left( \Gamma^\mu_{Z_1N_iN_j}\right)_L
\right]_{infinite \;part}&=&
ie\gamma^\mu
\left[
\left(
\Gamma_{CT}^{Z_1N_iN_j}
\right)_L
+
\left(
\Gamma_{D}^{Z_1N_iN_j}
\right)_L
\right]_{infinite \;part} \nonumber \\
&=&
ie\gamma^\mu
\left[
PL (CT)_L+
\left(
\Gamma_{D}^{Z_1N_iN_j}
\right)_L
\right]_{infinite\; part} =0.
\end{eqnarray}

The diagrams contributing to $\Gamma_{D}^{Z_1N_iN_j}$ are shown in Fig. 7.

\begin{figure}
\epsfig{file=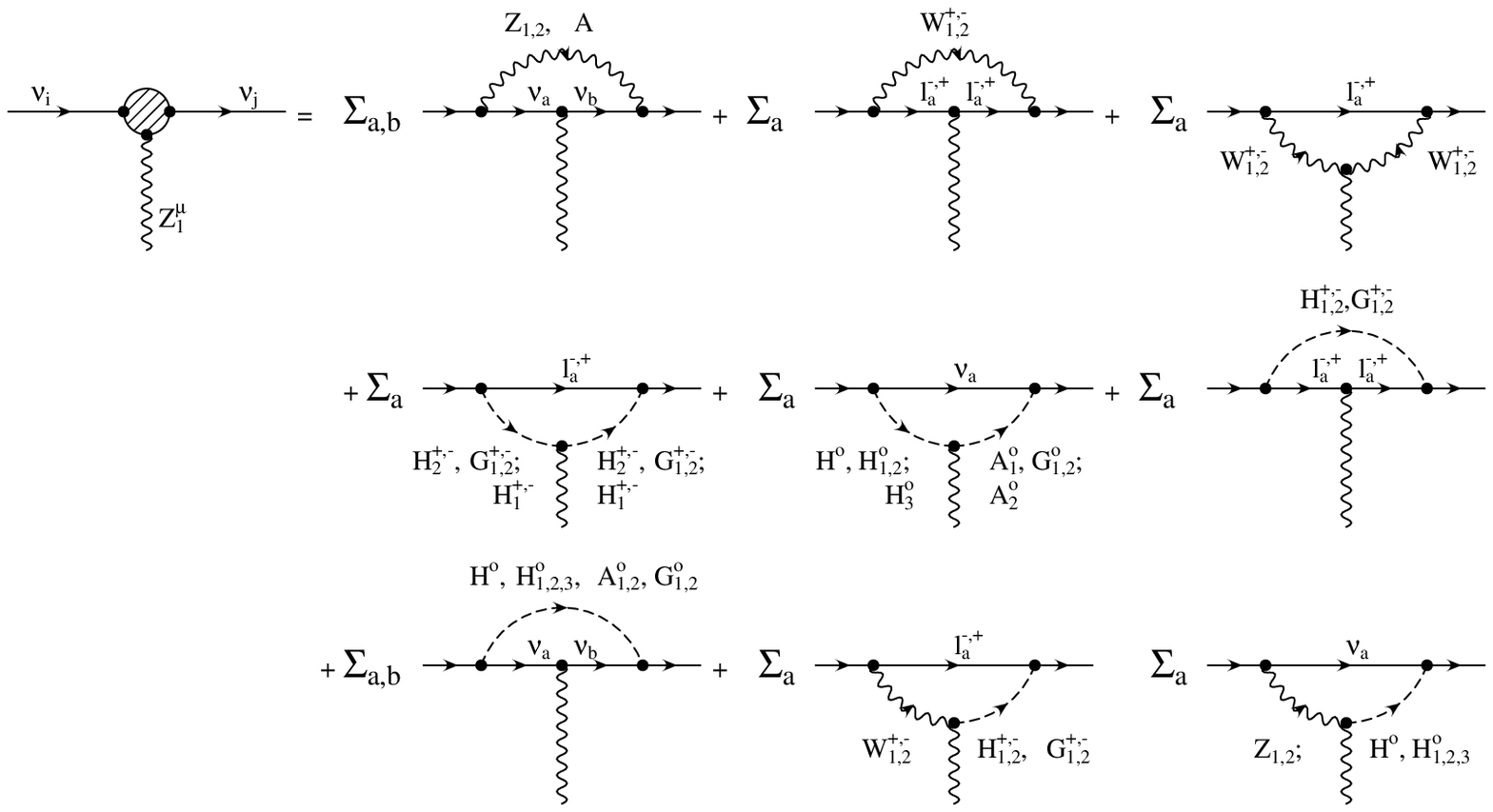,width=16cm}
\end{figure}

{\small Fig.7 \newline
The one-loop diagrams for the $Z_1N_iN_j$ vertex. } \newline
\vspace*{0.5cm}

As the first step of the calculations, the dimensional
regularization for all necessary two and three point functions has
been performed. Then, all needed couplings have been derived (see
Figs.1-7), and the infinite part of all necessary renormalization
constants and diagrams contributing to $Z_1N_iN_j$ vertex have
been calculated. Since neutrinos are Majorana particles, little
more complicated techniques must have been applied during the
calculation of the diagrams (see \cite{gl5}). Finally, a numerical
calculation (with the help of "Mathematica" program) has been
performed, proving the validity of Eq.(\ref{gctl1}). From the
technical side it is worth noting that the infinite part of the
renormalization constant for charge ($Y$) has been found to be
zero. Moreover, the calculations would not change if the
$Z^\frac{1}{2}_{Z_2Z_1}$ constant were held as a variable (i.e.
without putting its explicit value).

\newpage

\part*{VI Summary}

The on  mass shell scheme has been used to renormalize the manifest
left-right symmetric model. We have given general conditions to renormalize
the bulk of physical fields which enter the model. Also nonstandard
parameters ($\phi, \xi, \Omega_{L,R},K_{L,R}$),
connected with neutral and charged, left and right handed
currents have been discussed and renormalized. Consistency of the scheme
has been checked using
the $Z_1N_iN_j$ vertex (cancelations of infinities).

\section*{Acknowledgments}
This work was supported by Polish Committee for Scientific
Researches under Grants No. 2 P03B 033 15 and \\ 2 P03B 04215.
J.G. appreciates also the support of the Alexander von
Humboldt-Stiftung.

\end{document}